\documentclass{ws-rv9x6}
\usepackage{subfigure}   % required only when side-by-side / subfigures are used
\usepackage{ws-rv-thm}   % comment this line when `amsthm / theorem / ntheorem` package is used
\usepackage{ws-rv-van}   % numbered citation & references (default)
\makeindex
\begin{document}

\chapter[Two-dimensional systems of elongated particles]{Two-dimensional systems of elongated particles:\\ From diluted to dense}\label{chap:percandjam}

\author{Nikolai I. Lebovka}
\address{Department of Physical Chemistry of Disperse Minerals, F.~D.~Ovcharenko Institute of Biocolloidal Chemistry, NAS of Ukraine, Kyiv, Ukraine, 03142}
\address{Department of Physics, Taras Shevchenko Kyiv National University, Kyiv, Ukraine, 01033}

\author[N. I. Lebovka \& Y. Y. Tarasevich]{Yuri Yu. Tarasevich}
\address{Laboratory of Mathematical Modeling, Astrakhan State University, Astrakhan, Russia, 414056}

%\aindx{Lebovka, N.I.}
%\aindx{Tarasevich, Yu. Yu.}

\begin{abstract}
%This chapter is devoted to percolation and jamming of elongated particles on a plane. We consider both continuous and discrete space with the special attention to the square lattice. An overview of both methods and main results is presented.

This chapter is devoted to the analysis of jamming and percolation behavior of two-dimensional systems of elongated particles. We consider the problems both in continuous space and in discrete one (with the special attention to the square lattice), as well the systems with isotropically deposited and aligned particles. Overviews of different analytical and computational methods and main results are presented.

\end{abstract}
%\markright{Customized Running Head for Odd Page} % default is Chapter Title.

\body

\tableofcontents

\section{Introduction}\label{sec:intro}

During the past decades, the problems of particle packing have attracted growing both academic and practical interest. Systems composed of shape-anisotropic (elongated) particles are of special importance. The shape of the particles dictates the complex collective behavior, self-assembly and spontaneous orientational ordering\cite{Borzsonyi2013SM}. The phenomena including  jamming\cite{Behringer2018}, segregation\cite{Patra2018PRE,Tarasevich2017JSM}, and pattern formation\cite{Muller2014PRE} can be marked out in such systems.

In systems of interaction particles a \emph{jamming} transition refers the transition from some type of flowing (liquid-like) state to a stuck or rigid (solid-like) state. In contrast with typical crystallized solids, jammed systems are generally \emph{athermal}, \emph{i.e.}, there is no guarantee that this configuration corresponds to the equilibrium state with lowest energy of the system,  although a system may enter a jammed state at a given density\cite{Reichhardt2014SM}. Percolation is a kind of phase transition signaled by the emergence of a giant connected cluster of particles within a system under consideration\cite{Saberi2015PR}.

Our goal is consideration of different packing effects, jamming and percolation phenomena in the two-dimensional systems of elongated particles with different anisotropic shapes. Both continuous and discrete space problems (with special attention to the square lattices), and unoriented (isotropically oriented) and oriented along some preferable direction (anisotropically oriented) systems are analyzed.

\section{Some basic concepts and definitions}\label{sec:definitions}

A widely used method of modeling the monolayer adsorption of particles at a liquid---solid interface is \emph{random sequential adsorption} (RSA)\cite{Talbot2000CSA}. During RSA, particles are deposited randomly and sequentially onto a substrate, and their overlapping with previously placed particles is strictly forbidden, \emph{i.e.}, \emph{excluded volume} interaction is assumed. Excluded volume interaction (\emph{hard core interaction}, \emph{hard core exclusion}, \emph{hard core repulsion}) between the particles means that energy of interparticle interaction, $U_{i,j}$, is defined as
\begin{equation}\label{eq:HCI}
 U_{i,j} =
 \begin{cases}
 0, & \mbox{if } r_{i,j} > 0, \\
 \infty, & \mbox{if } r_{i,j} \leqslant 0,
 \end{cases}
\end{equation}
where $r_{i,j}$ is the shortest distance between $i$-th and $j$-th particles. Such the systems are treated as athermal.

Adhesion between deposited particles and the substrate is assumed to be very strong, so once deposited, a particle could not slip over the substrate or leave it (detachment is impossible). Adsorption of particles leads the deposit into a \emph{jammed state} where no additional particle can be added due to the absence of appropriate holes\cite{Evans1993RMP}. The corresponding concentration of particles is denoted as the jamming coverage. Noticeable, that RSA produces a non-equilibrium state. Diffusion of particles leads this state to the equilibrium (see, \emph{e.g.}, Refs.~\refcite{Scepanovic2013PhA,Grigera1997PLA,Eisenberg1998EPL,Fusco2001JCP,Fusco2002PhilMagB,Fusco2002PRE}).

Another kind of particle deposition is also possible. Particles are deposited randomly and sequentially, and their overlapping with previously placed particles is allowed.

Both continuous\cite{Feder1980JTB,Vigil1989JCP,Meakin1992PRA} and discrete space\cite{Kahlitz2012JCP,Pasinetti2019PRE} are used to simulate deposition of particles.
When a discrete space, \emph{e.g.}, a square lattice, is used, the fraction of occupied lattice sites, $p$, is a common quantity describing the system. When a deposition in a continuous space is considered, some other quantities are commonly used to characterize a deposit. The \emph{number density}, \emph{i.e.}, the number of particles, $N$, per unit area, $A$, is
\begin{equation}\label{eq:numdensity}
 n = \frac{N}{A}.
\end{equation}
Another useful quantity is the \emph{filling fraction},
\begin{equation}\label{eq:fillingfraction}
 \eta = n A_0,
\end{equation}
where $A_0$ is the area of one particle.
The total fraction of the plane covered by randomly deposited overlapping particles (coverage) is
\begin{equation}\label{eq:coveredfraction}
 \phi = 1 - \mathrm{e}^{-\eta}
\end{equation}
(see, \emph{e.g.}, Ref.~\refcite{Mertens2012PRE}).

\subsection{Common shapes of elongated particles in continuous-space simulations}\label{subsec:shape}
The shape of the particles can dictate their complex packing, percolation and phase behavior. Numerous experimental and simulation studies have been focused on studies of spherical and shape-anisotropic particles with both the convex and non-convex geometries\cite{Avendano2017COCIS}. For the convex geometry, the line segment connecting the two points inside the particle lies entirely inside this particle and for the non-convex geometry the situation is inverse.

\begin{figure}[!htb]
 \centering
 \subfigure[Ellipse.]{
 \includegraphics[height=0.10\textheight]{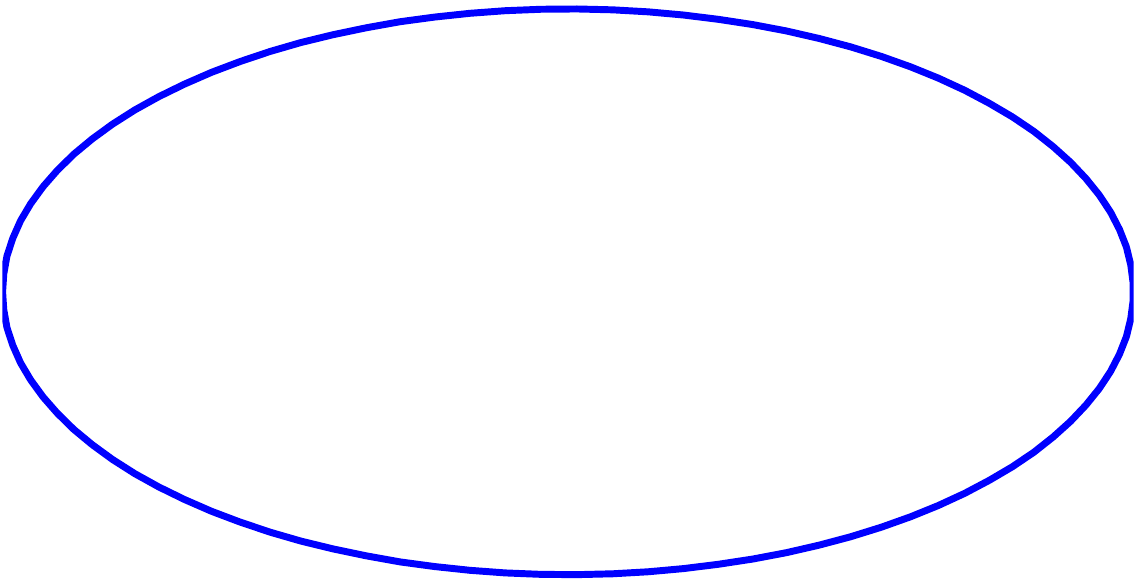}\label{fig:ellipse}}\hfill
 \subfigure[Superellipse, {$m = 3/2$}.]{
 \includegraphics[height=0.10\textheight]{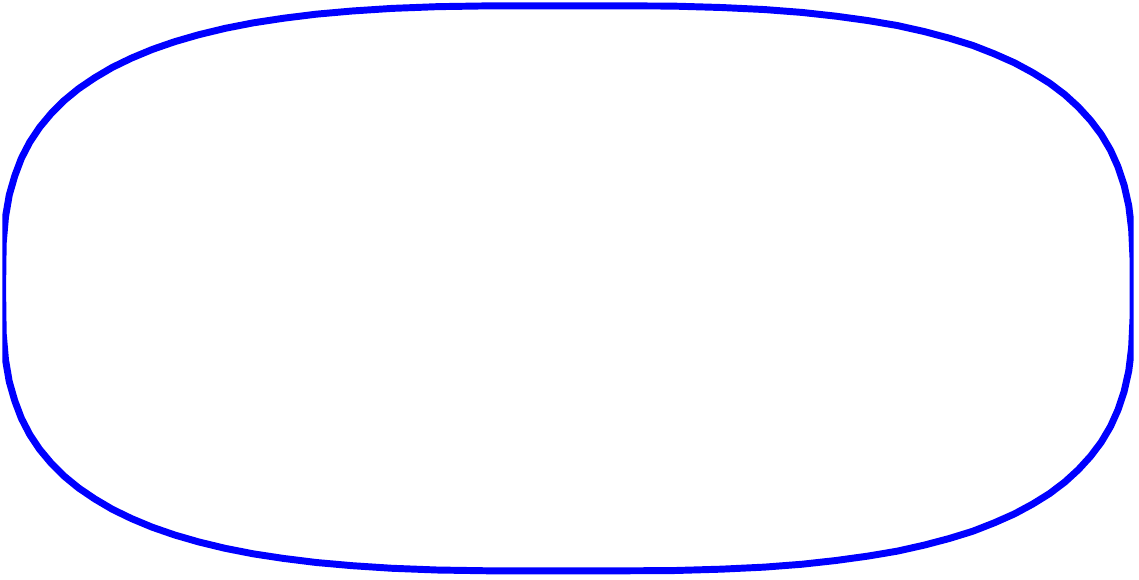}\label{fig:superellipse}}\hfill
 \subfigure[Stadium.]{
 \includegraphics[height=0.10\textheight]{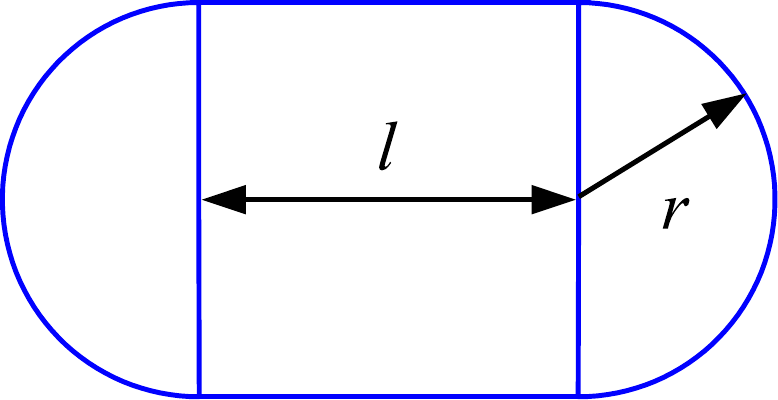}\label{fig:stadium}}
 \caption{Examples of some common shapes used in simulations to mimic elongated particles. In all three cases, the aspect ratio, \emph{i.e.}, the ratio of the longer dimension to the shorter one, is $\varepsilon = 2$.\label{fig:shapes}}
\end{figure}

The simplest convex shapes of $2d$ particle can be represented by square and by disk. To mimic the shape of elongated particles and, at the same time, simplify the simulations, different simple geometrical figures are used, \emph{e.g.}, sticks, rectangles, ellipses (\fref{fig:ellipse}), superellipses (\fref{fig:superellipse}), and stadia (\fref{fig:stadium}).
 In a Cartesian coordinate system, the equation of a superellipse (Lam\'{e} curve) is
\begin{equation}\label{eq:supereelips}
 \frac{|x|^{2m}}{a^{2m}} + \frac{|y|^{2m}}{b^{2m}} = 1,
\end{equation}
where $a$ and $b$ are the semimajor lengths in the direction of the $x$ and $y$ axes and $m$ is the shape parameter. $m=1$ corresponds to an ellipse while $m=+\infty$ corresponds to a rectangle.
A stadium is a rectangle with semicircles at a pair of opposite sides (\fref{fig:stadium}).
%\begin{equation}\label{eq:asperctratio}
% \varepsilon = 1 + \frac{l}{2r}.
%\end{equation}
A stadium (or ``discorectangle'') is a two-dimensional analog of a spherocylinder (a ``stadium of revolution'' or ``capsule''), \emph{i.e.}, a three-dimensional geometric shape consisting of a cylinder with hemispherical ends (\fref{fig:stadium}).

\begin{figure}[!htb]
 \centering
  \subfigure[Dimer, $\varepsilon = 1 + x$.]{
 \includegraphics[height=0.13\textheight]{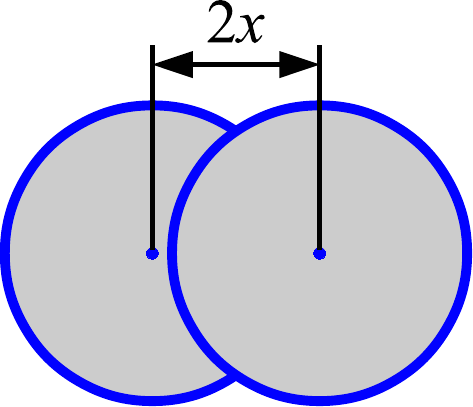}\label{subfig:dimer}}\quad
   \subfigure[Trimer, $\varepsilon = 1 + 2x$.]{
 \includegraphics[height=0.13\textheight]{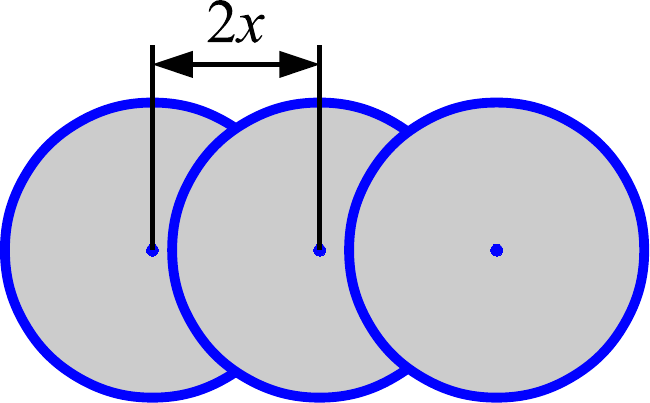}\label{subfig:trimer}}
 \caption{Examples of polymers. Disks have a unit radius, and parameter $x \in [0,1]$ corresponds to half the distance between closest disks\cite{Ciesla2015PCCP}.\label{fig:chain}}
\end{figure}

The linear polymers (dimer, trimer, etc.) are built of two, three, and more overlapping disks of
the same radius\cite{Ciesla2015PCCP,Ciesla2016JCP} (\fref{fig:chain}).
%All shapes, present in Fig. xx have the same length-to-width ratio $\varepsilon=L/d =a/b=2$, named as aspect ratio.

%======================================
\subsection{Anisotropy}\label{subsec:anisotropy}

Both isotropic and anisotropic deposition of particles has been studied. To characterize the anisotropy, different quantities are used. In the case of continuous space, the mean \emph{order parameter} is calculated as
\begin{equation}\label{eq:Scont}
 s = \frac 1 N\sum\limits_{i=1}^N \cos 2 \theta_i= 2\langle \cos^2\theta \rangle - 1,
\end{equation}
where $\theta_i$ is the angle between the axis of the $i$-th rod and the horizontal axis $x$, and $N$ is the total number of rods in the system (see, \emph{e.g.}, Ref.~\refcite{Frenkel1985}).
In a particular case of discrete space, \emph{viz.}, a square lattice, only two orientations of particles are allowed ($\theta =0$ and $\pi/2$). Hence, the order parameter is
\begin{equation}\label{eq:Sdiscr}
 s = \frac{N_y - N_x}{N_y + N_x},
\end{equation}
where $N_x$ ($N_y$) is the number of sites belonging to the horizontal (vertical) oriented particles.

The \emph{macroscopic anisotropy} of the system may be characterized by the value
\begin{equation}\label{eq:Anisotropy}
\mathcal{A} = \frac{P_{\parallel}}{P_{\perp}} = \frac{\langle |\cos\theta| \rangle}{\langle |\sin\theta| \rangle}.
\end{equation}
For the isotropic case, $\mathcal{A} = 1$, whereas for the highly anisotropic case, $\mathcal{A} \rightarrow \infty$\cite{Pike1974,Balberg1982a}. When the particles may be of different length, this quantity transforms into\cite{Balberg1983PRB,Mietta2014JCP}
\begin{equation}\label{eq:Anisotropyl}
\mathcal{A} = \frac{\langle l|\cos\theta| \rangle}{\langle l|\sin\theta| \rangle}.
\end{equation}

\begin{figure}[!htb]
 \centering
 \includegraphics[width=0.75\textwidth]{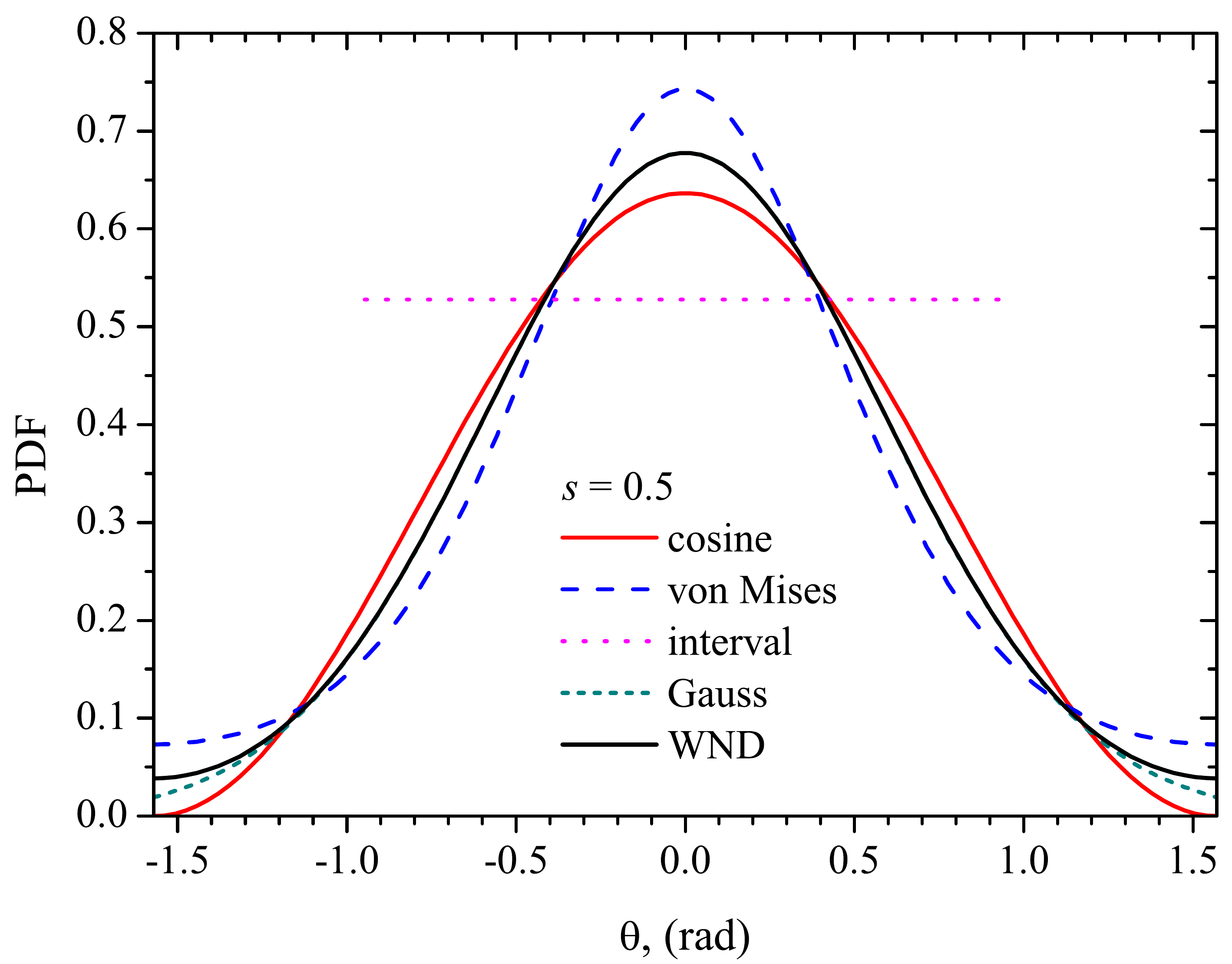}
 \caption{Different PDFs for $s=0.5$. Cosine \eref{eq:PDFcos}, von Mises \eref{eq:PDFvonMises}, interval \eref{eq:PDFinterval}, Gauss \eref{eq:PDFnormal}, WND \eref{eq:PDFwnd}.\label{fig:distributionss05}}
\end{figure}

Several ways have been used to prepare anisotropic samples. We start with the case of continuous space. The simplest way is to restrict the allowed orientation of particles\cite{Balberg1983PRB} as
\begin{equation}\label{eq:uniformangdistr}
-\theta_\text{m} \leqslant \theta \leqslant \theta_\text{m}.
\end{equation}
This distribution is uniform with the probability density function (PDF)
\begin{equation}\label{eq:PDFinterval}
f(\theta) = \frac{1}{2\theta_\text{m}}.
\end{equation}
In this case, \eref{eq:Anisotropy} transforms into
\begin{equation}\label{eq:Ainterval}
\mathcal{A} = \frac{\sin \theta_\text{m}}{1-\cos \theta_\text{m}}
\end{equation}
while the order parameter is
\begin{equation}\label{eq:sinterval}
s=\frac{\sin 2{\theta_\text{m}}}{2{\theta_\text{m}}}.
\end{equation}
Another kind of the PDF has been used in Ref.~\refcite{Horrmann2014AAM8,Klatt2017JSMTE}
\begin{equation}\label{eq:PDFcos}
f_\theta(\theta|\beta) = \frac{\Gamma\left( \frac{\beta}{2} + 1 \right)}{\sqrt{\pi} \Gamma \left( \frac{\beta + 1}{2}\right)} \cos^\beta \theta, \quad \theta \in \left[-\frac{\pi}{2}, \frac{\pi}{2}\right).
\end{equation}
In this case, the order parameter is
\begin{equation}\label{eq:scos}
s = \frac{\beta}{\beta + 2}.
\end{equation}
The parameter $\beta$ controls the degree of anisotropy of the system. $\beta = 0$ corresponds to a uniform distribution \eref{eq:PDFinterval} with $\theta_\text{m} = \pi$ and hence to an isotropic system ($s = 0$). The larger the value of $\beta$ the stronger the anisotropy. $\beta= \infty$ corresponds to a full alignment of the sticks along the $x$-axis ($s = 1$).
In Ref.~\refcite{Tomiyama2019340}, von Mises distribution has been used
\begin{equation}\label{eq:PDFvonMises}
 f_{\theta}(\theta |\kappa )=\frac{\exp \left( \kappa \cos 2\theta \right)}{\pi {I_0}(\kappa )}, \quad \theta \in \left[ -\frac{\pi }{2},\frac{\pi }{2} \right),
\end{equation}
where $I_n$ is the modified Bessel functions of $n$-th order.
In this case, the order parameter is
\begin{equation}\label{eq:svonMises}
s(\kappa)= \frac {I_1(\kappa )}{ I_0(\kappa )} = \frac {I_0^\prime(\kappa )}{I_0(\kappa )}.
\end{equation}
In  Refs.~\refcite{Tarasevich2018JAP,Lebovka2018PRE,Tarasevich2018PREanisotropy,Tarasevich2018PREsticks}, normal (Gauss) distribution has been used
\begin{equation}\label{eq:PDFnormal}
 f_\theta (\theta | \sigma )=\frac{1}{\sqrt{2\pi {{\sigma }^2}}}\exp \left( -\frac{{{\theta }^2}}{2{{\sigma }^2}} \right), \quad
 \theta \in (-\infty ,+\infty ).
\end{equation}
In this case, the order parameter is
\begin{equation}\label{eq:snormal}
s = \exp(-2\sigma^2).
\end{equation}
In fact, all above distributions can be treated as approximations of the wrapped normal distribution (WND)
\begin{equation}\label{eq:PDFwnd}
f_\theta(\theta | \sigma )=\frac{1}{\sqrt{2\pi {{\sigma }^2}}}\sum\limits_{k=-\infty }^{\infty }{\exp }\left[ \frac{-(\theta +2\pi k)^2 }{2\sigma^2} \right], \quad \theta \in [-\pi ,\pi ).
\end{equation}
\Fref{fig:distributionss05} demonstrate PDFs for one particular value of the order parameter.

In the case of discrete space (a square lattice), basically, an anisotropic deposition means deposition of particles that are strictly aligned along one direction, \emph{i.e.}, $s=1$\cite{Longone2012PRE,Vogel2017,Kundu2018PRE}. Deposit of arbitrary anisotropy can be produced using different methods\cite{Lebovka2011PRE}.

\begin{figure}
  \centering
  \subfigure[Permeable particles]{
  \includegraphics[width=0.3\textwidth]{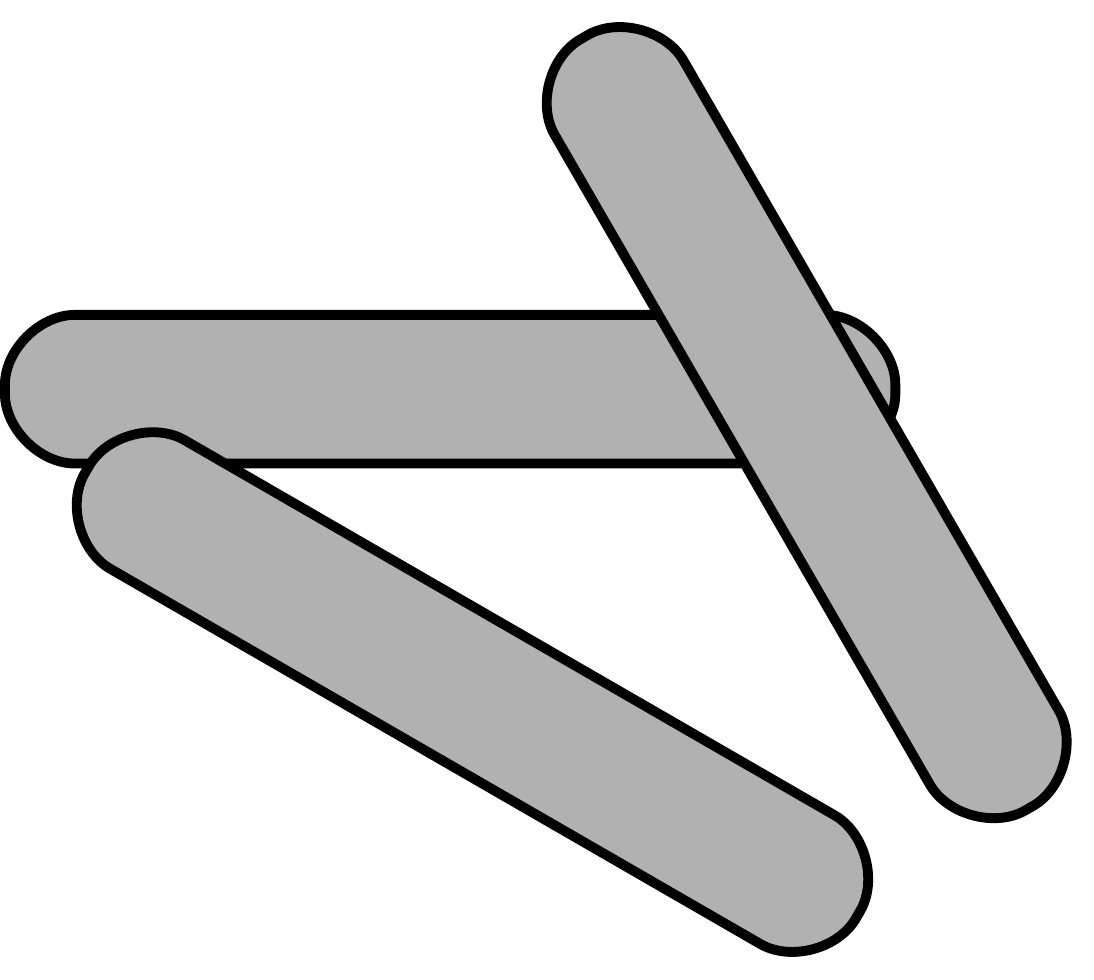}\label{fig:clustera}}\hfill
  \subfigure[Impermeable particles]{
  \includegraphics[width=0.35\textwidth]{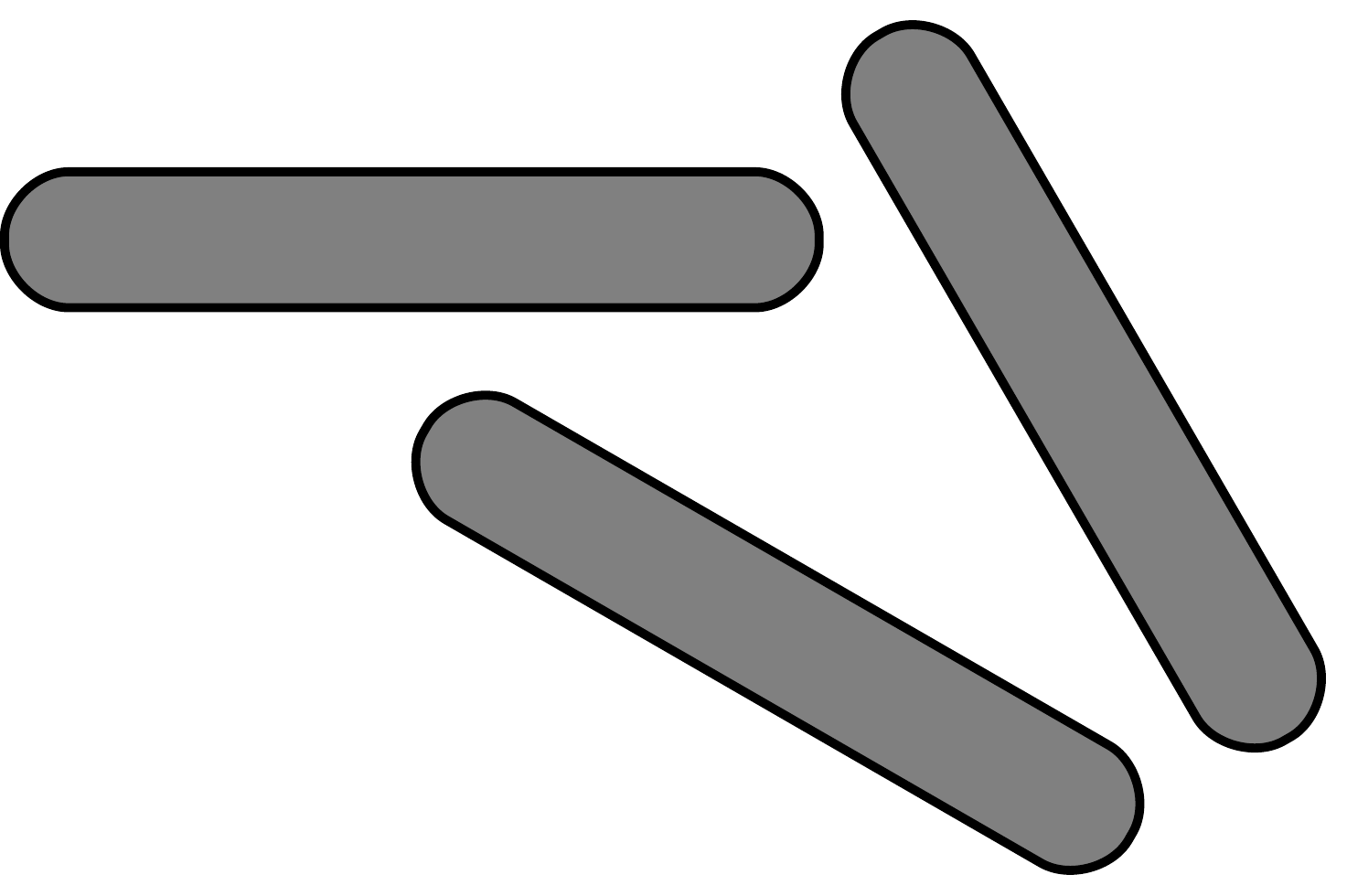}\label{fig:clusterb}}\hfill
  \subfigure[Hard--core--soft--shell particles]{
  \includegraphics[width=0.3\textwidth]{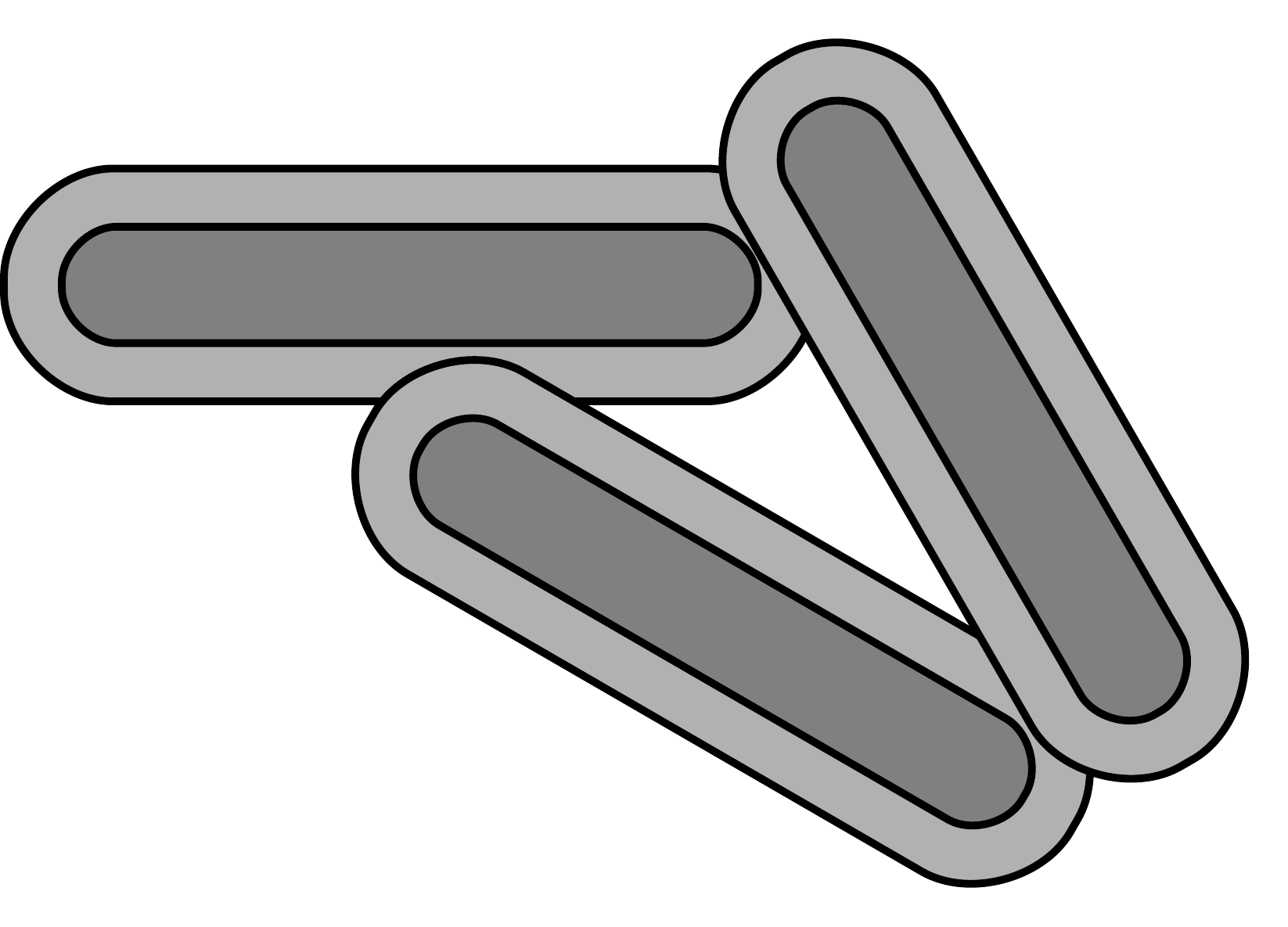}\label{fig:clusterc}}
  \caption{Permeable and hard--core--soft--shell particles can form a cluster while impermeable particles cannot.}\label{fig:clusters}
\end{figure}

When overlapping of particles is allowed, \emph{i.e.} particles are treated as permeable, overlapped particles form a cluster (\fref{fig:clustera}). When particles are impermeable, none cluster occurs (\fref{fig:clusterb}). Another possibility is so-called \emph{connectedness percolation} of non-overlapping particles\cite{Otten2011JCP,Drwenski2018JCP}. Two non-overlapping particles are assumed to be connected when the shortest distance between them does not a exceed a certain value, \emph{i.e.}, so-called cutoff distance\cite{Lee1988JCP}. This case can be also treated as a hard--core--soft--shell model (\fref{fig:clusterc}). In discrete space, when deposit was produced using RSA mechanism, adjoining particles are assumed to belong to a cluster. Deposited particles may form a spanning cluster, \emph{i.e.}, a set of neighboring particles which span between the opposite borders of the substrate. The occurrence of spanning clusters is associated with the concentration phase transition, where the corresponding concentration of particles is denoted as the percolation threshold. Since the middle of the last century, percolation theory has been applied to simulate disordered media and understand their physical properties\cite{Stauffer,Sahimi1994}. Some physical properties of the disordered media are fundamentally different below and above the \emph{percolation threshold}. Percolation theory is simple, yet productive, description of phase transitions.

\subsection{Excluded area}\label{subsec:exarea}

\begin{figure}
  \centering
  \subfigure[Rectangle.]{
    \includegraphics[width=0.32\textwidth]{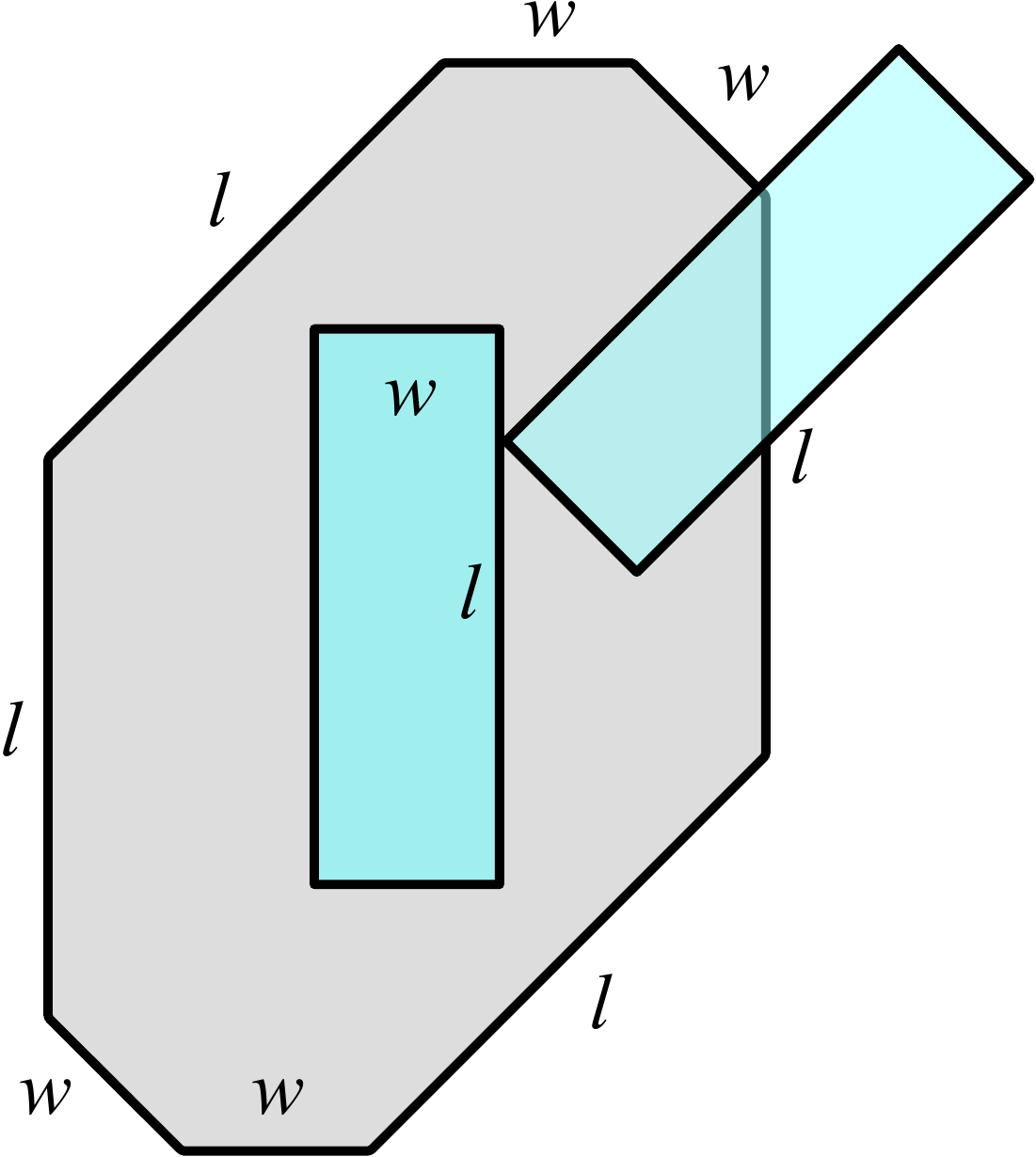}\label{fig:EArectangle}}\hfill
  \subfigure[Stadium.]{
  \includegraphics[width=0.33\textwidth]{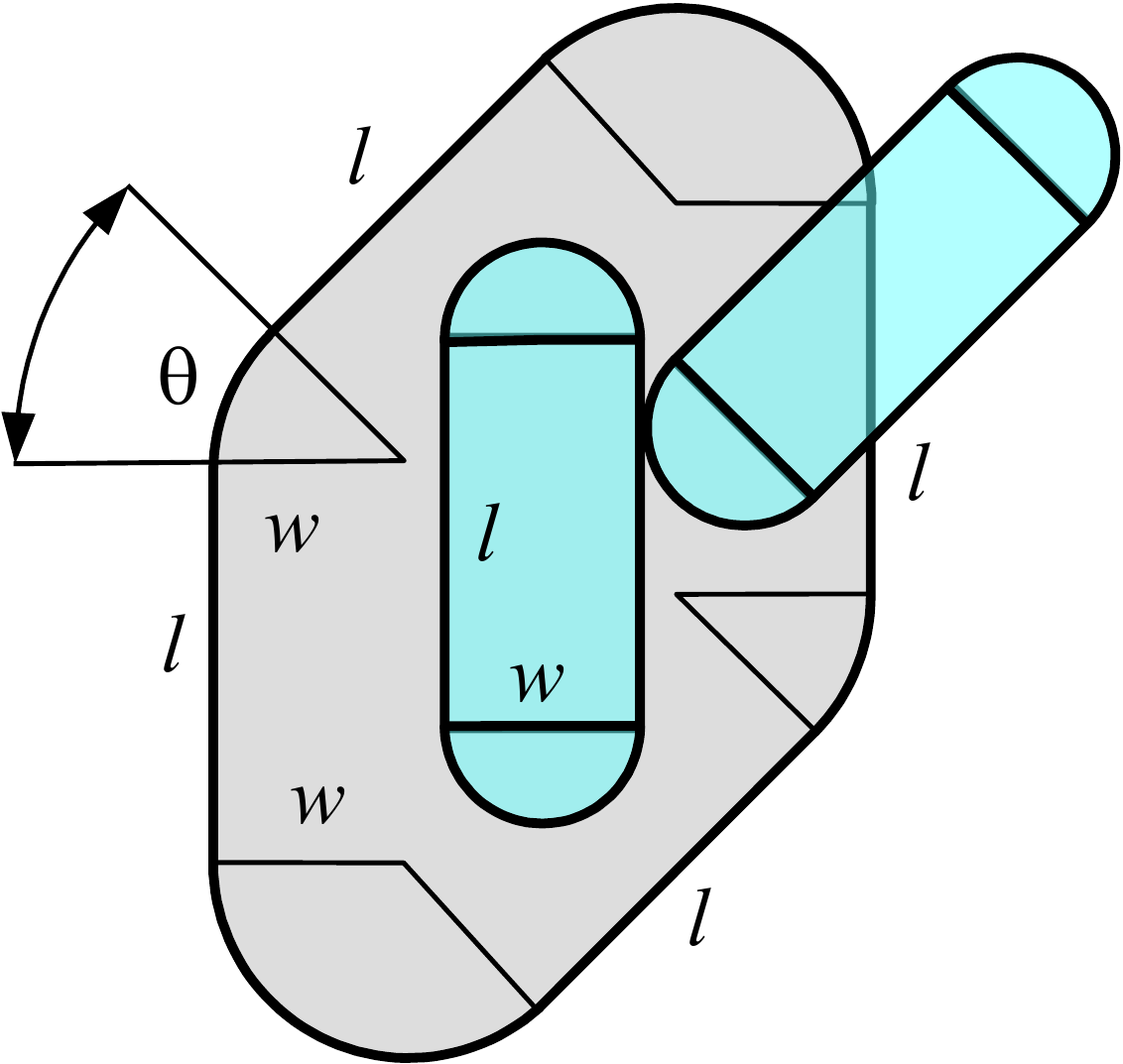}\label{fig:EAstadium}}\hfill
    \subfigure[Ellipse.]{
  \includegraphics[width=0.26\textwidth]{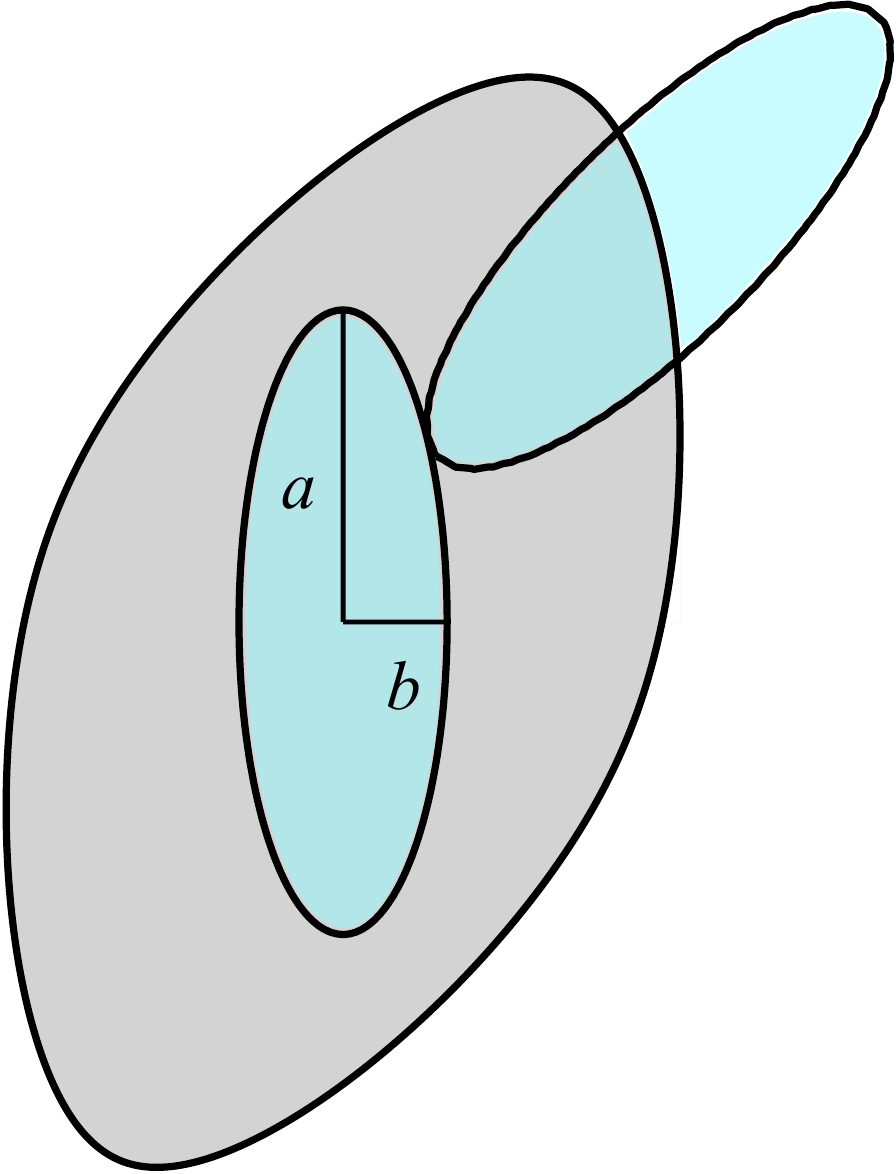}
  \label{fig:EAellipse}}\hfill
  \caption{Examples of excluded areas\cite{Balberg1984PRB,Zheng2007PRE}.}\label{fig:EA}
\end{figure}

When overlapping of objects is not allowed, the \emph{excluded area} (in 2D) or \emph{excluded volume} (in 3D) of an object is defined as the area (volume) around this object in which the center of another similar object cannot be located\cite{Onsager1949ANYAS} (\fref{fig:EA}). Generally speaking, the excluded volume depends on the relative orientation of the particles except the trivial case of disks (spheres). The concept was used by L.~Onsager in his analysis of effects of the shapes of colloidal particles on phase state of suspensions\cite{Onsager1949ANYAS}. It is useful to study packing and phase diagrams problems of two-dimensional particles. The excluded area is denoted as $A_\text{ex}$. Although the mathematical background of the excluded area computation is simple and clear\cite{Kihara1953RMP}, even  for the simplest objects, \emph{e.g.}, ellipses,  finding the excluded area can be a daunting task\cite{Zheng2007PRE,Martinez-Raton2011LC}.

%This area is originated from the hard-core repulsion between the two particles. The concept was used by L.~Onsager in his analysis of effects of the shapes of colloidal particles on phase state of suspensions\cite{Onsager1949ANYAS}. It is useful to study packing, percolation, and phase diagrams problems of two-dimensional particles it is useful to introduce concept of excluded area denoted as $A_\text{ex}$.
%The presence of one hard-core particle creates some surrounding area from which the second particle is excluded. The excluded area is defined as the area around the first particle that is excluded for the centre of mass of another one.

The normalized excluded area, $A_\text{ex}/\ell^2$, where $\ell$ is the largest dimension of the particle, depends only on the aspect ratio, $\varepsilon$, and inter-particle angle, $\theta$.  They were evaluated analytically for different variety of shapes and they are
\begin{equation}\label{eq:Aexrect}
A_\text{ex}/{\ell^2}=\left| \sin \theta  \right|(1 + \varepsilon^{-2})+2\varepsilon^{-1}(1+\left| \cos \theta  \right|)
\end{equation}
for the rectangles\cite{Balberg1984PRB},
\begin{equation}\label{eq:Aexstad}
A_\text{ex}/{\ell^2}=4\varepsilon^{-1} \left(1-\varepsilon^{-1}\right) + \pi \varepsilon^{-2} +{{\left(1-\varepsilon^{-1}\right)}^2}\left| \sin \theta  \right|.
\end{equation}
for the stadia\cite{Balberg1984PRB}, and
\begin{equation}\label{eq:Aexell}
A_\text{ex}/{\ell^2}=\left[ \pi \varepsilon^{-1} +\left( {{\Delta }_{+}}(\theta )+{{\Delta }_{-}}(\theta ) \right)E(\theta ) \right]/2,
\end{equation}
for the ellipses\cite{Martinez-Raton2011LC}. Here,
$$
E(\theta )=\int\limits_{0}^{\pi /2}{\sqrt{1-\kappa (\theta ){{\sin }^2}t} \,\mathrm{d}t}
$$
is the complete elliptic integral of the second kind,
$$
\kappa (\theta )=4\frac{\Delta_+\Delta_-}{(\Delta_+ + \Delta_-)^2},
$$
 and
$$\Delta_\pm = \sqrt{\left(1 \pm \varepsilon^{-2}\right)^2 - \left(1 - \varepsilon^{-2}\right)^2 \cos^2 \theta}.$$

\begin{figure}
  \centering
  \subfigure[Rectangle.]{
  \includegraphics[width=0.48\textwidth]{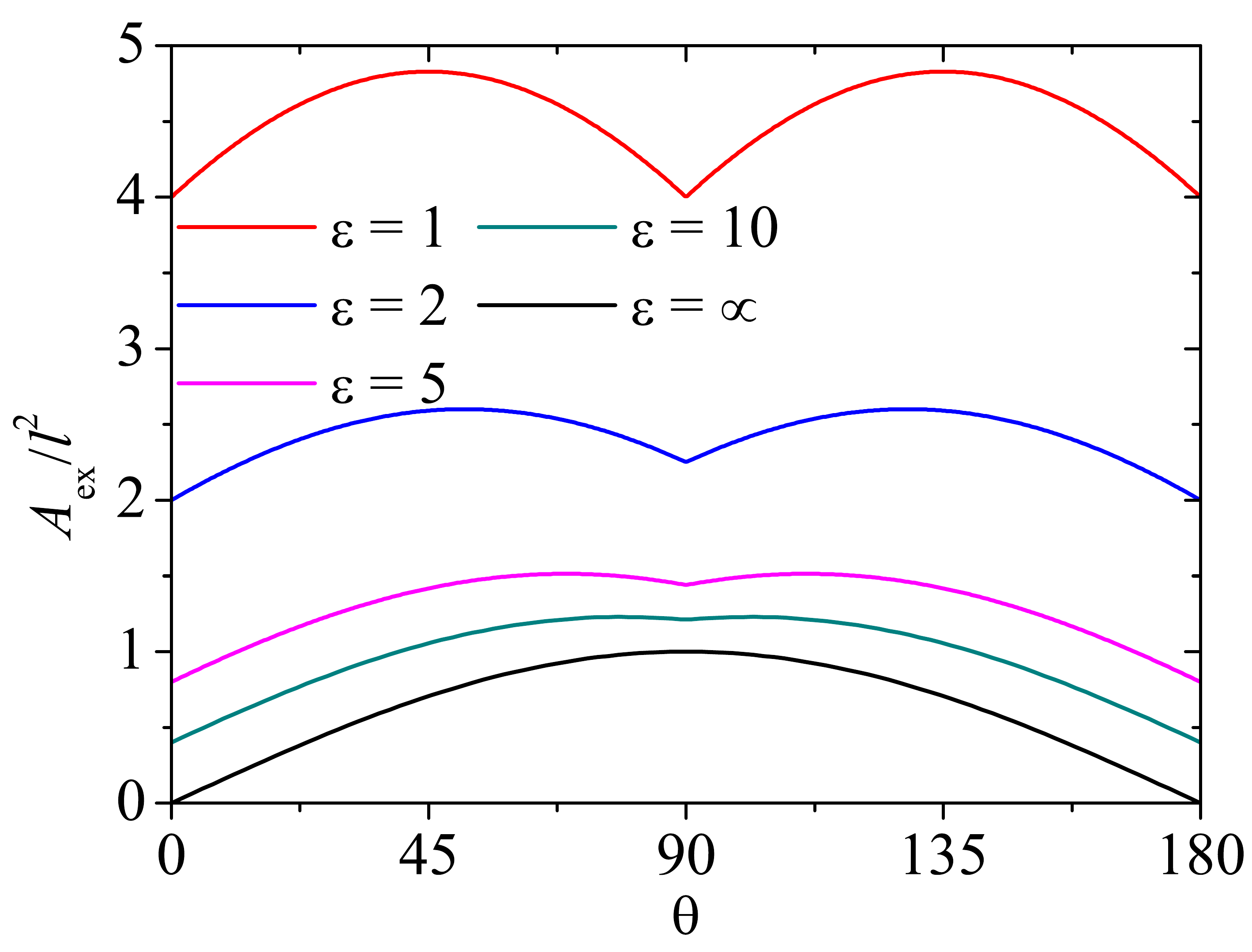}\label{fig:agularrect}}\hfill
  \subfigure[Stadium.]{
  \includegraphics[width=0.48\textwidth]{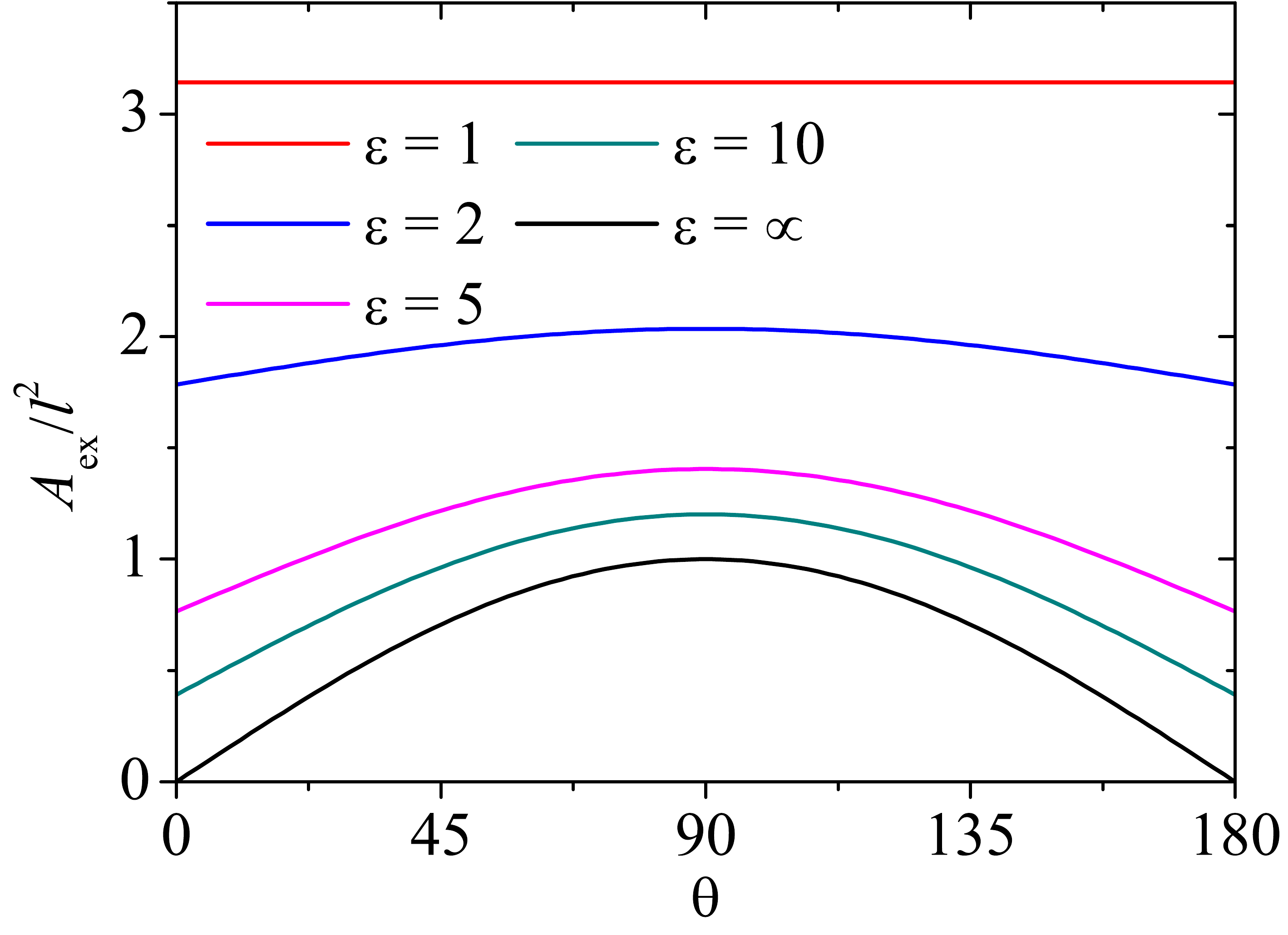}\label{fig:agulardiscorect}}
  \caption{Angular dependencies of the normalized excluded areas, $A_\text{ex}(\theta)/\ell^2$,  for different values of the aspect ratio, $\varepsilon$. }\label{fig:angular}
\end{figure}

Example of $A_\text{ex}(\theta)/\ell^2$ dependencies for the rectangles and stadia are presented in \fref{fig:angular}. Both these functions exhibit the minimums at $\theta=0$ and $\pi$. However, for rectangles, the function $A_\text{ex}(\theta)/\ell^2$ has secondary minimum at $\theta=\pi/2$ at sufficiently small aspect ratio ($\varepsilon < 10$). The appearance of the tetratic phases for the rectangle systems is related with this minimum.

The formulas for oriented rectangles and stadia were also evaluated. For uniform distribution of particles with PDF~\eref{eq:PDFinterval}, the following equations were obtained\cite{Balberg1984PRB}
\begin{equation}\label{eq:Aexaverrect}
  \frac{\langle A_\text{ex}\rangle_{\theta_\text{m}}}{\ell^2}=2\varepsilon^{-1}\left( 1 + \frac{1-\cos 2\theta_\text{m}}{2\theta_\text{m}^2} \right) + ( 1 + \varepsilon^{-2})\frac{4\theta_\text{m}-2\sin 2\theta_\text{m}}{4\theta_\text{m}^2}
\end{equation}
for rectangles,
\begin{equation}\label{eq:Aexaverstad}
\frac{\langle A_\text{ex} \rangle_{\theta_\text{m}}}{\ell^2}=4\varepsilon^{-1} \left(1 - \varepsilon^{-1}\right)+\pi \varepsilon^{-2}+{{(1-\varepsilon^{-1})}^2}\frac{ 4{\theta_\text{m}}-2\sin (2{\theta_\text{m}})}{4{\theta_\text{m}}^2}
\end{equation}
for stadia.

Particularly, for fully oriented particles, $\theta_\text{m}=0$, these formulas give $\langle A_\text{ex}\rangle/{\ell^2}=2\varepsilon^{-1}$ for rectangles, and $\langle A_\text{ex}\rangle/{\ell^2}=4\left(1 - \varepsilon^{-1}\right)\varepsilon^{-1} + \pi \varepsilon^{-2}$ for stadia.

In isotropic case ($\theta_\text{m}=\pi/2$), the orientational averages of $\langle A_\text{ex} \rangle/\ell^2$ over all allowed relative orientations are
\begin{equation}\label{eq:Aexisotrrect}
  \langle A_\text{ex} \rangle/{\ell^2}=2\varepsilon^{-1}(1+4\pi^{-2}) + \left(1 + \varepsilon^{-2}\right)2\pi^{-1}
\end{equation}
for rectangles\cite{Balberg1984PRB},
\begin{equation}\label{eq:Aexisotrstad}
\langle A_\text{ex} \rangle/\ell^2=4 \varepsilon^{-1} \left(1 - \varepsilon^{-1}\right) + \pi \varepsilon^{-2}+2{{\left(1 - \varepsilon^{-1}\right)}^2}\pi^{-1} \end{equation}
for stadia\cite{Balberg1984PRB}, and
\begin{equation}\label{eq:Aexisotrell}
\langle A_\text{ex} \rangle/\ell^2=\pi \left[ 1+\zeta  \right]/(2\varepsilon),
\end{equation}
for ellipses\cite{Xia1988PRA,Li2016PhysA}.
Here,
\begin{equation}\label{eq:zeta}
  \zeta=\frac{C^2}{4\pi A},
\end{equation}
is the shape factor\cite{Ciesla2016JCP}, where $C$ and $A =\pi a b = \pi \ell^2/(4\varepsilon)$ are the ellipse perimeter and area, respectively.
The ellipse perimeter can be evaluated using the following formula (Chandrupatla \& Osler, 2010):
$P=(2\ell/\varepsilon)\int\limits_{0}^{\pi /2}{\sqrt{1-(1-{\varepsilon^2}){{\sin }^2}t}\, \mathrm{d} t}$. 	

Note that for the circle (disk), the shape factor is $\zeta=1$ and $ A_\text{ex}(\theta)/\ell^2 = \pi$.

\begin{figure}
  \centering
  \includegraphics[width=0.75\textwidth]{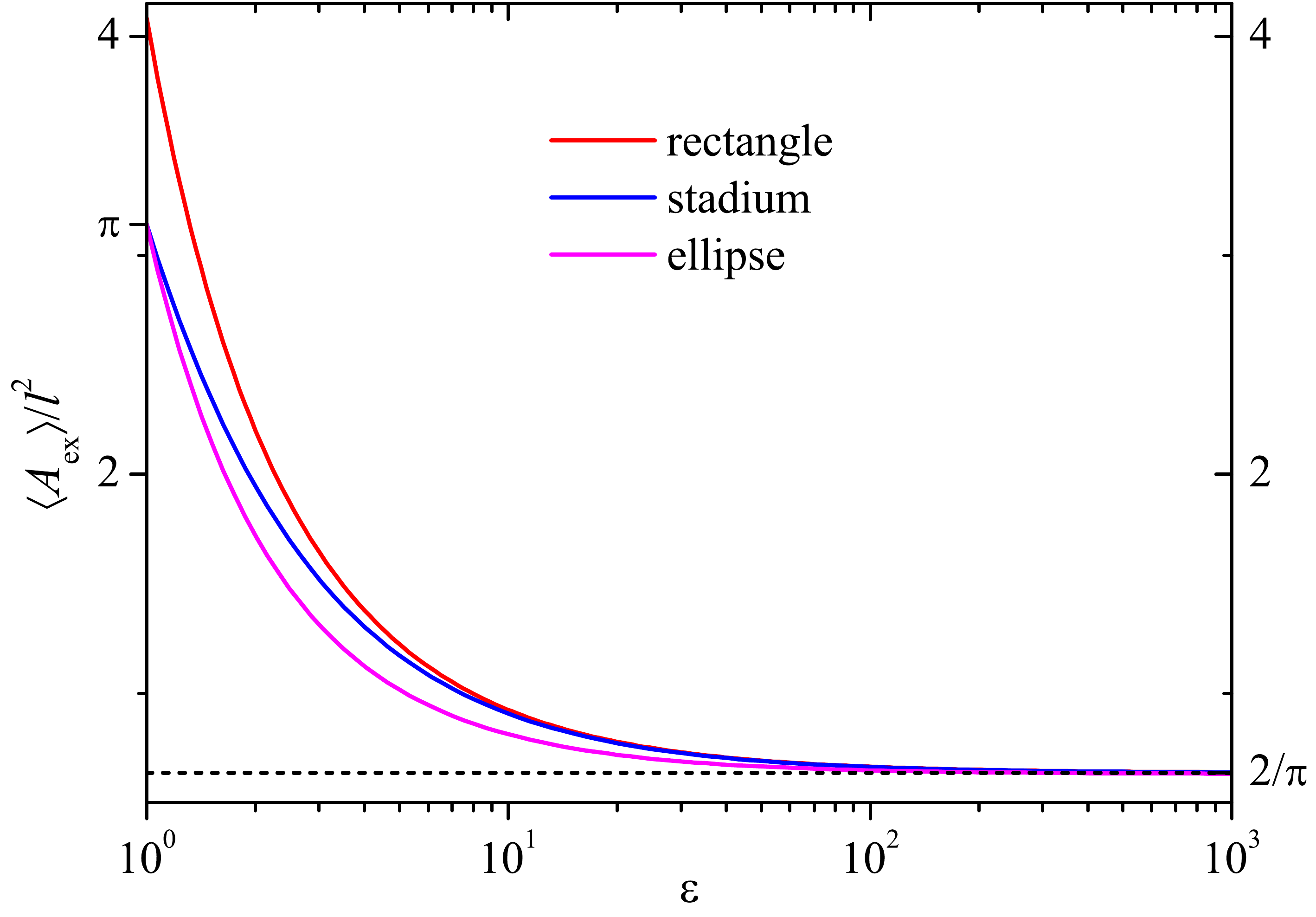}
  \caption{Averaged normalized excluded area $\langle A_\text{ex}(\theta)/\ell^2 \rangle$ versus the aspect ratio, $\varepsilon$, for isotropically deposited rectangles~\eref{eq:Aexisotrrect}, stadia~\eref{eq:Aexisotrstad}, and ellipses~\eref{eq:Aexisotrell}. }\label{fig:Aexaver}
\end{figure}

\Fref{fig:Aexaver} presents the averaged normalized excluded area $\langle A_\text{ex}(\theta)/\ell^2 \rangle$ versus the aspect ratio, $\varepsilon$, for the rectangles, stadia, and ellipses. At $\varepsilon=1$ the rectangle turns to the square ($\langle A_\text{ex}(\theta)/\ell^2 \rangle = 2\left(1+4\pi^{-2}+2\pi^{-1}\right) \approx 4.08$), where as the discorectangle and ellipse transform to the circle ($\langle A_\text{ex}(\theta)/\ell^2 \rangle = \pi$). In another limiting case of $\varepsilon \to \infty$, all shapes transform to the needles with  $\langle A_\text{ex}(\theta)/\ell^2 \rangle = \pi$.

%\subsection{The tessellation and Voronoi excluded area }
%The Voronoi statistics can be also sensitive to structural changes in the system. The Voronoi tessellation and Voronoi excluded surface were used for analysis random close packings of axisymmetric particles\cite{Baule2013NC}.

\section{Packings}\label{sec:packings}

The problems of hard-particle packings have been actively treated in considerations of structure of crystals liquids, glasses, colloids, amorphous and granular materials\cite{Torquato2010RMP,Baule2014SM}.  The most comprehensive information was obtained for packing problem of two-dimensional hard circular disks, and similar packings of hard spheres in $d$-dimensional Euclidean space ($d \geqslant 3$)\cite{Torquato2010PRE}.
Among the main two-dimensional packing models the saturated packing without direct contacts between particles and mechanically stable jammed packings with direct contacts between particles can be distinguished (see, \emph{e.g.}, Refs.~\refcite{Lubachevsky1990JSP,Torquato2010PRE,Ramos2018JCP}).

\subsection{Saturated packings}\label{subsec:saturpack}

In saturated packing, the particles cover the $2d$ surface without any overlapping. The most popular, and is also the simplest is the model of RSA model which accounts for excluded volume effects\cite{Evans1993RMP}. In this model the particles randomly (in the case of anisotropic shape the random particle position and orientation are selected) and irreversibly adsorb on a surface without overlapping with previously deposited particles. After being placed the particles are not able to move or reorient. With the coverage of the surface the adsorption probability decreases and above some limiting coverage $\phi_\text{j}$, known as the jamming or saturation limit, there is no free room for the deposition of a new incomer.

Historically, the RSA was proposed for the first time by Flory in the 1939 in his studies of the attachment of blocking pendant groups on a linear polymer\cite{Flory1939JACS}. Later, in the 1958 the analytical solution for RSA of unitary segments on a line (for so-called car parking problem) was obtained\cite{Renyi1958}.  For $1d$ RSA, the saturation density is $\phi_\text{j} = 0.7475979202 \dots$.

Nowadays, the different RSA models  are actively used for modelling of adsorption of colloids and proteins, Polyelectrolyte adsorption, and formation of nanoparticle monolayers\cite{Evans1993RMP,Adamczyk2012COCIS,Szilagyi2014SM,Morga2017PCCP,Talbot2000CSA,Cadilhe2007JPhysCM}.

\subsubsection{Continuous models}\label{subsec: Contmodels}
In continuous models, the positions and orientations of elongated particles can take real values. For $d \geqslant 2$, the RSA has been estimated through numerical simulations. The pioneering simulations of RSA of disks on flat surface gave for the jamming coverage $\phi_\text{j} = 0.547 \pm 0.002$\cite{Feder1980JTB}. The experimental estimations gave the values  $\phi_\text{j} = 0.625 \pm 0.100$ (adsorption of ferritin proteins on polycarbonate and carbon surfaces)\cite{Feder1980JCIS}, and $0.55 \pm 0.01$ (latex spheres on a surface). The values of $\phi_\text{j}$  for the RSA packing of hard hyperspheres in $d$-dimensional Euclidean space ($d=2-8$) were also numerically estimated\cite{Zhang2013PRE}.

The kinetic of jamming was approximated using the power equation, known as the Feder’s law\cite{Onoda1986PRA,Swendsen1981PRA,Hinrichsen1986}
$$\phi (t)={{\phi }_{j}}-c t^{-1/d},$$	
where $t$ is a dimensionless time, $c$ is a positive constant. Parameter $d$ is interpreted as a number of degrees of freedom of deposited particle.

For spherically symmetric particles the value of d is equal to the collector dimension, \emph{e.g.}, $d=2$ for the flat surface collector. This relation was proven to be correct for one- to eight-dimensional collectors\cite{Zhang2013PRE}. Typically for particles with anisotropic shape in two dimensions, $d \approx 3$\cite{Viot1992PA,Viot1990}. This power law was confirmed numerically for particles with different shapes\cite{Ciesla2013PRE,Shelke2007PRE,Viot1990} and in some cases it can be violated\cite{Baule2017PRL,Verma2018PA}.

For RSA model the value of $\phi_\text{j}$ is well defined, and it allows precise investigations of the influence of packing shape on the jamming coverage\cite{Torquato2010RMP}. The problems have also been investigated in details for oriented\cite{Brosilow1991PRA} and unoriented squares\cite{Viot1990}, rectangles, polygons\cite{Zhang2013PRE,Ciesla2014PRE}, stadia, and ellipses\cite{Vigil1989JCP,Vigil1990JCP,Sherwood1990JPA}, superdisks\cite{Gromenko2009PRE}, sphere dimers\cite{Ciesla2013SS,Ciesla2015PCCP}, and sphere polymers\cite{Ciesla2013PRE}.

\begin{figure}
  \centering
  \includegraphics[width=0.75\textwidth]{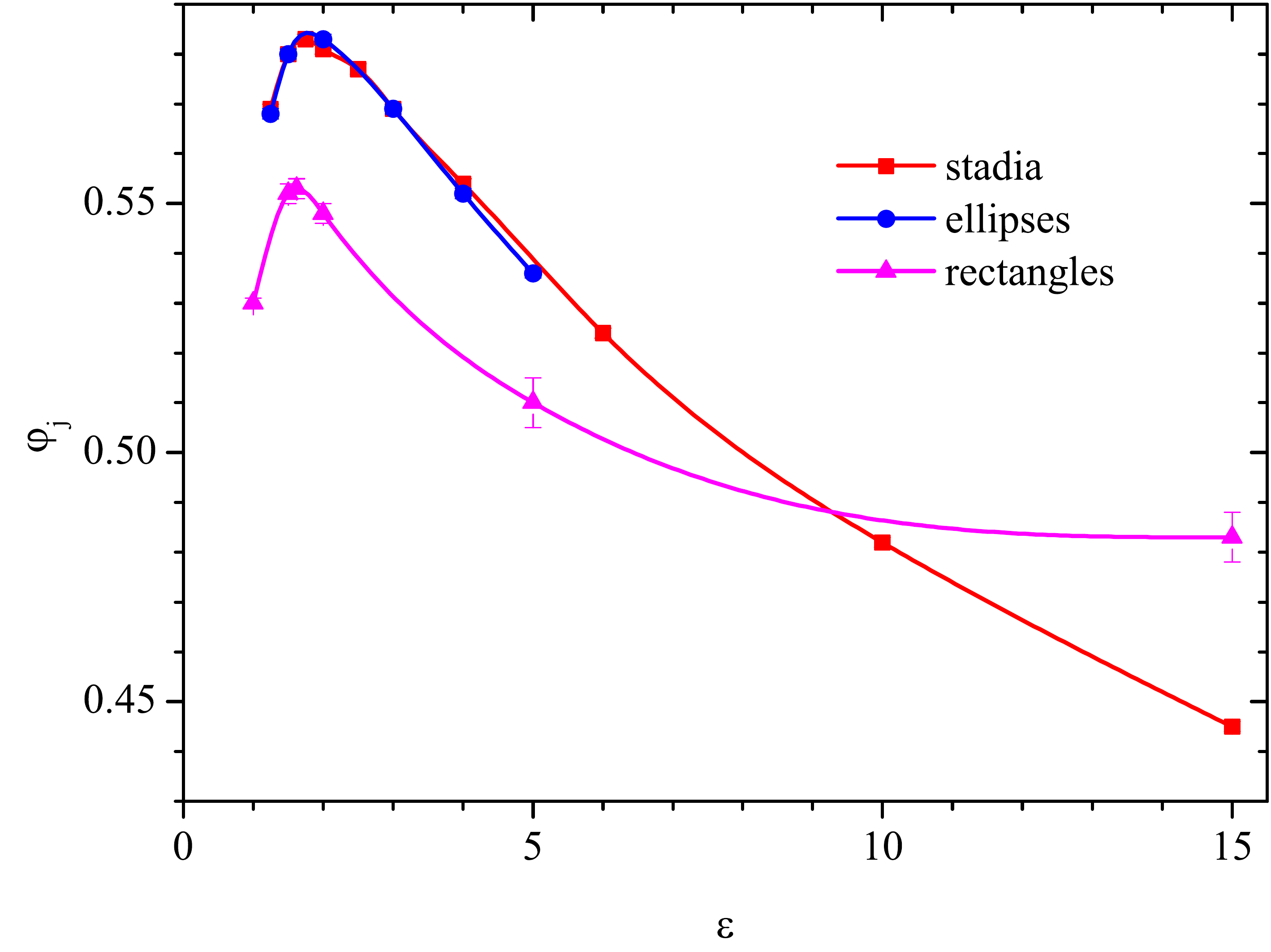}
  \caption{Jamming coverage $\phi_\text{j}$ versus the aspect ratio $\varepsilon$ for unoriented rectangles, stadia, and ellipses\cite{Viot1992JCP}). (Drawn from the data presented in the cited works).}\label{fig:jamcovvsas1}
\end{figure}

\Fref{fig:jamcovvsas1}  compares the jamming coverage $\phi_\text{j}$ versus the aspect ratio $\varepsilon$ for unoriented rectangles, stadia, and ellipses\cite{Viot1992JCP}. For all investigated shapes the obtained the jamming coverages gone through the maximums and the highest values of the mean packing fractions were observed at some aspect ratios $\varepsilon \approx 1.7$ for rectangles, and $\varepsilon \approx 1.85$ for stadia and ellipses. It is expected that for very elongated shapes (for infinite aspect ratio) the $\phi_\text{j}$ goes to zero according to a power law
\begin{equation}\label{eq:phijvseps}
\phi _\text{j}(\varepsilon) = \phi_\text{j}(\infty) + a \varepsilon^{-1/(1+\sqrt{2})}.
\end{equation}
where $\phi_\text{j}(\infty)=0$
%\phi_\text{j} \propto \varepsilon^{-p},
%\end{equation}	
%where $p= 1/(1+\sqrt{2}$)
\cite{Viot1992JCP}.

Later, the more effective algorithms allowed reducing the statistical errors were developed for estimation of the saturated packings of convex anisotropic objects under RSA protocols\cite{Haiduk2018PRE,Ciesla2016JCP,Ciesla2015PCCP}.

\begin{figure}
  \centering
  \includegraphics[width=0.75\textwidth]{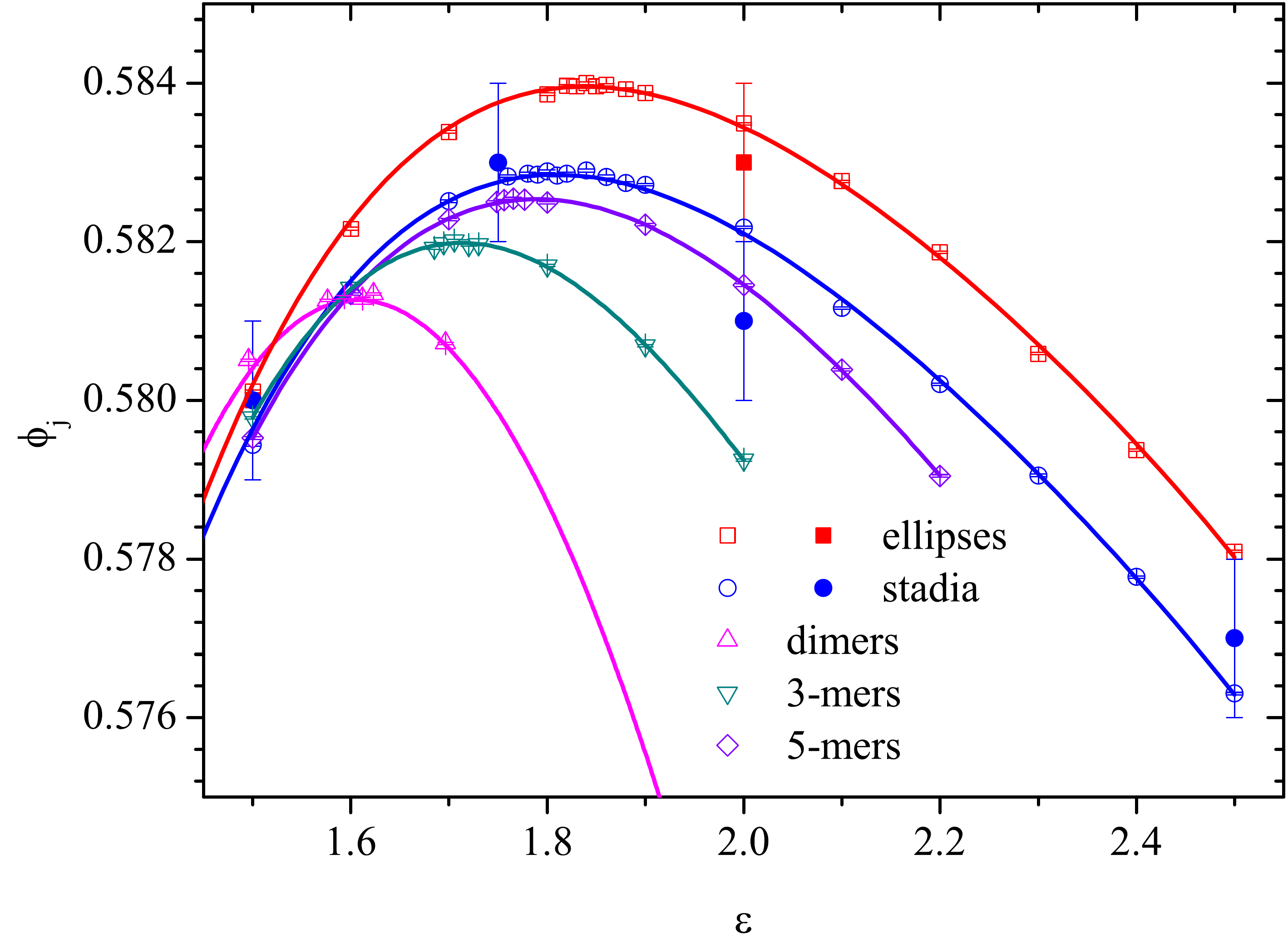}
  \caption{Jamming coverage $\phi_\text{j}$ versus the aspect ratio $\varepsilon$ for different unoriented shapes: ellipses and stadia\cite{Haiduk2018PRE}, disco $n$-mers\cite{Ciesla2016JCP} (open symbols). For comparison the data points (filled symbols) for ellipses and stadia from\cite{Viot1992JCP} are also presented. (Drawn from the data presented in the cited works).}\label{fig:jamcovvsas}
\end{figure}

\Fref{fig:jamcovvsas} collects the $\phi_\text{j} (\varepsilon)$ dependencies obtained stadia, ellipses\cite{Haiduk2018PRE} and disco $n$-mers ($n=2-10$)\cite{Ciesla2015PCCP,Ciesla2016JCP}. Among these shapes the highest jamming coverage $\phi_\text{j} =0.583999 \pm 0.000017$ was observed for ellipses with aspect ratio of $\varepsilon = 1.84$. For stadia, maximal packing fraction was $\phi_\text{j} =0.582896 \pm 0.000019$ at approximately the same aspect ratio\cite{Haiduk2018PRE}.

\begin{figure}
  \centering
  \includegraphics[width=0.75\textwidth]{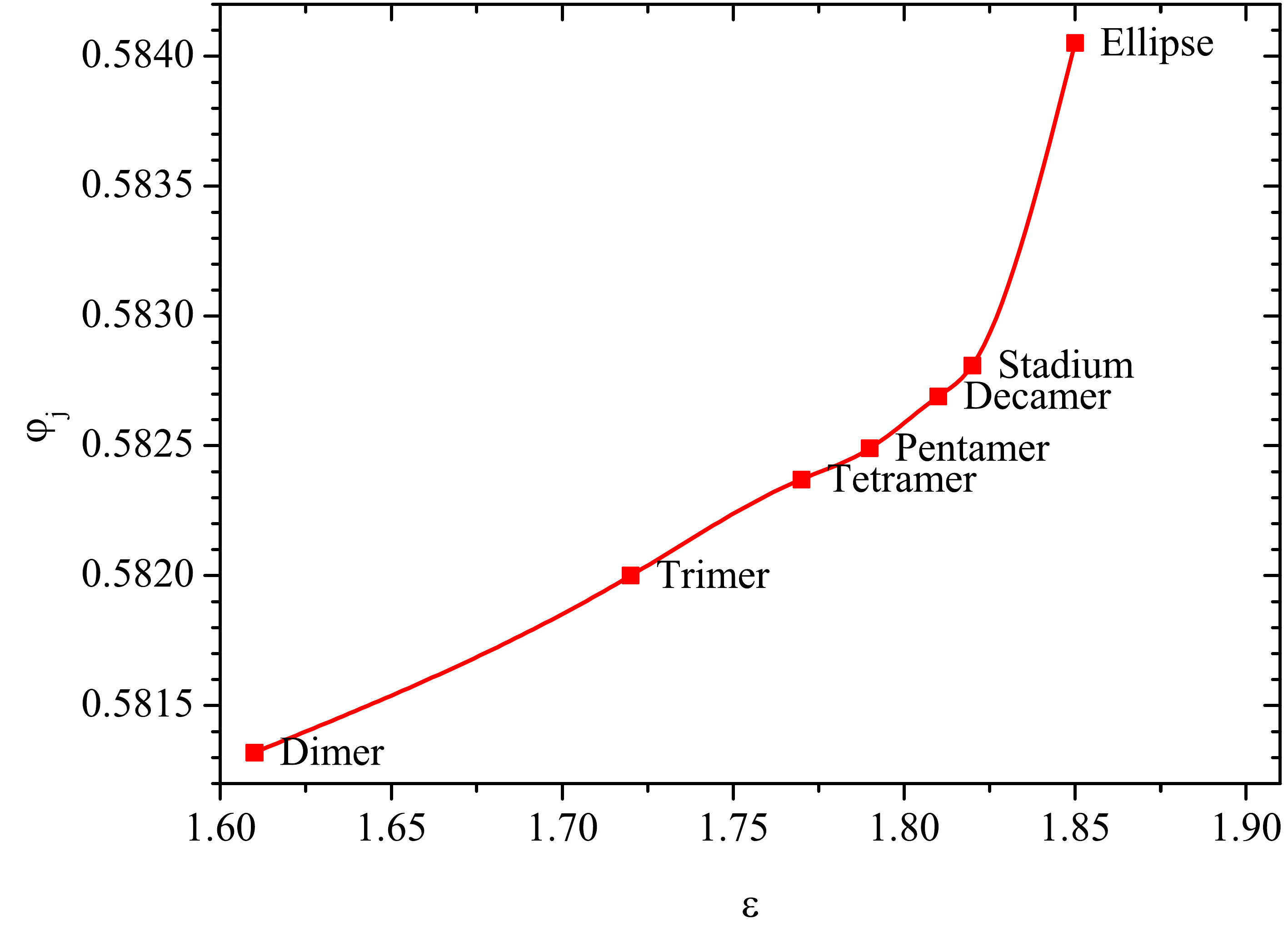}
  \caption{Jamming coverage, $\phi_\text{j}$, versus the aspect ratio, $\varepsilon$, at the point of maximum jamming for different shapes\cite{Ciesla2016JCP}. (Drawn from the data presented in the cited work).}\label{fig:jamcov}
\end{figure}

It is interesting that at the point of maximum coverage (point of maximum jamming) the values of $\phi_\text{j}$ correlates with the corresponding values of $\varepsilon$ for different shapes. \Fref{fig:jamcov} presents $\phi_\text{j}$ versus $\varepsilon$ dependence collected from the data for disco $n$-mers ($n=2-10$), stadia and ellipses.

For shape-anisotropic particle practically all studies treated random orientation RSA problems. The particular interest represents investigation of the RSA packing with preferred orientation.

%==================================================

The different classes of continuous and lattice models have been developed for description of saturated packing.
The continuous models with complete orientational and positional freedom are more realistic for modeling of practical packing and phase states. The discrete space (lattice) models are simpler in computations and they can be used for incorporation the flexibility, polydispersity, interactions, heterogeneity and presence of defects\cite{Chick1991}. The strict statistical analysis shown that the discrete space (lattice) model overestimates the entropy of short rods and of disordered phases\cite{Chick1991}.

\subsubsection{\label{sec: Latmodels} Lattice models}
In lattice models, the positions and orientations of the elongated particles can be fixed accounting for the lattice structure of a discrete 2D substrate. Many efforts have been concentrated on the study of deposition of linear $k$-mers on a discrete 2D substrates. The linear $k$-mers represents the particle occupying $k$ adjacent adsorption sites along a line on a lattice. For linear $k$-mers aspect  ratio is $\varepsilon=k$ and for jamming concentration $\phi_\text{j}$ we use the notation $p_\text{j}$.

In one dimensional (1D) RSA of $k$-mers for jamming concentration $p_\text{j}$ the following exact result has been reported\cite{Krapivsky2010}
\begin{equation}\label{eq:ue1Djamm}
p_\text{j} = k
\int\limits_{0}^{\infty}\exp\left(-u-2\sum\limits_{i=1}^{k-1}\frac{1-\exp(-iu)}{i}\right)\,\mathrm{d}u.
\end{equation}
In particular, the jamming concentration for dimers ($k=2$) placed along one line is $p_\text{j}=1-\mathrm{e}^{-2} \approx 0.86466$ and for trimers ($k=3$) is $p_\text{j} = 3 D (2) - 3 \mathrm{e}^{-3} D (1)\approx 0.82365$, where $D(x) = \mathrm{e}^{-x^2}\int_{0}^{x} \mathrm{e}^{t^2} \,\mathrm{d}t$ is Dawson's integral.
The continuous limit  corresponds to the infinite aspect ratio ($\varepsilon=k \to \infty$), and in this case the jamming threshold tends to R{\'e}nyi's Parking Constant
${p_\text{j}}\to {c_{R}} \approx 0.7475979202\dots$\cite{Renyi1958}.

In two dimensions the analyzed RSA problems include deposition of $k$-mers on different discrete lattices:
square\cite{Manna1991,Bonnier1994,Bonnier1996,Leroyer1994PRB,Budinski-Petkovic1997,Budinski-Petkovic1997a,Lee1998,Vandewalle2000epjb,Ramirez-Pastor2000,Kondrat2001PRE,Fusco2001JCP,Fusco2002PhilMagB,Cornette2007,Lebovka2011a}, triangular\cite{Ghaskadvi2000,Budinski-Petkovic2002,Budinski-Petkovic2005PRE,Budinski-Petkovic2008PRE,Budinski-Petkovic2011,Loncarevic2007a,Loncarevic2010,Perino2017JSMTE,Scepanovic2013PhA}
%, honeycomb\cite{Manna1991,Bonnier1994,Bonnier1996,Budinski-Petkovic1997},
and fractal\cite{Nazzarro1997,Chang2008,Cornette2011,Pasinetti2019PRE} lattices.

\begin{figure}%Figure 01
\centering
\subfigure[Square lattice.]{
\includegraphics[width=0.4\linewidth]{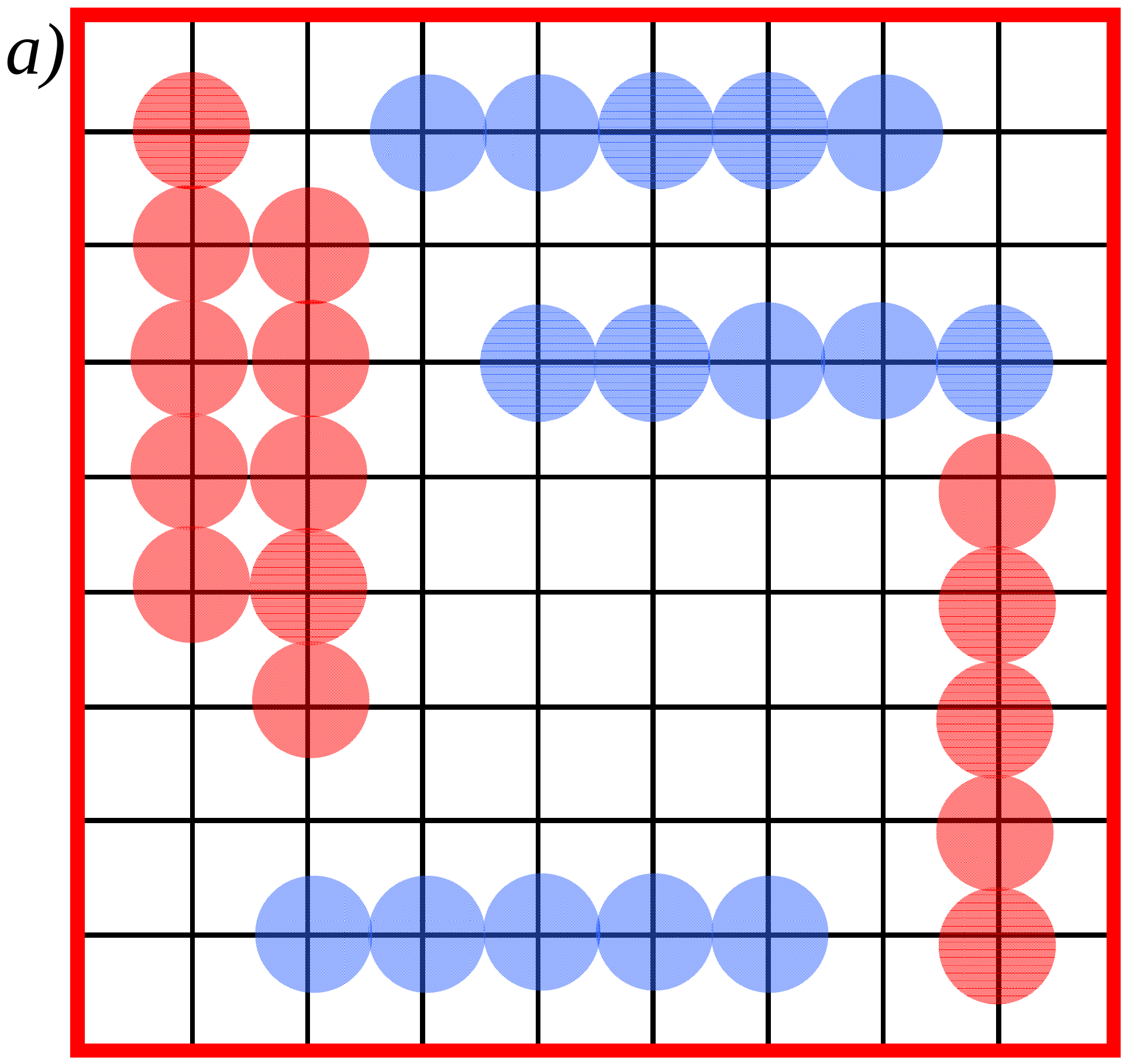}\label{subfig:suare}}\hfill
\subfigure[Triangular lattice.]{
\includegraphics[width=0.4\linewidth]{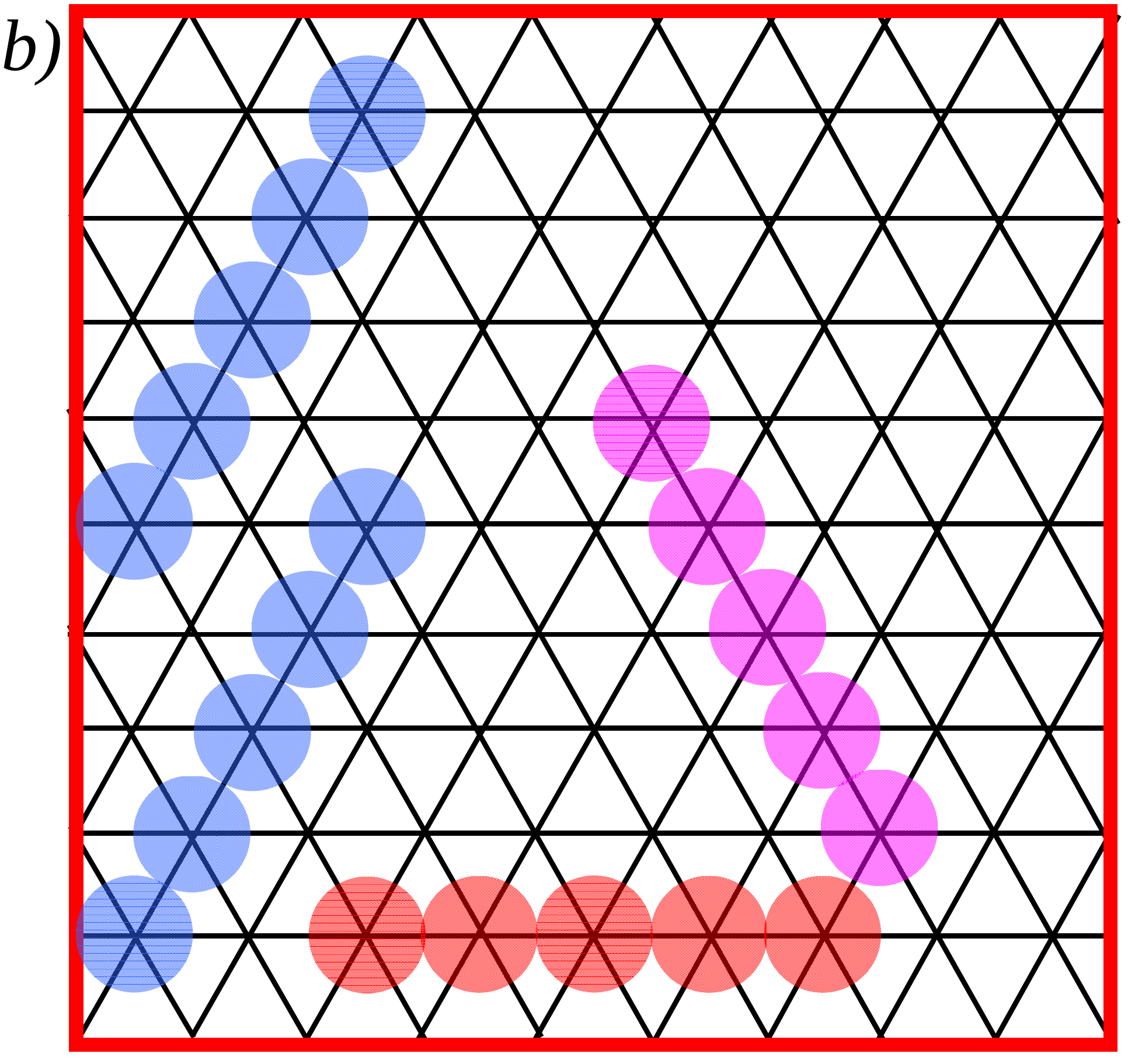}}
\caption{\label{fig:SqTri}Examples of linear pentamers adsorbed on square and triangular lattices.}
\end{figure}

In square lattice, the number of possible orientations of $k$-mers is restricted to 2, and in triangular lattice it is restricted to 3 (\fref{fig:SqTri}).

\begin{figure}%Figure 01
\centering
\includegraphics[width=0.3\linewidth]{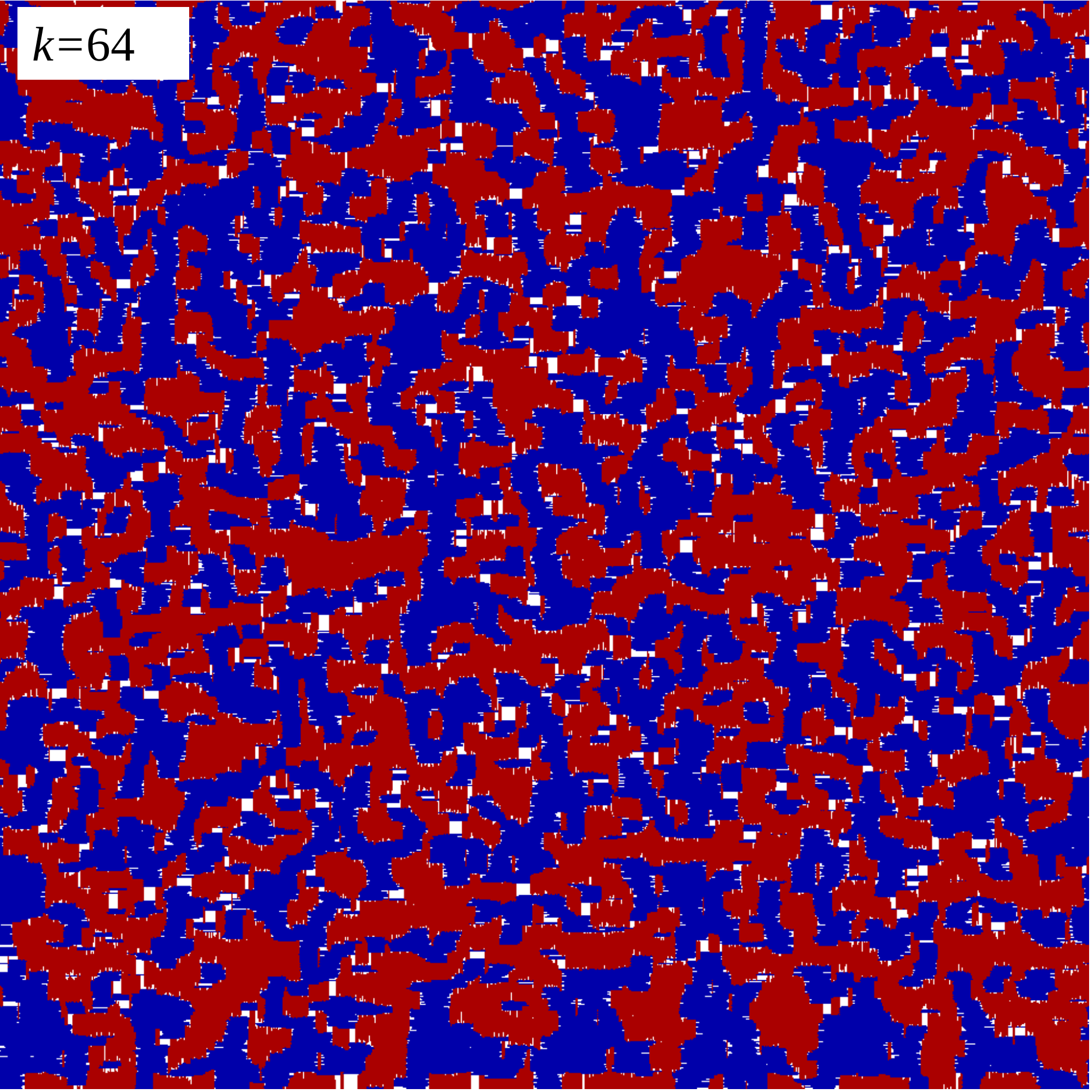}\hfill
\includegraphics[width=0.3\linewidth]{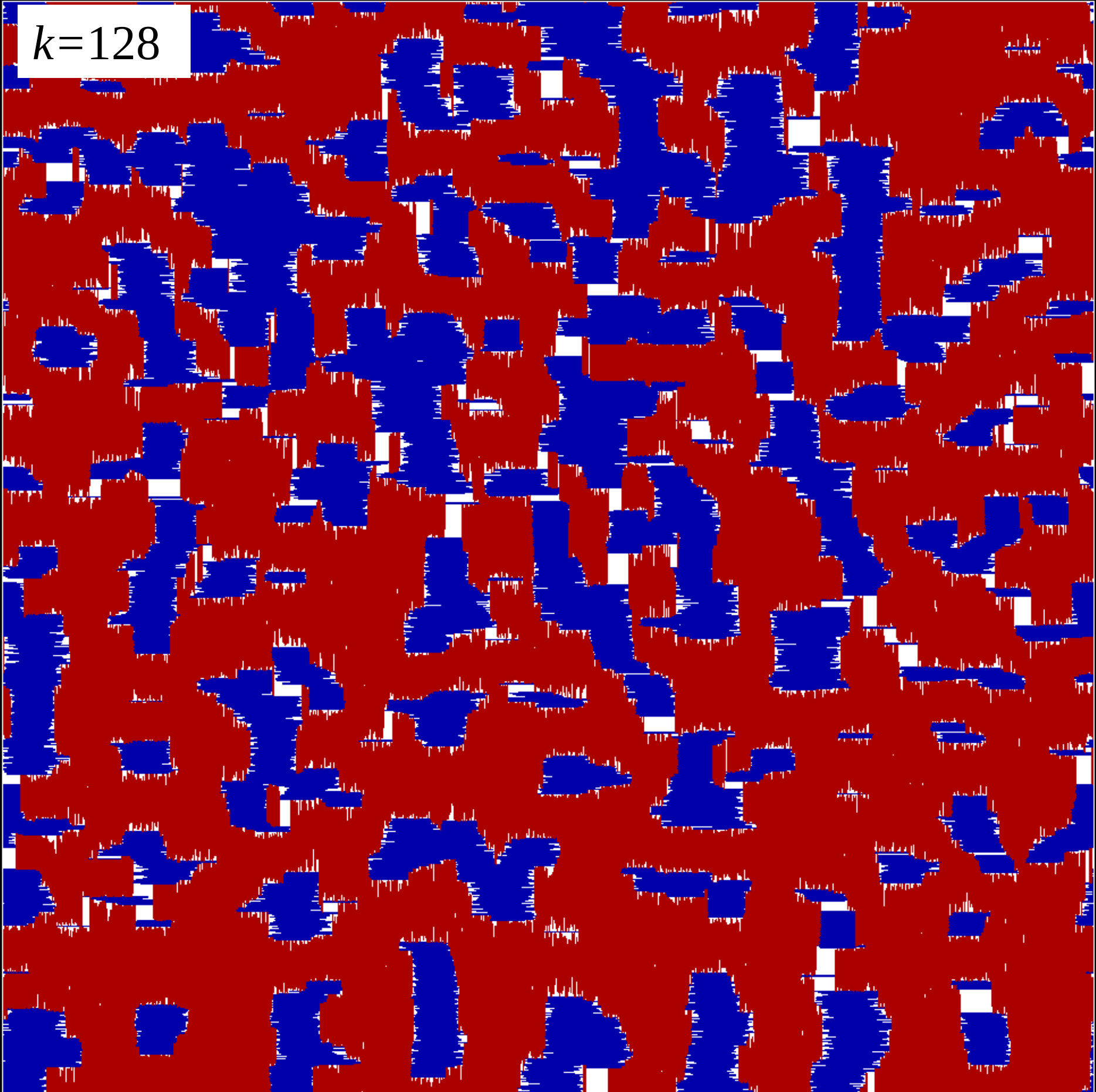}\hfill
\includegraphics[width=0.3\linewidth]{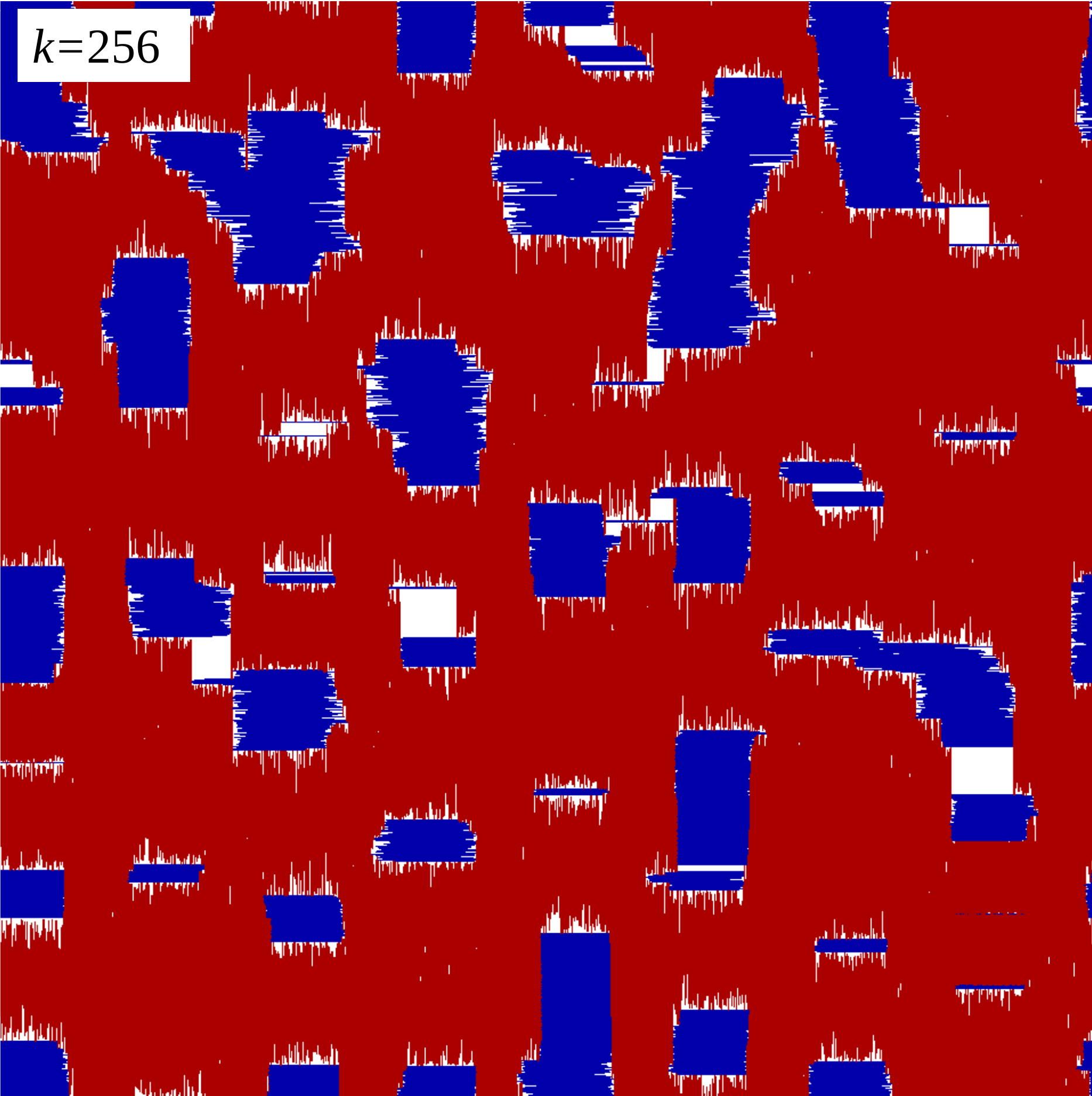}\hfill
\caption{\label{fig:Stacks}Examples of jamming configurations of randomly oriented $k$-mers ($k=64, 128$, and $256$) on a square lattice of size $L=4096$. Horizontal $k$-mers are shown in red, vertical $k$-mers are shown in blue,empty sites are shown in white.}
\end{figure}

For randomly oriented $k$-mers on a 2D square lattice deposited $k$-mers tend to align parallel to each other and form large domains (stacks) with voids ranging from a single site up to the length of $k$-mer\cite{Manna1991,Vandewalle2000epjb,Lebovka2011a}. \Fref{fig:Stacks} presents examples of jamming configurations for randomly oriented $k$-mers with different length on a square lattice of size $L=4096$. The domain structures in form of blocks of parallel oriented $k$-mers was observed. These blocks can be represented as the squares of size $k \times k$\cite{Lebovka2011a}. Therefore, a jamming configuration can be presented as a combination of:
\begin{itemize}
  \item blocks of vertically oriented $k$-mers ($v$-blocks);
  \item blocks of horizontally oriented $k$-mers ($h$-blocks);
  \item empty sites (voids).
\end{itemize}
The relative area occupied by $v$- and $h$-blocks was approximately the same.

The similar stacking and void formation was also observed for continuous RSA model of deposition of infinitely-thin line segments\cite{Ziff1990,Viot1992PA}. However, for continuous and lattice models the principally different behavior of the jamming coverage $p_\text{j}$ versus the aspect ratio $\varepsilon$ ($k=\varepsilon$ for $k$-mers) was observed. For example, in the off-lattice case, in the limit of infinite aspect ratio, the asymptotic jamming concentration $\phi_\text{j}(\infty)$ is 0 (see \eref{eq:phijvseps})\cite{Viot1992PA}.

For lattice model, the asymptotic jamming concentration $p_\text{j}(\infty)$ remains finite\cite{Manna1991,Bonnier1994,Bonnier1996,Kondrat2001PRE,Lebovka2011a}. The presence of finite coverage by the infinite $k$-mers has been interpreted as a consequence of the alignment constraint\cite{Bonnier1994, Bonnier1996,Lebovka2011a}.

Note that limiting jamming coverage,  $p_\text{j}(\infty)$, can depend upon the symmetry of the lattice and details of RSA deposition mechanism\cite{Manna1991, Bonnier1994,Bonnier1996,Kondrat2001PRE,Lebovka2011a}. For square and triangular lattice the dependencies $p_\text{j}(\varepsilon)$ were rather similar\cite{Budinski-Petkovic1997b}, but the jamming coverages had greater values on the square lattice. This difference was explained the orientational freedom of depositing particles. On a triangular lattice there is a greater number of possible orientations ($3$) and an enhanced probability for blocking of lattice sites. For RSA on a triangular lattice the domains of parallel $k$-mers and of clusters of
blocked sites were also observed in the late stages of deposition\cite{Budinski-Petkovic1997b}.

The mechanism of deposition is also important. In simple RSA model the vacant site is randomly selected and any unsuccessful attempt of deposition of $k$-mer is rejected. For this RSA model the different estimations for randomly oriented $k$-mers on square lattice gave $p_\text{j}(\infty)\approx0.66$\cite{Bonnier1994,Kondrat2001PRE},
$\approx 0.655(9)$\cite{Lebovka2011a}. However, for modified ``end-on'' RSA model of randomly oriented $k$-mers on square lattice
(for this model, once a vacant site has been found, the deposition is randomly attempted in all the directions until the segment is adsorbed or rejected) the noticeably smaller value of $p_\text{j}(\infty)=0.583\pm 0.010$ was obtained\cite{Manna1991}.

For RSA model of randomly oriented $k$-mers on square lattice, the $p_\text{j}(k)$ were fitted using the series expansion
\begin{equation}\label{eq:Bonnier}
p_\text{j}(k) = p_\text{j}(\infty)+a/k+b/k^2,
\end{equation}
with parameters of $p_\text{j}(\infty)=0.660\pm 0.002$, $a \approx 0.83$ and $b \approx -0.70$ ($2\leqslant k\leqslant 512$)\cite{Bonnier1994, Bonnier1996},
and the power law
\begin{equation}\label{eq:power}
p_\text{j}(k) = p_\text{j}(\infty)+a/k^\alpha.
\end{equation}
with parameters of $p_\text{j}(\infty)=0.66\pm 0.01$, $a \approx 0.44$ and $\alpha \approx 0.77$\cite{Kondrat2001PRE} ($1\leqslant k \leqslant 2000$), and $p_\text{j}(\infty)=0.655(9)$, $a \approx 0.416(0)$ and $\alpha \approx 0.720(7)$ ($1\leqslant k \leqslant 256$)\cite{Lebovka2011a}.

For non-randomly oriented linear $k$-mers the different lattice directions are selected with  different probabilities. For examples, on a square lattice the vertical and horizontal orientations are selected according to the predefined order parameter $s$ \eref{eq:Sdiscr}. The complete ordering ($s=1$), corresponds to the 1d problem with jamming coverage defined by \eref{eq:ue1Djamm}.

The non-randomly oriented linear $k$-mers on a square lattice the conventional RSA model and relaxation RSA (RRSA) model have been compared\cite{Lebovka2011a}. In conventional RSA model, the lattice site was randomly selected and an attempt of deposition of a $k$-mer with orientation defined by order parameter $s$ was done. Any unsuccessful attempt was rejected and $k$-mer with a new orientation was selected. In the RRSA model, the unsuccessful attempt was not rejected and a new lattice site was randomly selected until deposition of the given $k$-mer.
In both models, the deposition terminates when a jamming state is reached along one of direction. The RSA and RRSA models can reflect different binding ability of $k$-mers near the flat substrate. In RSA model, the binding is weak and the $k$-mer may return to the bulk after an unsuccessful attempt to precipitate. In RRSA model, the selected $k$-mer is localized near the adsorbing plane is strong, and it has an additional possibility of joining the surface after an unsuccessful attempt.

The different mechanisms in RSA and RRSA model deposition strongly affected the deposition behavior\cite{Lebovka2011a}. First of all, in the RSA model the order parameter $s$ is not conserved. In this model, the substrate ``selects'' the $k$-mer with appropriate orientation, and it results in deviation of predetermined order parameter $s$ from the actually obtained one, $s_0$. In two particular cases of $s=0$ and $s=1$ the RSA and RRSA models gave practically equivalent results, but in other cases for RSA model the value of $s_0$ noticeably exceeded the value of $s$.  On the contrary, the RRSA model better preserves the predetermined anisotropy, and $s_0 \approx s$.

\begin{figure}%Figure 01
\centering
\includegraphics[width=0.3\linewidth]{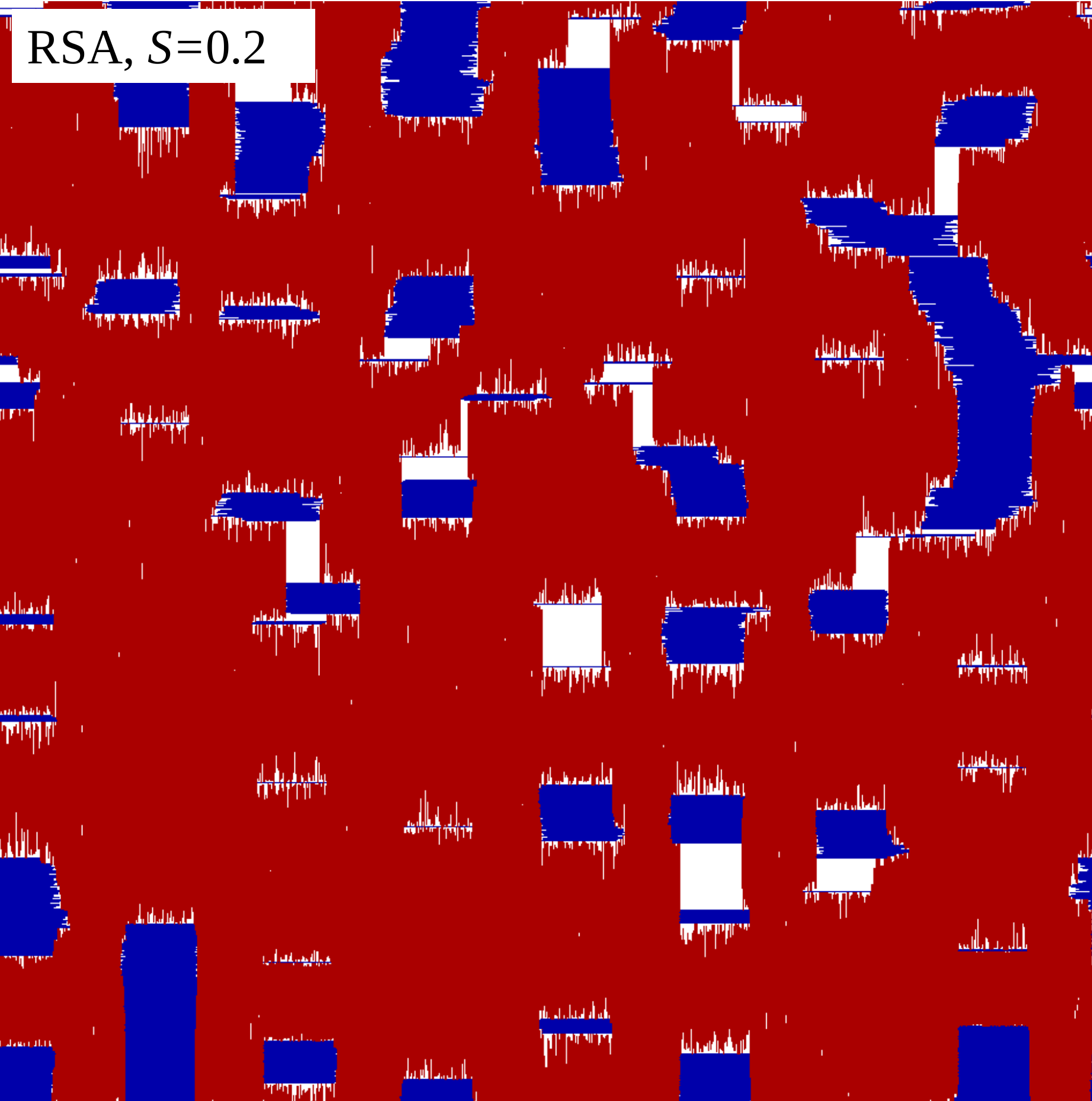}\hfill
\includegraphics[width=0.3\linewidth]{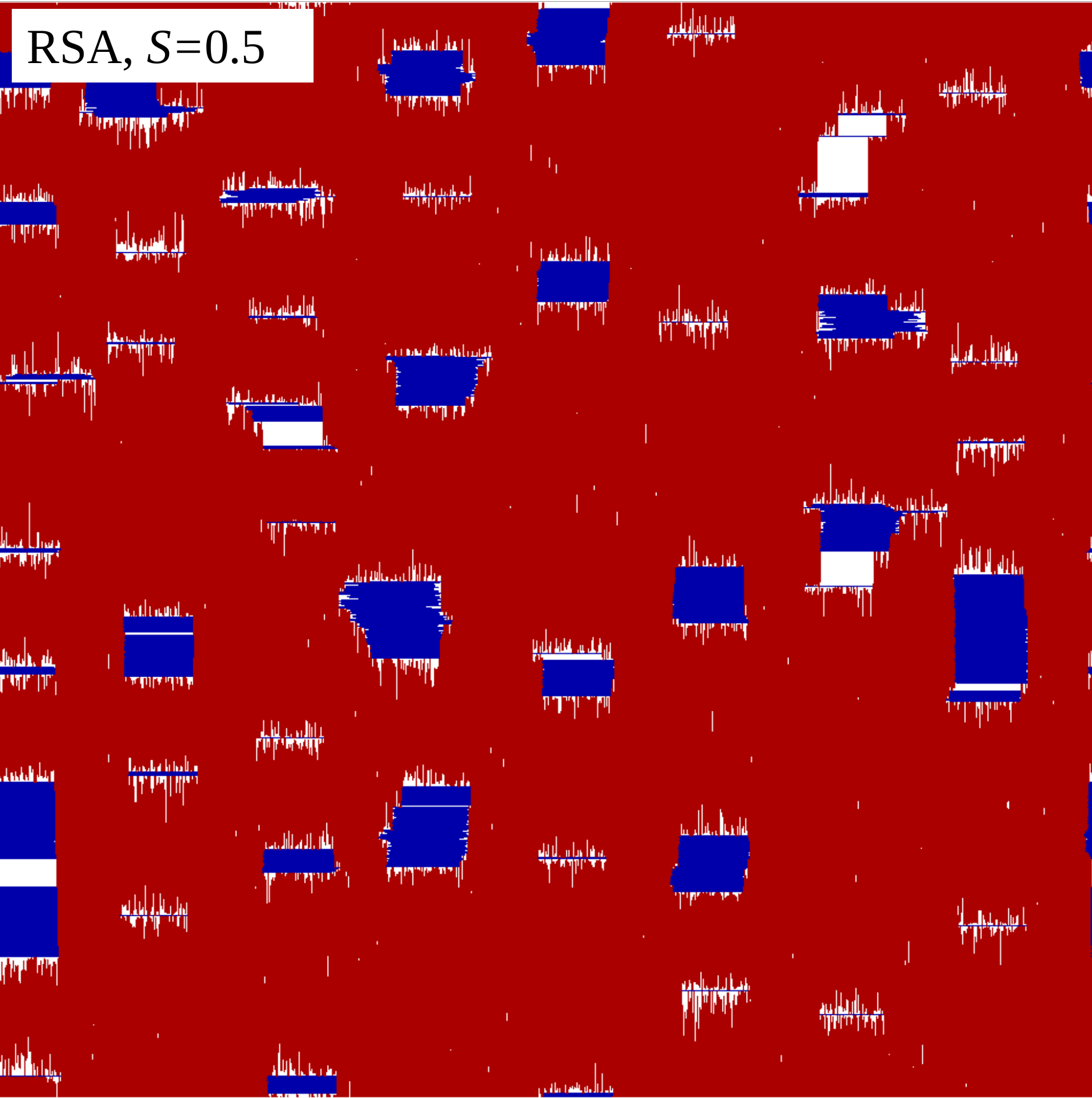}\hfill
\includegraphics[width=0.3\linewidth]{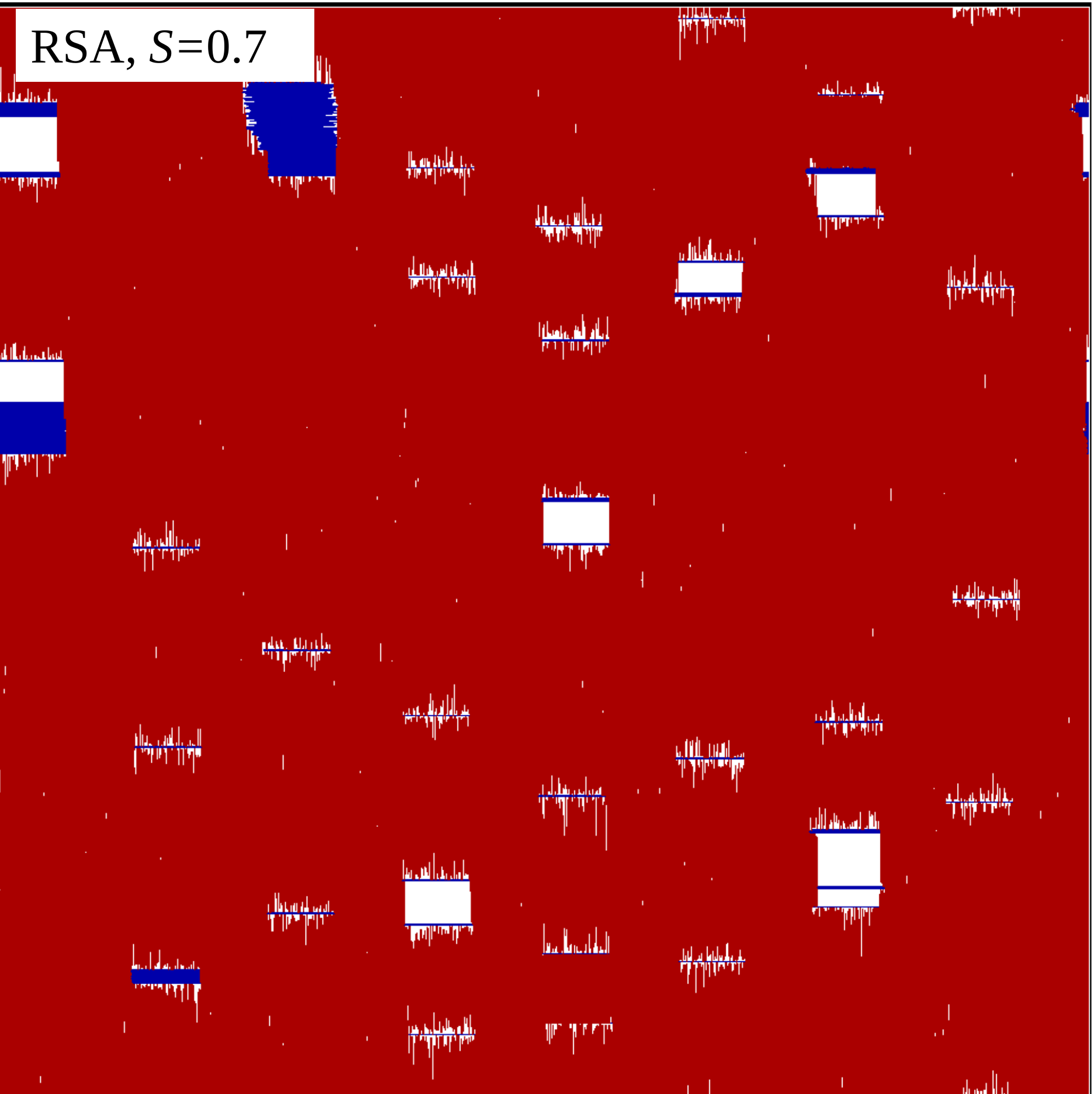} \\
\includegraphics[width=0.3\linewidth]{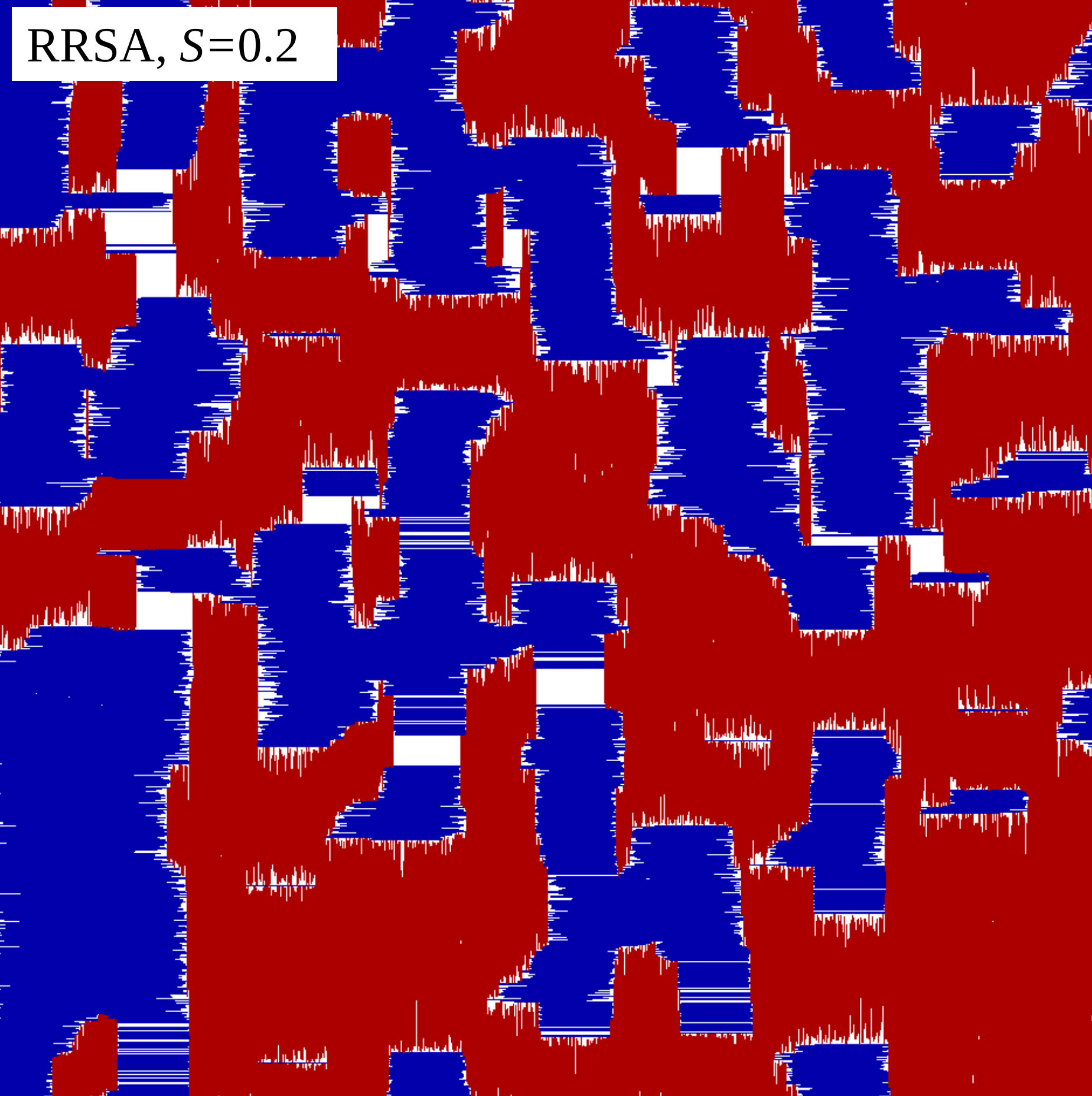}\hfill
\includegraphics[width=0.3\linewidth]{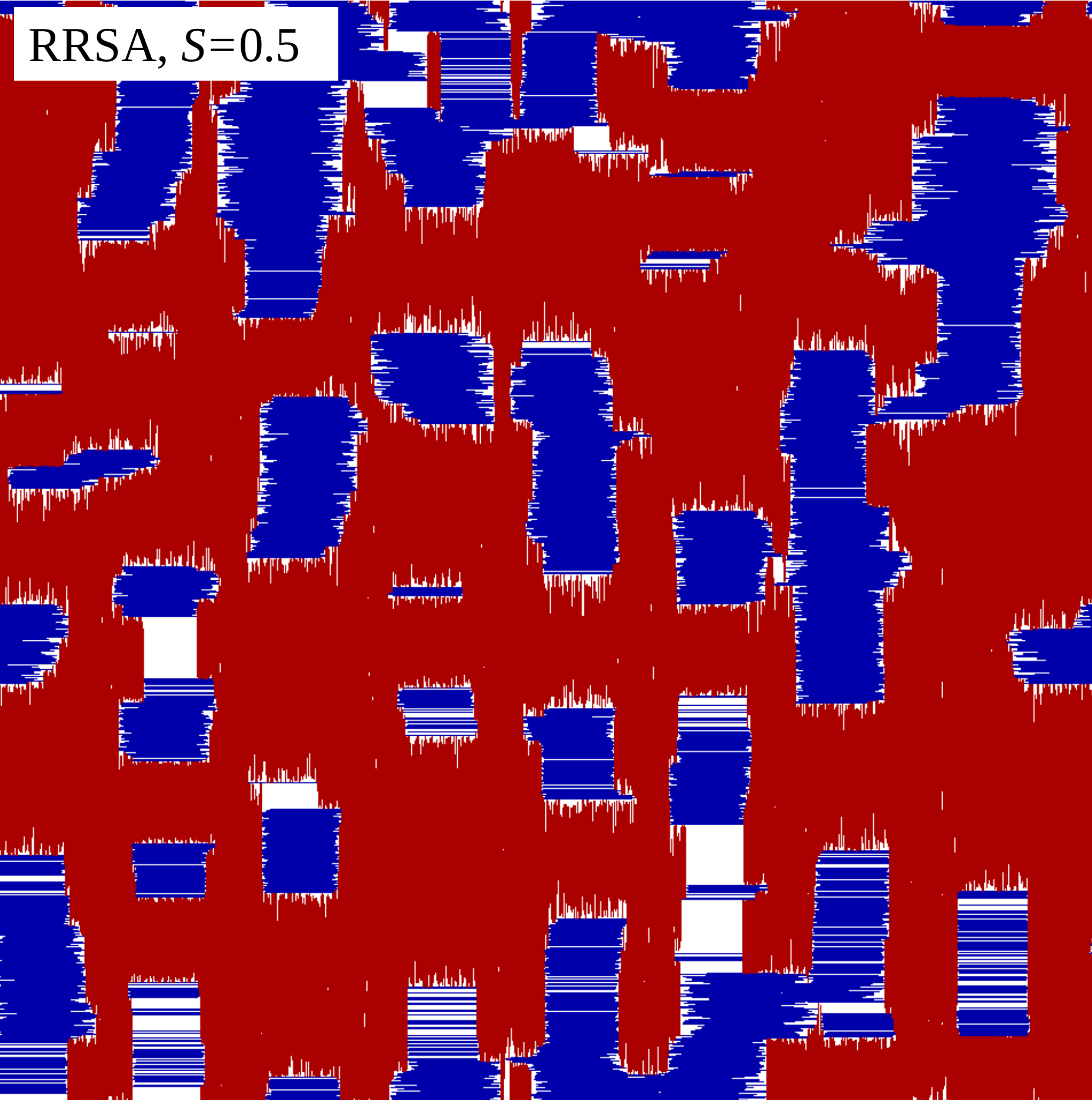}\hfill
\includegraphics[width=0.3\linewidth]{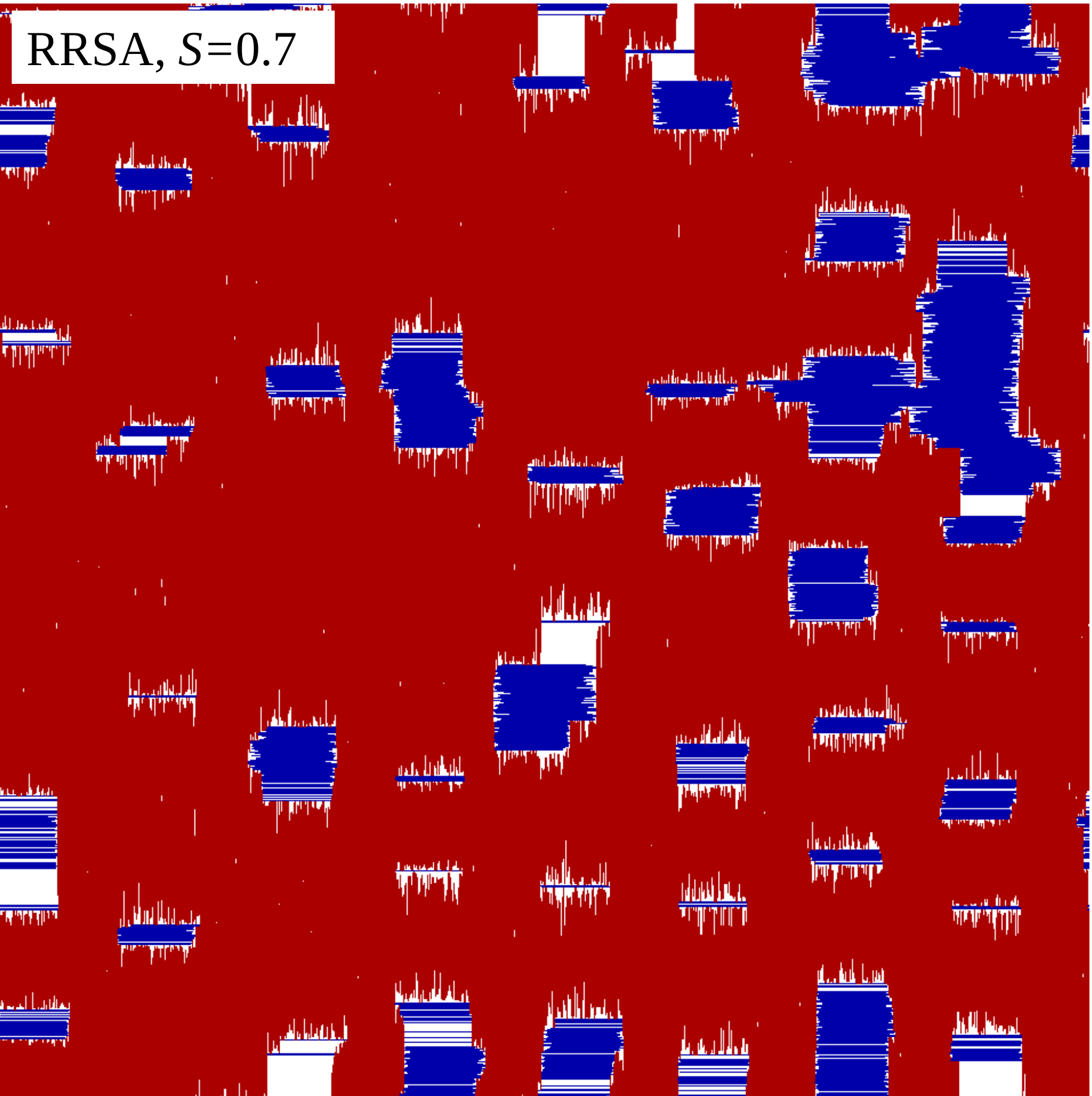}\hfill
\caption{\label{fig:RSA_RRSA}Examples of jamming configurations of oriented along vertical direction $k$-mers ($k=256$) on a square lattice of size $L=4096$ for RSA and RRSA models at different order parameters, $s$. Horizontal $k$-mers are shown in red, vertical $k$-mers are shown in blue,empty sites are shown in white.}
\end{figure}

\Fref{fig:RSA_RRSA} compares typical jamming patterns for RSA and RRSA models obtained at different values of order parameter, $s$, for $k=256$. The domain structures in form of blocks of vertically oriented $k$-mers ($v$-blocks), blocks of horizontally oriented $k$-mers ($h$-blocks), and empty sites (voids) were observed. The relative area occupied by $v$- and $h$-blocks was approximately the same at $s=0$, but increase in $s$ resulted in decrease of the relative area occupied by $h$-blocks.  The vertically oriented $v$-blocks became dominating in jamming patterns at $s>0$. Moreover, at the same values of $s$ the relative area occupied by $v$-blocks was visually larger for RSA model than for RSA model.  For example, for well oriented systems (at $s\approx 0.5-0.7$), the jamming structures in RSA model present large 1d-like $v$-domains with small inclusions of $h$-domains and voids, and the jamming structures of RRSA model included higher number of $h$-domains with large voids between them than those of RSA model. Finally, at $s=1$, the jamming configuration transferred into the 1d-like domains of interconnected $v$-blocks.

The analysis of the connectivity between the similar $v$- or $h$-blocks was performed\cite{Lebovka2011a}. For randomly oriented species (\emph{i.e.}, at $s=0$), the infinite connectivity (\emph{i.e.},  percolation) was not observed. However, connectivity between $v$-blocks increased with increase of $s$ at some threshold value of $s_\text{c}$ the formation of the infinite (spanning) clusters of vertically oriented $k$-mers was observed. For the same $k$, the RSA model gave lower value $s_\text{c}$ than RRSA model, \emph{e.g.}, for $k=16$,  $s_\text{c}\approx 0.126$ (RSA model) and $s_\text{c}\approx 0.240$ (RRSA model). The value of $s_\text{c}$ decreased as the length of $k$-mer increased for RSA model, and opposite behavior was observed for RRSA model. Finally, in continuous limit $k\to\infty$, the difference between the threshold order parameters of RSA and RRSA models becomes rather large, $\triangle s_\text{c}\approx 0.12$.

\begin{figure}%Figure 01
\centering
\subfigure[Limiting jamming coverage, $p_\text{j}(\infty)$.]{
\includegraphics[width=0.48\linewidth]{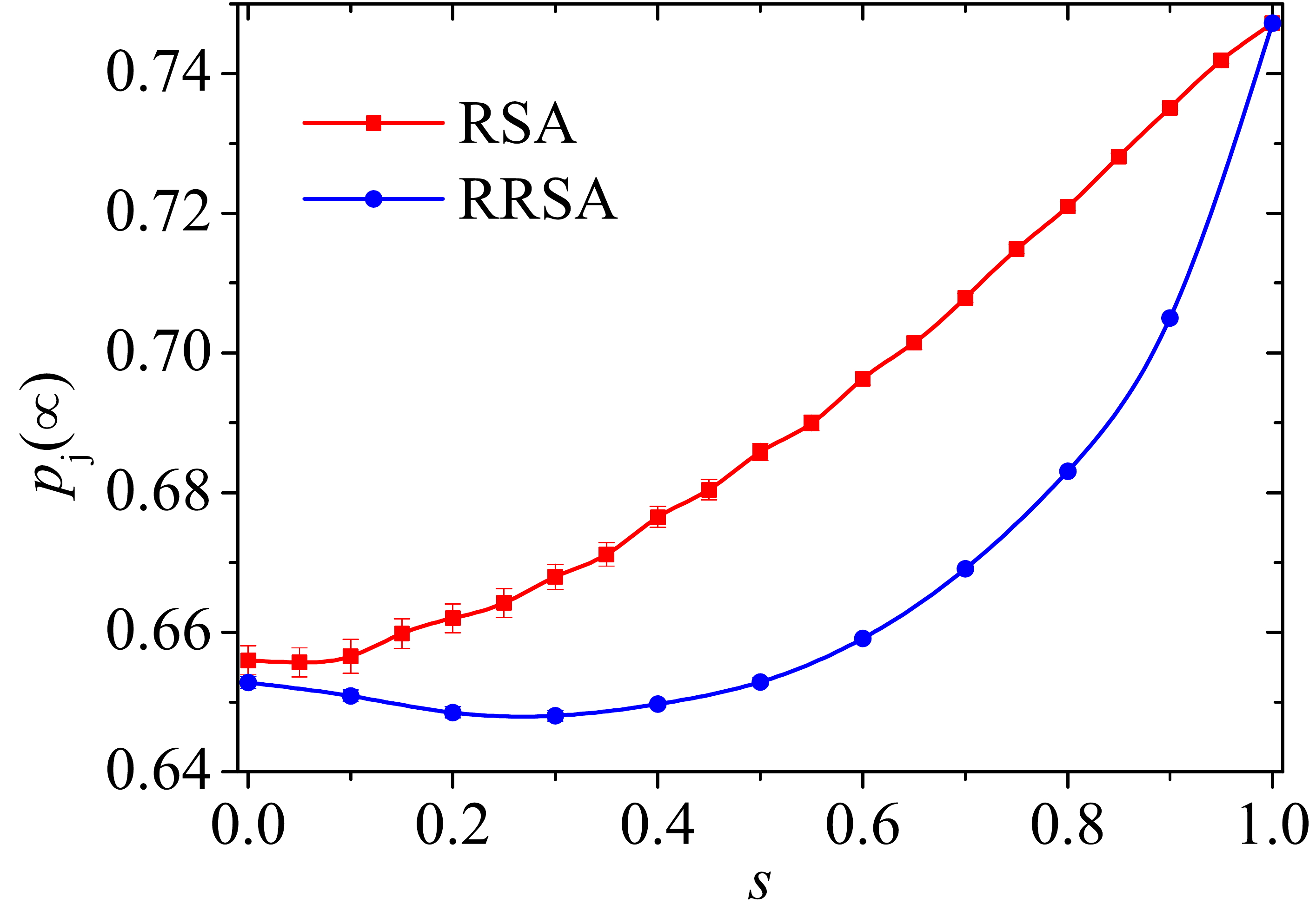}\label{subfig:ljc}}
\subfigure[Power exponent, $\alpha$.]{
\includegraphics[width=0.48\linewidth]{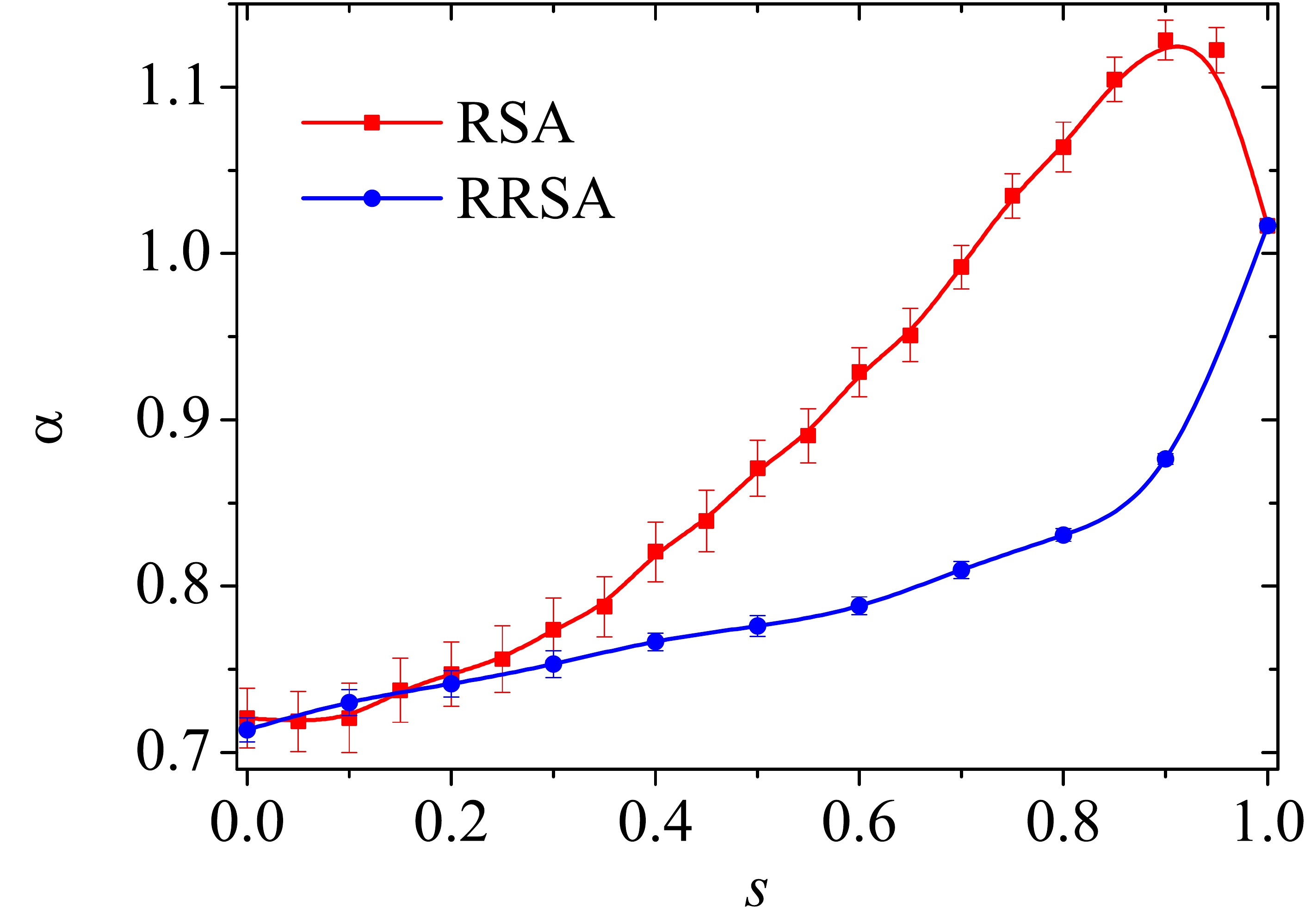}\label{subfig:alpha}}
\caption{Limiting jamming coverage, $p_\text{j}(\infty)$, and power exponent, $\alpha$, versus the order parameter $s$ for RSA and RRSA models of deposition of $k$-mers on square lattice (collected from the data presented in Ref.~\refcite{Lebovka2011a}). \label{fig:PowerLaw}}
\end{figure}

Analysis $p_\text{j}(k)$ dependencies had shown that power law \eref{eq:power} can be satisfactory applied\cite{Lebovka2011a}. \Fref{fig:PowerLaw} shows the limiting jamming coverage, $p_\text{j}(\infty)$, and power exponent, $\alpha$, versus the order parameter $s$ for RSA and RRSA models of deposition of $k$-mers on square lattice. With increase of $s$, the value of $p_\text{j}(\infty)$ continuously increased for RSA model, and gone through the minimum (at $s\approx 0.25$) for RRSA model (\fref{subfig:ljc}). In the intermediate region of $s$, it was observed $p_\text{j}(s)$ (RSA)$>p_\text{j}(s)$ (RRSA). This behavior can be  explained by the amplification of the actual order parameter, $s_0>s$, in the RSA model and stability of it $s_0\approx s$ in the RRSA model. The power exponents, $\alpha$, for RSA and RRSA model were not universal (\fref{subfig:alpha}). The power exponent $\alpha$ grown when the order parameter $s$ increases from $\alpha\approx0.72$ at $s=0$ to $\alpha\approx 1$ at $s=1$. The simple relation $d_\text{f}=2-\alpha$ was suggested for estimation of the fractal dimension $d_\text{f}$ of the jamming networks\cite{Ziff1990}. The jamming RSA and RRSA networks displaced fractal properties with $d_\text{f}\approx 1$ for completely ordered $k$-mers, ($s=1$) and value of $d_\text{f}$ increased for RRSA model and gone through  minimum for RSA model when the order parameter $s$ decreases.

Different variants of the more general lattice RSA models for deposition of $k$-mers, taking account of the heterogeneity of substrates, interactions between the deposited particles and the possibility of surface diffusion have been proposed\cite{Talbot2000CSA,Privman2000,Privman2000a,Senger2000}. %,Adamczyk2005,Weronski2005.
These models are more realistic in their description of the experimental results for colloid particle adsorption on substrates characterized by a wide spectrum of binding energies. The RSA problems for $k$-mers deposited on disordered (or heterogeneous) substrates with defects, or for $k$-mers with defects, have also attracted great attention\cite{Ben-Naim1994,Lee1996,Budinski-Petkovic2002,Kondrat2005,Kondrat2006,Cornette2003epjb,Cornette2006,Cornette2011,Budinski-Petkovic2011,Budinski-Petkovic2012,Tarasevich2015PRE,Lebovka2015PRE}.

Different models of lattices with impurities and non-ideal deposited objects were analyzed\cite{Tarasevich2015PRE,Lebovka2015PRE,Centres2015JSM}.
The jamming and percolation of linear $k$-mers on square lattice with impurities have been studied for $k=2-64$\cite{Centres2015JSM}. The presence of defects in the system was simulated by introducing a fraction of imperfect bonds, which are considered forbidden for deposition. In the LK$_\text{d}$  model, some fraction of the sites of square lattice,  $f_\text{d}$, was occupied by point defects (impurities) and the occupied sites were forbidden for deposition of the objects\cite{Tarasevich2015PRE}. In more general model the defects can be present both on the lattice and inside the $k$-mers\cite{Lebovka2015PRE}.

Another variant is the cooperative RSA model where the adsorption probability depends upon the local environment\cite{Evans1993RMP,Zuppa1999}. In this model strong near-neighbor (NN) lateral repulsive interactions were taken into account. For RSA deposition of monomers on a square lattice with complete NN exclusion, the jamming is observed at $p_\text{j}=0.3641$. In the C$_\text{d}$ model for deposition of $k$-mers a  restricted number of lateral contacts with previously deposited particles is allowed\cite{Lebovka2015PRE}. In this model, deposition occurs  when $z\leqslant (1 - f_\text{d}) z_\text{m}$  where $z$ is the number of contacts of the $k$-mer,  $z_\text{m}=2(k+1)$ is the maximum numbers of contacts of the $k$-mer. The fraction of forbidden NN contacts, $f_\text{d}$, may be identified with the fraction of defects that influence the process of deposition. The defect-free variant of this model ($f_\text{d}=0$) corresponds to the classical RSA model.

\begin{figure}%Figure 01
\centering
\includegraphics[width=0.3\linewidth]{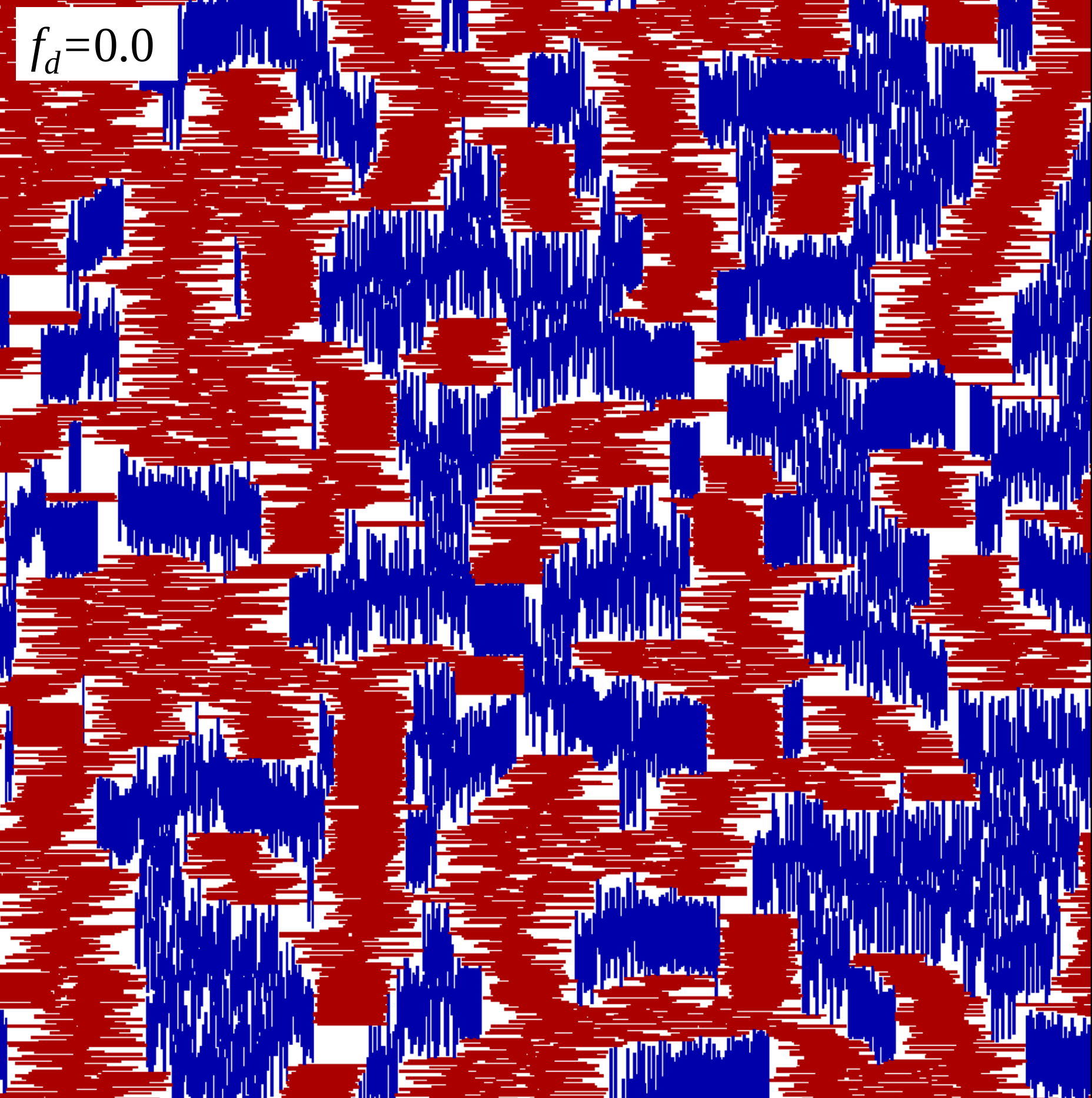}\hfill
\includegraphics[width=0.3\linewidth]{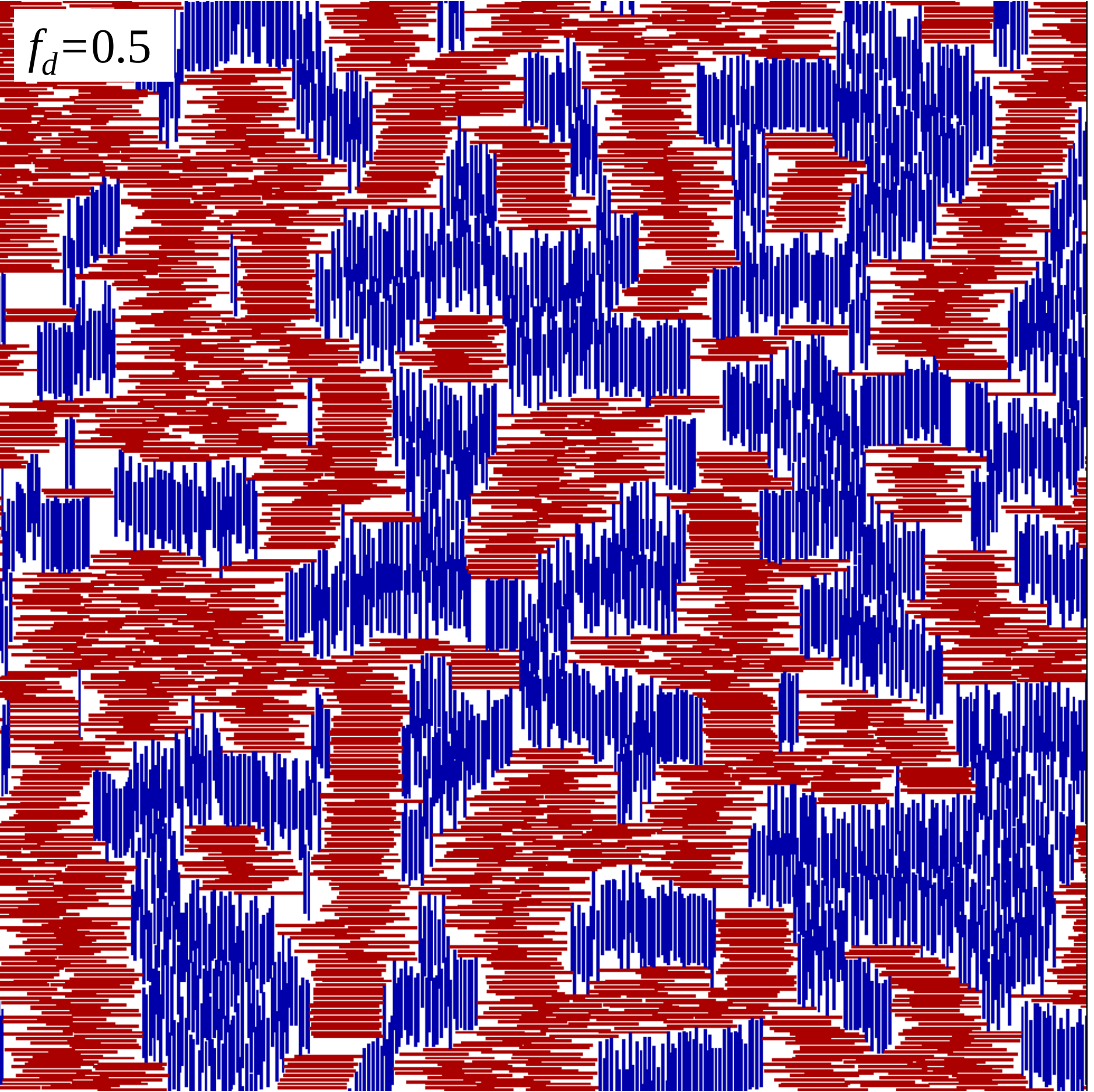}\hfill
\includegraphics[width=0.3\linewidth]{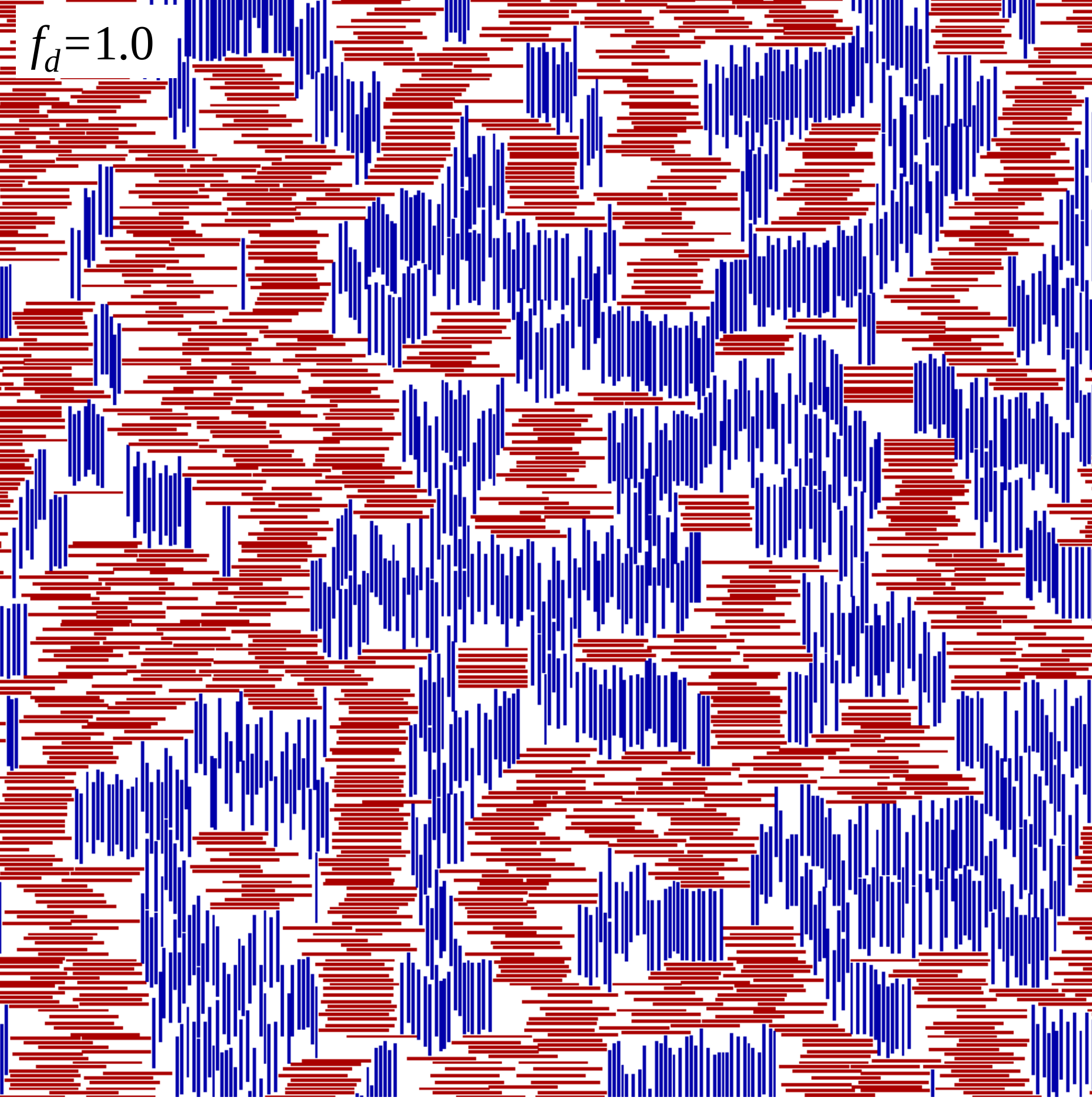} \\
\caption{\label{fig:Ex_Cont}Examples of jamming patterns for C$_\text{d}$ variant of RSA model for deposition of unoriented $k$-mers ($s=0$) on a square lattice of size $L=2048$ for $k=32$ and different values of $f_\text{d}$. Horizontal $k$-mers are shown in red, vertical $k$-mers are shown in blue,empty sites are shown in white. Here, the part of the lattice sized $512 \times 512$ is shown.}
\end{figure}

\Fref{fig:Ex_Cont} demonstrates typical examples of the jamming patterns for the C$_\text{d}$ model with restricted numbers of lateral contacts for $k=32$ ($z_\text{m}=66$) and different values of $d$\cite{Lebovka2015PRE}.

\begin{figure}%Figure 01
\centering
\includegraphics[width=0.75\linewidth]{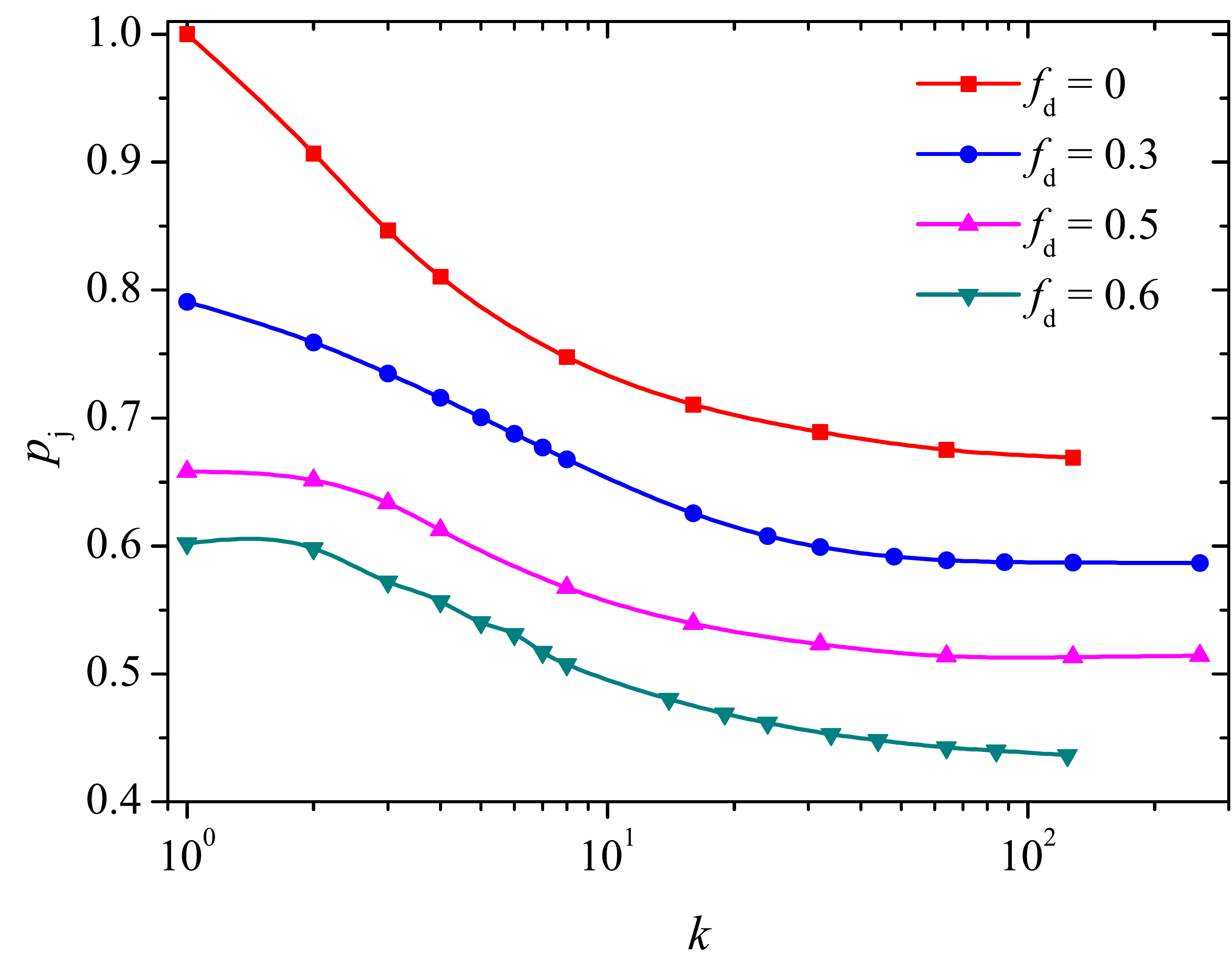}
\caption{Jamming coverage, $p_\text{j}$, versus the aspect ratio, $\varepsilon$, for deposition of unoriented $k$-mers  ($s=0$) using the C$_\text{d}$ variant of RSA model at different fractions of forbidden contacts, $f_\text{d}$ (collected from the data presented in Ref.~\refcite{Lebovka2015PRE}). \label{fig:Fi_Contact}}
\end{figure}

\Fref{fig:Fi_Contact} shows the jamming coverage, $p_\text{j}$, versus the aspect ratio, $\varepsilon$, for deposition of unoriented $k$-mers  ($s=0$) using the C$_\text{d}$ variant of RSA model at different fractions of forbidden contacts, $f_\text{d}$.  In absence of forbidden contacts, $f_\text{d}=0$ the model is equivalent to the simple RSA model\cite{Lebovka2015PRE}. For the given value of $\varepsilon$ increase of $f_\text{d}$ resulted in decrease of $p_\text{j}$.
The presented LK$_\text{d}$ and  C$_\text{d}$ variants of RSA model were more extensively studied for percolation behavior\cite{Lebovka2015PRE}.

%====================================================

\subsection{Dense jammed packing}\label{subsec:jampack}

In dense jammed packing neighboring particles have to touch themselves. In early studies for dense jammed packings of spheres, the three main models were proposed: the ordered close packing, random close packing, and random loose packing.
The ordered close packing of disks corresponds to the densest possible packing in two dimensions  with packing density (surface fraction) of $\phi  = \pi /\sqrt{12} \approx 0.9069$ in triangular lattice. The random close and random loose jammed packings were introduced less definitively and they can be experimentally obtained by random filling of container with particles using the shaking or without it and extrapolating the measured densities eliminating finite-size effects\cite{Berryman1983PRA}. The numerous early studies of random close packing in two dimensions estimated that values of $\phi$ fall in the range of $0.82-0.89$\cite{Stillinger1964JCP,Kausch1971JCIS,Visscher1972,Quickenden1974JCIS,Sutherland1977JCIS,Kenichi1980IJES,Shahinpoor1980PT,Schreiner1982JCSFT2,Shahinpoor1982JCIS}. However, most $2d$ packing protocols could produce highly ordered arrangements of disks. For example, the molecular dynamic simulations with growing in size particles allowed obtaining the covering of $\phi \approx 0.90$ and demonstrated existence of dense polycrystalline textures with irregular grain boundaries and linear shear fractures\cite{Lubachevsky1990JSP}. Recently the more general concepts of the locally jammed, collectively jammed, strictly jammed packings\cite{Torquato2010RMP} and ``maximally random jammed'' (MRJ) state was introduced\cite{Atkinson2014PNAS}.

\section{Percolation}\label{sec:percolation}
\subsection{Exact result: Percolation probability on torus}
Although essential part of results in percolation has been obtained by using computer simulations, they are supported by some important rigorous results and reliable assumptions. Using conformal field theory, Pinson proved\cite{Pinson1994JSP} that the probability, $R$, of finding a wrapping cluster on a torus at the percolation threshold is $R(p_\text{c}) = R^\ast$, where $0 < R^\ast < 1$. The value of $R^\ast$ depends on the way how the wrapping cluster is found (\tref{tab:R} presents $R^\ast$ with different number of exact digits). Here, the indices $h$ and $v$ correspond to the probabilities of wrapping horizontally or vertically around the system, respectively; the index $e$ corresponds to the probability of wrapping around either direction; the index $b$ corresponds to the probability of wrapping around both directions simultaneously; the index 1 corresponds to the probability of wrapping around one direction but not the other.
\begin{table}[ht]
\tbl{Percolation probability $R^\ast$.}
{
\begin{tabular}{@{}ccc@{}}
\toprule
Criterion & Refs.~\refcite{Newman2000PRL,Newman2001PRE} & Ref.~\refcite{Mertens2012PRE} \\
\colrule
$h,v$ & 0.521\,058\,290 & 0.521\,058\,289\,248\,821\,787\,848 \\
$b$ & 0.351\,642\,855 & 0.351\,642\,853\,927\,474\,898\,465 \\
$e$ & 0.690\,473\,725 & 0.690\,473\,724\,570\,168\,677\,230 \\
1 & 0.169\,415\,435 & 0.169\,415\,435\,321\,346\,889\,383\\
\botrule
\end{tabular}
}
\label{tab:R}
\end{table}

For any fixed value of the system size, the probabilities are connected with each other as follows\cite{Newman2001PRE,Mertens2012PRE}
\begin{align}\label{eq:relationsR}
R^\text{h} & = R^\text{v},\\
R^\text{e} & = 2R^\text{h} - R^\text{b} = 2R^1 + R^\text{b},\\
R^1 & = R^\text{h} - r^\text{b} = R^\text{e} - R^\text{h} = \frac12 \left( R^\text{e} - R^\text{b} \right)\\
R^\text{b} & \leqslant R^\text{h} \leqslant R^\text{e},\\
R^1 & \leqslant R^\text{h}.
\end{align}

For $L \to \infty$,
\begin{equation}\label{eq:R}
 R =
 \begin{cases}
 0, & \text{if } p<p_\text{c}, \\
 R^\ast, & \text{if } p=p_\text{c}, \\
 1, & \text{if } p>p_\text{c}.
 \end{cases}
\end{equation}

The union--find algorithm\cite{Newman2000PRL,Newman2001PRE} is widely used to check for any occurrences of wrapping clusters. This algorithm has initially been intended for a discrete space\cite{Newman2000PRL,Newman2001PRE} and then adapted for continuous percolation\cite{Li2009PRE,Mertens2012PRE}. For each given system size, $L$, and number of deposited particles, $N$, numerous independent runs should be performed to obtain the probability of percolation, $R^{(\text{c})}_{N,L}$. Here, the superscript $c$ means a used criterion, \emph{viz.}, $h, v$, or $b$ mean that the cluster winds the torus in the horizontal, or vertical direction, or in both directions, respectively.

\subsection{From microcanonical ensemble to canonical one}
To obtain the probability $R^{(\text{c})}(\eta,L)$ of percolation in the grand canonical ensemble, $R^{(\text{c})}_{N,L}$ should be convolved with the binomial distribution\cite{Newman2000PRL,Newman2001PRE} or the Poisson one\cite{Li2009PRE,Mertens2012PRE}, in the case of discrete and continuous space, respectively.

\subsubsection{Discrete space: Convolution with binomial distribution}
For a canonical percolation ensemble, the probability of there being exactly $N$ occupied sites on the lattice is given by the binomial distribution\cite{Newman2000PRL,Newman2001PRE}
\begin{equation}\label{eq:binom}
 B(N_0,N,p) = \binom{N_0}{N}p^N(1-p)^{N_0-N},
\end{equation}
where $N_0 = L^2$ is the total number of sites in the system under consideration.
Let an observable, \emph{e.g.}, a probability of percolation, is measured within the microcanonical ensemble for all values of $N$, giving a set of measurements $\{R_N\}$, then the value in the canonical ensemble will be given by
\begin{equation}\label{eq:binom2}
 R(p) =\sum_{N=0}^{N_0} B(N_0,N,p) R_n = \binom{N_0}{N}p^N(1-p)^{N_0-N} R_N.
\end{equation}
To find $R(p)$ for all $p$, $R_N$ is needed for all values of $N$.

\subsubsection{Continuous space: Convolution with Poisson distribution}
To obtain the probability $R(\eta,L)$ of percolation in the grand canonical ensemble with filling fraction $\eta$, the probabilities $R_{N,L}$ should be convolved with the Poisson distribution
\begin{equation}\label{eq:convolution}
 R(\eta,L)= \sum_{N=0}^\infty \frac{\lambda^N \mathrm{e}^{-\lambda}}{N!} R_{N,L},
\end{equation}
where $\lambda = \eta L^2 /a = n L^2$ is the mean\cite{Li2009PRE,Mertens2012PRE}.
The weights in \eref{eq:convolution}
$w_N (\lambda)={\lambda^N}/{N!}$
can be calculated using the recurrent relations\cite{Mertens2012PRE},
\begin{equation}\label{eq:weight1}
 w_{\overline{N}-i} =
 \begin{cases}
 1, & \mbox{for } i=0, \\
 (\overline{N}- i + 1)\lambda^{-1} w_{\overline{N} -i +1}, & \mbox{for } i=1,2,\dots,
 \end{cases}
\end{equation}
and
\begin{equation}\label{eq:weight2}
 w_{\overline{N}+i} =
 \begin{cases}
 1, & \mbox{for } i=0, \\
 \lambda(\overline{N}+ i)^{-1} w_{\overline{N} + i - 1}, & \mbox{for } i=1,2,\dots,
 \end{cases}
\end{equation}
herewith the relation
$$
\sum_{N=0}^\infty \frac{\lambda^N }{N!} = \sum_{N=0}^\infty w_N (\lambda) = \mathrm{e}^\lambda, \quad \forall \lambda > 0
$$
should be borne in mind. Here, $\overline{N} = \lfloor \lambda \rfloor$.
Therefore, the convolution can be calculated as
\begin{equation}\label{eq:RNL}
R(\eta,L)= \sum_{N=0}^\infty w^\ast_N(\lambda) R_{N,L},
\end{equation}
where
\begin{equation}\label{eq:wstar}
 w^\ast_N(\lambda) = \frac{w_N(\lambda)}{\sum\limits_{N=0}^\infty w_N(\lambda)}.
\end{equation}
The factor $\mathrm{e}^{-\lambda}$ is absent in the master equation \eref{eq:RNL}, since
$$
\sum_{N=0}^\infty w_N (\lambda) = \mathrm{e}^\lambda \sum_{N=0}^\infty w^\ast_N (\lambda).
$$
For a given system size $L$, the percolation threshold $\eta_\text{c}(L)$ is estimated by numerical solution of the equation
$$
R(\eta_\text{c},L) = R^\ast.
$$

\subsection{Scaling}
While rigorous results relate to the \emph{thermodynamic limit}, \emph{i.e.}, to infinitely large systems, any simulation deals with a finite-size system. The tool to build a bridge between results of computer simulation for a finite-size system ($L \times L$) and the thermodynamic limit ($L \to \infty$) is the \emph{scaling analysis} that allows eliminating a finite-size effect. A widely used approach to reduce a finite-size effect is elimination of boundaries by applied \emph{periodic boundary conditions} (PBCs), in fact, jamming and percolation are considered on a torus instead of a plane.

Conformal field theory gives exact values for the wrapping probabilities at the
transition in the thermodynamic limit\cite{Pinson1994JSP}. This theory provides the most efficient method to estimate the percolation threshold\cite{Newman2000PRL,Newman2001PRE,Li2009PRE,Mertens2012PRE} since
\begin{equation}\label{eq:scaling}
 \eta_\text{c}(\infty) - \eta_\text{c}(L) \propto L^{-2-1/\nu}, \text{ where } \nu= 4/3.
\end{equation}
The estimation of the standard deviation suggest
\begin{equation}\label{eq:std}
 \sigma \approx M^{-1/2} L^{-3/4},
\end{equation}
where $M$ is the number of independent runs\cite{Mertens2012PRE}. Since the error rapidly decreases as the system size increases, the insufficient number of the independent runs may be the main source of errors. For example, in Ref.\refcite{Newman2000PRL}, to obtain the value of the percolation threshold accurate to six decimal places, the number of independent runs was at least $3 \cdot 10^8$ while the system size was fairly small $L = 32, 64, 128,$ and 256.
%Typically, we used systems of sizes $L = 8, 16, 32, 64$ to perform the scaling analysis. The number of independent runs was $10^5$.

Unfortunately, conformal field theory gives exact values for the wrapping probabilities at the
transition in the thermodynamic limit only for isotropic systems\cite{Pinson1994JSP,Newman2000PRL,Newman2001PRE}. When a system is anisotropic, a different, less efficient, approach should be applied. For each particular value of $L$, the equation $R(n_\text{c},L) = 0.5$ should be solved numerically, e. g., using bisection or other appropriate method. Then, the scaling relation\cite{Stauffer}
\begin{equation}\label{eq:scaling43}
 n_\text{c}(\infty) - n_\text{c}(L) \propto L^{-1/\nu}
\end{equation}
can be applied to find the percolation threshold in the thermodynamic limit.

\begin{figure}[!htb]
 \centering
 \includegraphics[width=0.75\textwidth]{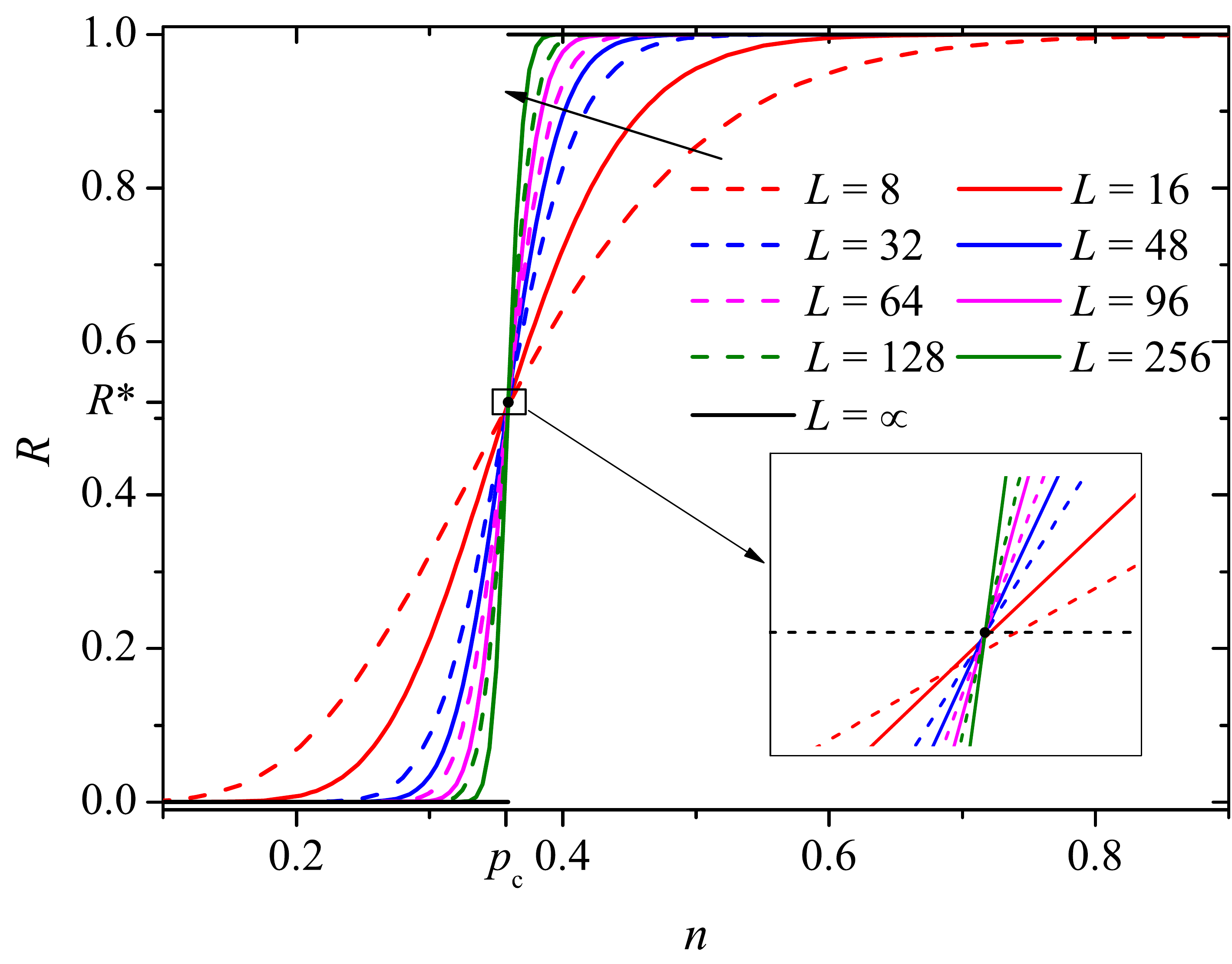}
 \caption{The percolation probability $R^\text{h}$ vs. number of discs per unit area, $n$. The percolation threshold in the thermodynamic limit, $p_\text{c}$, and the corresponding percolation probability $R^\ast$ are also shown.\label{fig:comparison}}
\end{figure}

\Fref{fig:comparison} demonstrates the behavior of $R(n)$ for discs with unit radius.

A scaling relation is also valid for jamming concentration
\begin{equation}\label{eq:scalingjam}
 n_\text{j}(\infty) - n_\text{j}(L) \propto L^{-1/\nu}, \quad \nu = 2/d,
\end{equation}
 where $d$ is the dimension of the space\cite{Vagberg2011PRE,Goodrich2016,Coniglio2017}.

\subsection{Continuous space}
\subsubsection{Isotropic systems}

\begin{figure}[!htb]
 \centering
 \includegraphics[width=0.75\textwidth]{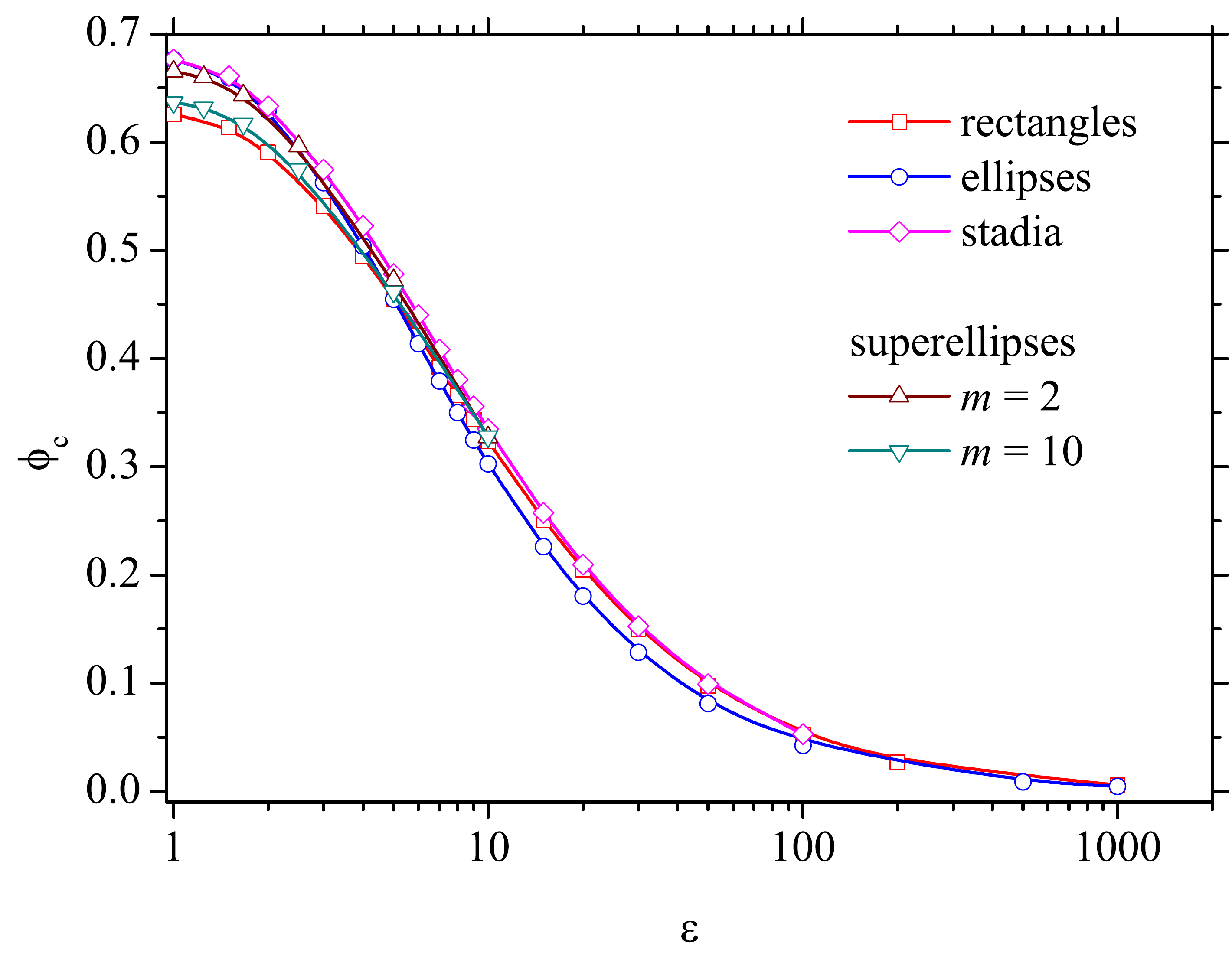}
 \caption{Comparison of the critical coverage $\phi_\text{c}$ vs the aspect ratio for different shapes, viz, rectangles\cite{Li2013PRE}, ellipses\cite{Li2016PhysA}, stadia\cite{Tarasevich2019Percolation}, and superellipses\cite{Lin201917PT}.\label{fig:phicvseps}}
\end{figure}

In this section, we consider square regions and isotropically oriented particles ($s=0$).
Percolation thresholds of two-dimensional continuum systems of rectangles\cite{Li2013PRE} and ellipses\cite{Li2016PhysA} for a wide range of aspect ratios from $\varepsilon = 1$ to $\varepsilon = 1000$ have been reported. Both ellipses and rectangles transform into sticks when $\varepsilon = \infty$. When $\varepsilon = 1$, a rectangle is simply a square, while an ellipse is a disc. Currently, the best known value of the percolation threshold of zero-width sticks of equal length that are randomly oriented and placed onto a plane, is
\begin{equation}\label{eq:ncstick}
 n_\text{c}^\times = 5.637\,285\,8(6)
\end{equation}
sticks per unit area\cite{Mertens2012PRE}. By convention, the value of $A_0$ in \eref{eq:fillingfraction} for sticks is taken as equal to $l^2$, where $l$ is the length of the stick. The best known value of the percolation threshold of discs, \emph{i.e.}, ellipses with $\varepsilon =1$, is $\eta_\text{c}^\circ = 1.12808737(6)$\cite{Mertens2012PRE}. Recently, percolation thresholds of superellipses have been reported\cite{Lin201917PT}. Percolation thresholds as the total fractions of the plane covered by the particles, $\phi_\text{c}$, have been presented for 14 shapes, for each of 6 aspect ratios\cite{Lin201917PT}. Recently, percolation thresholds for stadia with aspect ratios up to 100 have been reported\cite{Tarasevich2019Percolation}.
\Fref{fig:phicvseps} presents comparison of the critical coverage $\phi_\text{c}$ vs the aspect ratio for different shapes. \Fref{fig:phicvseps} evidences a weak dependence of the percolation threshold on particular particle shape.

\subsubsection{Anisotropic systems}
In this section, we consider square regions and anisotropically deposited particles ($s \geqslant 0$). Sometimes, anisotropic systems are assumed to mean rectangular systems\cite{Klatt2017JSMTE}.
A computer study of the percolation threshold in a two-dimensional anisotropic system of conducting sticks has been performed\cite{Balberg1983PRB}. Here, two kinds of angle distributions were taken into consideration, \emph{viz.}, uniform distribution within an interval \eref{eq:PDFinterval}, and the normal distribution \eref{eq:PDFnormal}; log-normal distribution of lengths was assumed. An analytical relationship between the critical density of sticks and anisotropy \eref{eq:Anisotropy} has been proposed. This relation predicts that the percolation threshold will increase with increasing anisotropy (the order parameter) from its isotropic value. Obviously, in a system of completely aligned, \emph{i.e.}, parallel, sticks, no percolation can occur. An important result of this article is that for an anisotropic two-dimensional system one should expect an isotropic percolation threshold but an anisotropic conductance, which disappears as the threshold is approached.

The conductivity of stick percolation clusters with anisotropic alignments has been studied by means of computer simulation and finite-size scaling analysis\cite{Yook2012JKPS}. The angular distribution of the sticks corresponded to \eref{eq:uniformangdistr}. The critical number density, $n_\text{c}$, does not vary much for $\theta_\text{m} \in (5\pi/18,\pi/2]$ while it changes rapidly as
\begin{equation}\label{eq:ncfit}
 n_\text{c} \sim \theta_\text{m}^{-0.9}
\end{equation}
for $\theta_\text{m} < 5\pi/18$. The percolation threshold (critical number density) increases rapidly as the anisotropy is increased. Taking into account \eref{eq:sinterval}, relation \eref{eq:ncfit} can be rewritten as
\begin{equation}\label{eq:asymp}
n_\text{c} \sim (1 - s)^{-0.45}.
\end{equation}

The finite continuum percolation of rectangles with different aspect ratios has been studied using the angular distribution \eref{eq:PDFcos}\cite{Klatt2017JSMTE}. The percolation in anisotropic systems, both for finite simulation boxes and in the limit of infinite system size has been investigated. The authors confirmed that there is no difference in effective percolation thresholds for different directions. This behavior is universal. Although, in finite systems, the difference between the effective percolation thresholds is observed, it vanishes in the thermodynamic limit. The value of the percolation threshold is non-universal, \emph{i.e.}, it depends on the anisotropy, \emph{viz.}, the more anisotropic the orientation distribution, the larger is the percolation threshold.

The percolation threshold for penetrable rectangles has been examined using the excluded area
between particles\cite{Chatterjee2015JSP}. The percolation threshold was found to rise with increases in the degree of particle alignment.
\begin{equation}\label{eq:Chatterjee}
n_\text{c} \ell^2 = \frac{ \varepsilon^2 }{ (  \varepsilon^2 + 1 ) \langle | \sin \theta | \rangle  +  \varepsilon ( 1 + 2 \langle | \cos \theta | \rangle) }.
\end{equation}

\begin{figure}[!htb]
  \centering
  \includegraphics[width=0.75\textwidth]{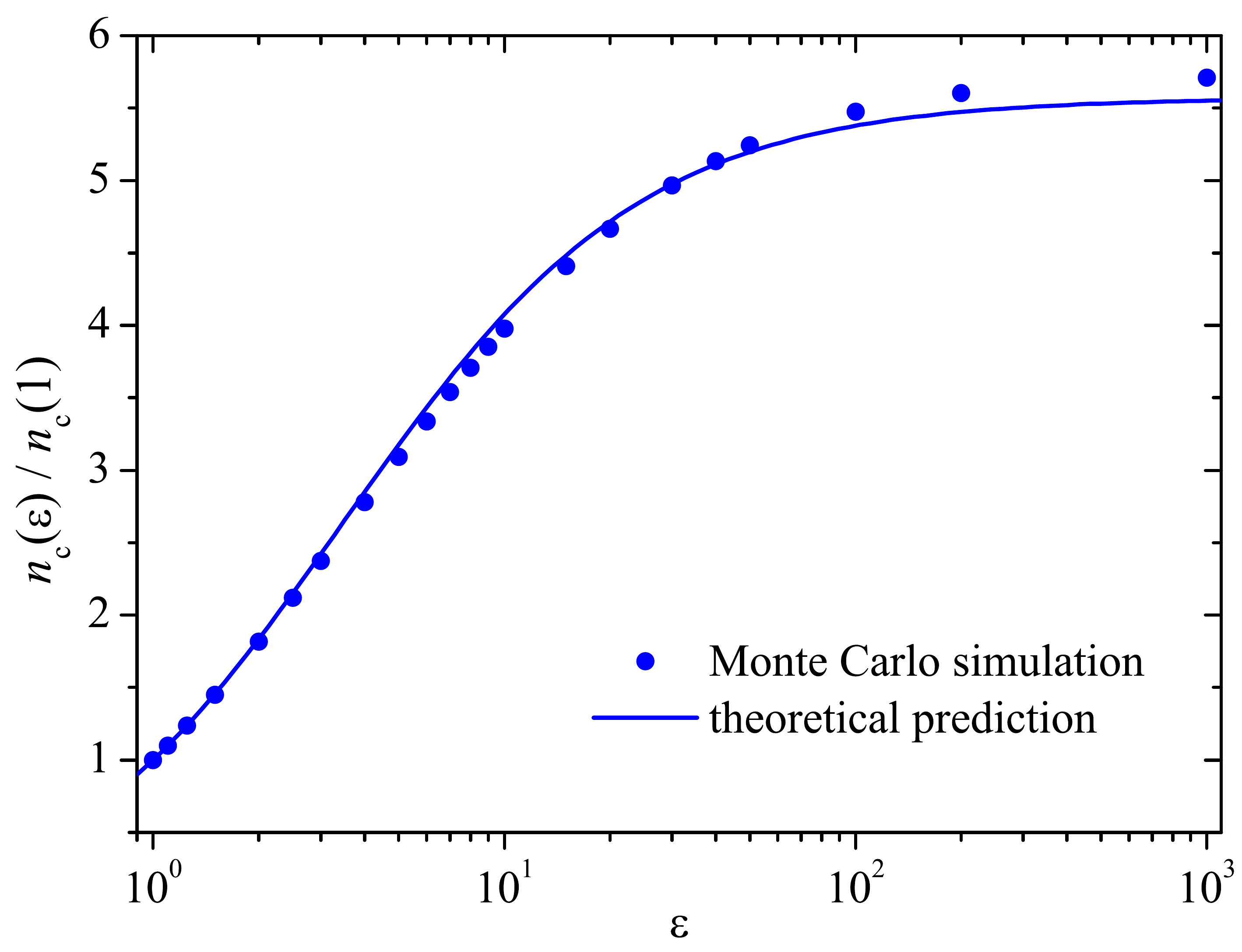}
  \caption{Normalized critical number density, $n_\text{c}/n_\text{c}(1)$ versus the aspect ratio, $\varepsilon$. Comparison of theoretical prediction\cite{Chatterjee2015JSP} and Monte Carlo simulation\cite{Li2013PRE}. The dependence of the percolation threshold on the aspect ratio $\varepsilon$ is shown for monodisperse, isotropically (randomly) oriented, penetrable rectangles.}\label{fig:rectangles-comp}
\end{figure}

\Fref{fig:rectangles-comp} compares the theoretical prediction \eref{eq:Chatterjee} and Monte Carlo simulation\cite{Li2013PRE} of the dependence of the percolation threshold on the aspect ratio $\varepsilon$ of monodisperse, isotropically (randomly) oriented, penetrable rectangles.

%Furthermore, the effect of the length dispersity of sticks on the percolation threshold has also been studied in several works. For instance, sticks with log-normal distributions of lengths were considered in\cite{Balberg1983PRB}.
The effects of length distribution, angular anisotropy, and wire curvature have been investigated both numerically and experimentally\cite{Langley2018NH}. Each of these quantities was assumed to be normally distributed. The percolation threshold decreases as either the length or the angle dispersity increases.
%Furthermore, the cooperative influence of both effects, simultaneously, on the percolation threshold may be of special interest.

Percolation in systems of aligned rods with different aspect ratios has been studied\cite{Chatterjee2014JCP}. Both systems of rods of equal length and systems consisting of mixtures of short and long rods were considered. For systems of equally-sized rods, the percolation threshold was found to be a monotonically increasing function of the orientational order parameter.

Numerical simulations of stick percolation have been performed\cite{Mietta2014JCP} with uniform angular distributions of the sticks within a given interval as well as with normal distributions, while the stick lengths corresponded to a log-normal distribution. The probabilities of percolation were presented for different values of the parameters.

The percolation of zero-width sticks on a plane paying special attention to the cooperative effects of both the alignment of sticks and their length dispersity on the percolation threshold has been studied using computer simulation\cite{Tarasevich2018PREsticks}. The critical number density increases as the order parameter increases. The curves were fitted by
\begin{equation}\label{eq:fitsticks}
 n_\text{c}(s) = \frac{n_\text{c}^\times}{\sqrt{1 - s^\alpha}},
\end{equation}
where the fitting coefficient $\alpha= 1.8449 \pm 0.0026$. Since percolation of parallel zero-width sticks is impossible for any finite value of the number density, $n_\text{c}(1) = \infty$. The asymptotic behavior of \eref{eq:fitsticks} $n_\text{c}(s \to 1)$ is similar to \eref{eq:asymp}.

\subsection{Discrete space}

One of the possible ways to simulate the adsorption of particles is based on the use of a discrete space. The simplest example of such the discrete space is a square lattice. In this case, the simplest example of an elongated particle is linear $k$-mer (also denoted as needle\cite{Vandewalle2000epjb}, line segment\cite{Leroyer1994PRB}, linear segment\cite{Kondrat2001PRE}, stiff-chain\cite{Becklehimer1992,Adamczyk2008JCP}, rod\cite{Longone2012PRE}, or stick\cite{Becklehimer1992}), \emph{i.e.}, a rectangular ``molecule'', which occupy $k$ successive faces of the lattice.

\subsubsection{Percolation of linear $k$-mers isotropically deposited on square lattice}\label{subsec:longkmers}
Study of percolation of long linear $k$-mers on square lattice has a long history. For $k \leqslant 20$,
the power-law dependence of the percolation threshold on the value of $k$ had been reported
\begin{equation}\label{eq:Becklehimer}
 p_\text{c}(k) \propto k^{-1/2},
\end{equation}
while the jamming coverage decreases\cite{Becklehimer1992}. Particle deposition was performed using a so-called ``end-on'' mechanism of RSA\cite{Evans1993RMP}

By contrast, non-monotonic dependence of the percolation threshold on $k$ was found for $k \leqslant 40$ deposited using conventional RSA\cite{Leroyer1994PRB}. For small values of $k$, the percolation threshold decreases as \begin{equation}\label{eq:Leroyer}
p_\text{c}(k) = k^{-1} + \mathrm{const}
\end{equation}
while, for larger values of $k$, it increases.

\begin{figure}[htbp]
  \centering
  \includegraphics[width=0.75\textwidth]{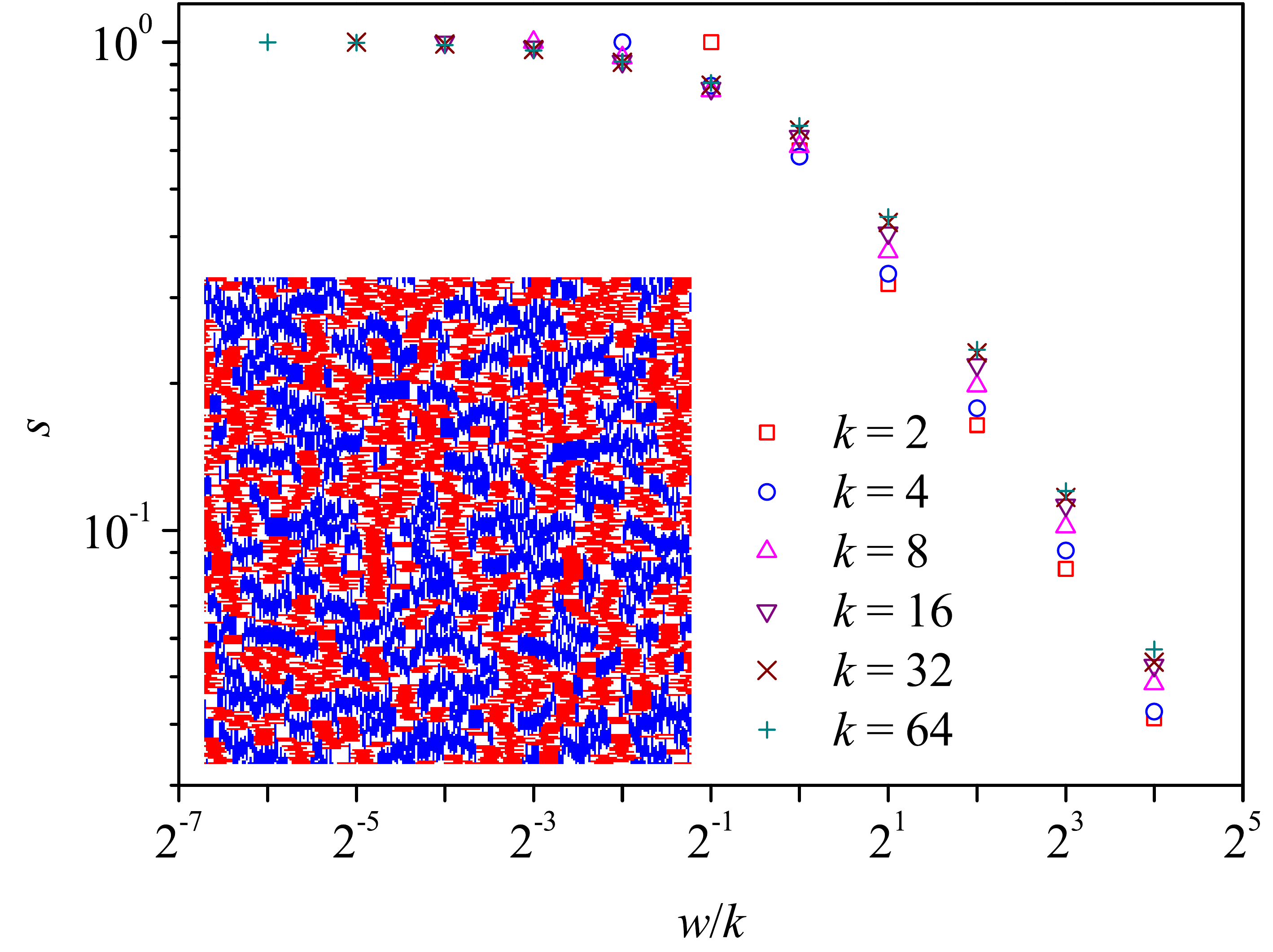}
  \caption{Dependency of the local order parameter, $s$, on the ratio $w/k$, where $w$ is the size of the window in which the local order parameter was calculated, $k$ is the needle length.\cite{Tarasevich2018MC} Inset: Example of a stack structure of a jammed state obtained using RSA of 10-mers onto a square lattice.}\label{fig:stacks}
\end{figure}

Minimal value of the percolation threshold correspond to $k \approx 15$. This behavior was argued in terms of the stack structure of the percolation cluster, \emph{viz.}, the percolation cluster is built of regions (stacks) of the same orientation, the typical size of which is $k$. The authors suggested that, in contrast to rectangles and squares where jamming saturation occurs before percolation\cite{Nakamura1987PRA}, the percolation threshold can be reached for any (finite) size of the $k$-mers. The local order parameter has been applied to study the internal structure of the percolation cluster. Recently, the stack structure of the jammed state was studied using the order parameter\cite{Tarasevich2018MC}. The characteristic size of each stack has been reported to be of the order of $k\times k$ (\fref{fig:stacks}). \Eref{eq:Becklehimer} and \eref{eq:Leroyer} describe quite different behaviors due to the use of different kinds of RSA\cite{Evans1993RMP,Leroyer1994PRB}.

The jamming and percolation of needles and squares on a square lattice have been investigated\cite{Vandewalle2000epjb}. The ratio $p_\text{c}/p_\text{j}$ was found to be a constant of $0.62 \pm 0.01$ up to $k=20$.
The authors suggested that both quantities scale as
\begin{equation}\label{eq:Vandewalle}
 p(k) = C\left[ 1 - \gamma\left( \frac{k - 1}{k}\right)^2 \right],
\end{equation}
where $\gamma = 0.31 \pm 0.01$ (except for the case $k=1$). The presented results evidence that the percolation threshold is significantly smaller than the jamming coverage when $k \to \infty$ if \eref{eq:Vandewalle} is valid for any values of $k$. Both at the percolation threshold and in the jammed state, the stack structure was observed.

The jamming and percolation of needles with lengths up to $k=2000$ have been investigated\cite{Kondrat2001PRE}. For $k>15$, an increase in the percolation threshold was found.
 For $15 \leqslant k \leqslant 45$, the fitting formula
\begin{equation}\label{eq:Kondrat}
 p_\text{c}/p_\text{j} \propto 0.50 + 0.13 \log_{10} k
\end{equation}
was proposed. A monotonic increase in $p_\text{c} /p_\text{j}$ holds over a wide range of values of $k$ even up to $k =2000$. The dependence of the jamming concentration on the value of $k$ was fitted by
\begin{equation}\label{eq:jamming}
 p_\text{j}(k) = p_\text{j}(\infty) + a / k^\alpha,
\end{equation}
where $p_\text{j}(\infty) = 0.66 \pm 0.01$, $a = 0.44$, $\alpha = 0.77$. The parameters of the fitting formula were refined in Ref.~\refcite{Lebovka2011PRE} as $p_\text{j}(\infty) = 0.655 \pm 0.009$, $a = 0.416$, $\alpha = 0.720 \pm 0.007$ based on simulations and scaling analysis up to $k=256$. It was also found that
\begin{equation}\label{eq:Tarasevichpc}
 p_\text{c}(k) = a_0 / k^{\alpha_0} + b \log_{10} k + c,
\end{equation}
where $a_0 = 0.36 \pm 0.02$, $\alpha_0 =0.81 \pm 0.12$, $b = 0.08 \pm 0.01$, and $c =0.33 \pm 0.02$\cite{Tarasevich2012PRE}. For the ratio $p_\text{c}/p_\text{j}$, a fitting formula was proposed
\begin{equation}\label{eq:Tarasevichpcpj}
p_\text{c}/p_\text{j} = B \log_{10} k + C,
\end{equation}
where $B= 0.119 \pm 0.003$ and $C = 0.513 \pm 0.006$\cite{Tarasevich2012PRE}.

The monotonic behavior of the percolation threshold as a function of $k$ for $k \leqslant 15$ was reported and the following fitting formula was proposed
\begin{equation}\label{eq:Cornette}
 p_\text{c}(k) = p_\text{c}^\ast + \Omega \exp\left( -\frac{k}{\kappa}\right),
\end{equation}
$p_\text{c}^\ast = 0.461 \pm 0.001$, $\Omega = 0.197 \pm 0.02$, and $\kappa = 2.775 \pm 0.02$\cite{Cornette2003epjb}.

\begin{figure}[htbp]
\centering
\includegraphics[width=0.75\textwidth]{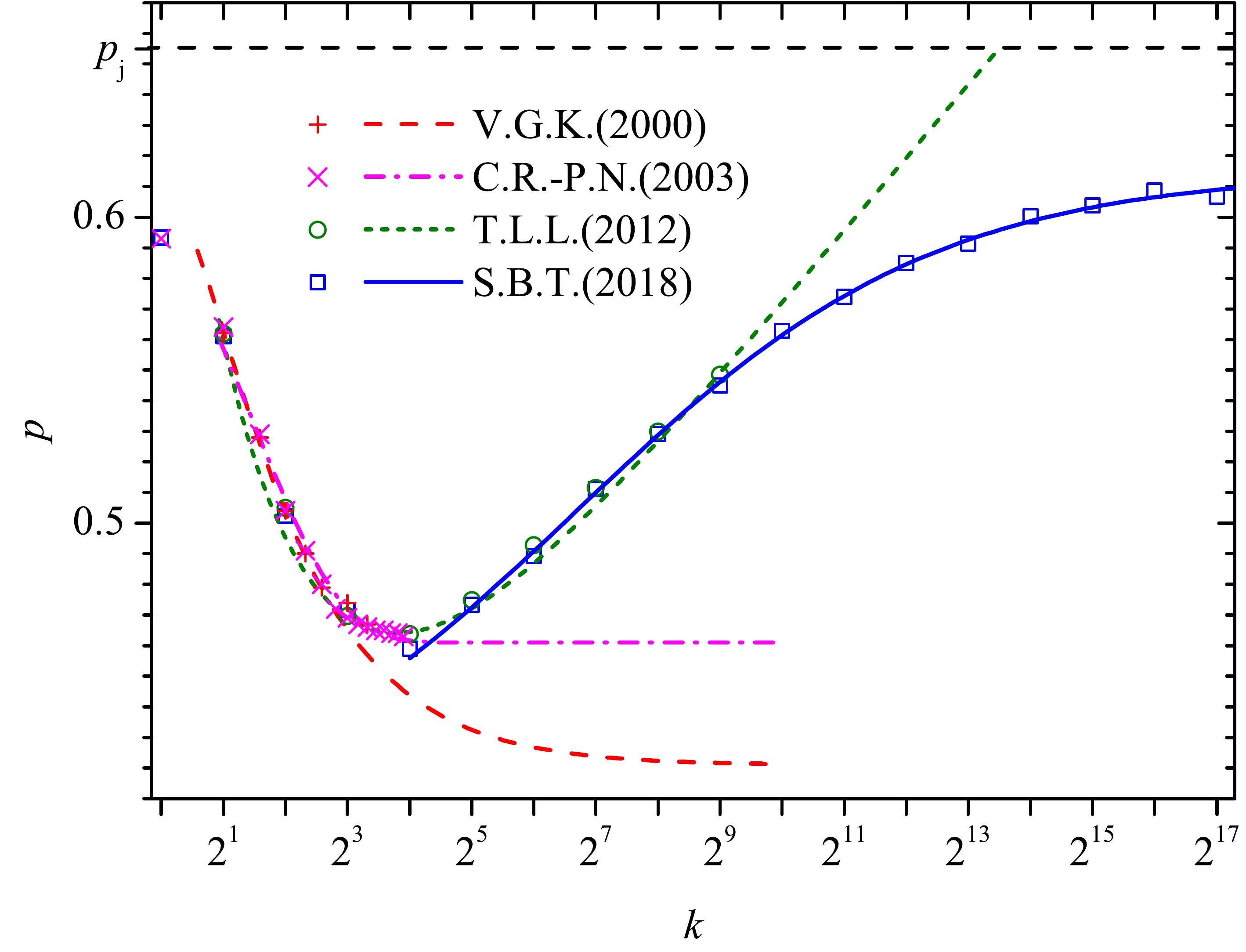}\\
\caption{Dependence of $p_\text{c}(k)$ for linear $k$-mers isotropically deposited using RSA onto a square lattice with PBCs. The data were collected from different sources, \emph{viz.},
 V.G.K.(2000)\cite{Vandewalle2000epjb}, C.R.-P.N.(2003)\cite{Cornette2003epjb},
 T.L.L.(2012)\cite{Tarasevich2012PRE}, S.B.T.(2018)\cite{Slutskii2018PRE}.
\label{fig:densities}}
\end{figure}

The percolation threshold and the jamming coverage for lengths of $k$-mers up to $2^{17}$ has been studied by means of numerical simulations\cite{Slutskii2018PRE}. A parallel algorithm has been developed and applied to the isotropic RSA samples of large linear $k$-mers on a square lattice with PBCs.
For $2^4 \leqslant k \leqslant 2^{17}$, the data can be fitted by
\begin{equation}\label{eq:fit}
 p_\text{c} = A + \frac{B}{ C + \sqrt{k} },
\end{equation}
where $A = 0.615 \pm 0.001$, $B = -2.26 \pm 0.09$, $C = 10.2 \pm 0.6$, the adjusted coefficient of determination is $R^2 =0.999$\cite{Slutskii2018PRE}. The percolation thresholds are shown in \fref{fig:densities}. Known approximations \eref{eq:Vandewalle}, \eref{eq:Tarasevichpc}, \eref{eq:Cornette} and \eref{eq:fit} are also shown for comparison.

A proof was presented that there will be a percolating cluster in any jammed configuration of nonoverlapping fixed-length, horizontal, or vertical needles on a square lattice\cite{Kondrat2017PRE}. The theorem ensures that the ratio of the percolation threshold and jamming coverage $p_\text{c}/p_\text{j}$ is well defined for all needle lengths. The theorem was proved only for the case of rigid boundaries. Shortly, the theorem was also proved for the case of PBCs\cite{Slutskii2018PRE}. The theorem refutes the conjecture\cite{Tarasevich2012PRE,Tarasevich2015PRE,Centres2015JSM} that in the RSA of such needles on a square lattice, percolation does not occur if the needle length exceeds a critical value $k^*$ of the order of several thousands.

\subsubsection{Percolation of linear $k$-mers anisotropically deposited on square lattice}\label{subsubsec:anislatt}

\begin{figure}
  \centering
  \includegraphics[width=0.75\textwidth]{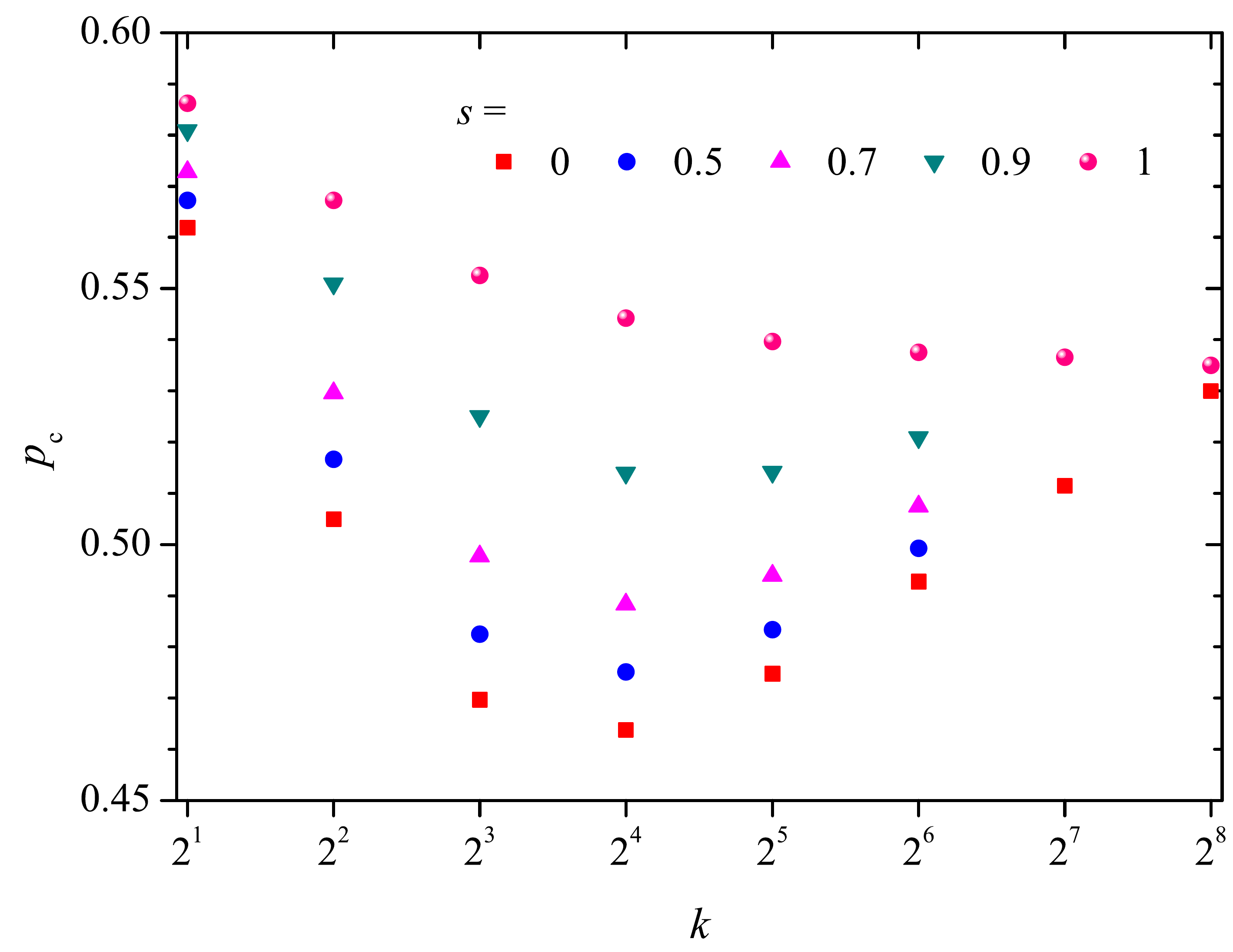}
  \caption{Dependence of the percolation threshold $p_\text{c}$ on the order parameter $s$ for $k$-mers of different length.\cite{Tarasevich2012PRE}}\label{fig:pcvss}
\end{figure}

The percolation of partially aligned linear $k$-mers on a square lattice with PBCs has been investigated by using computer simulations\cite{Tarasevich2012PRE}. The increase of the system ordering always led to an increase of the percolation threshold. For completely ordered systems ($s=1$), the percolation threshold $p_\text{c}$ monotonically decreased as $k$ increased (\fref{fig:pcvss}). For isotropic and partially ordered systems ($0 \leqslant s < 1$), the percolation threshold $p_\text{c}$ was a nonmonotonic function of $k$, \emph{viz.}, a minimum of $p_\text{c}$ has been observed for a certain length of $k$-mers, $k \approx 2^4$ (\fref{fig:pcvss}).

Jamming and percolation in anisotropic systems of elongated objects up to $k=100$ have been studied.\cite{Romiszowski2013} As the needle length increases, the values of the percolation threshold decreases and after that stage increases depending on the values of the order parameter. For a strongly ordered system containing needles, the ratio of jamming and percolation thresholds is almost independent on the needle length.

\section{Conclusion}

Our consideration of jamming and percolation is restricted only to elongated stiff rod-like particles. The real-world systems, including the polymers, biomolecules, nanotubes or nanowires the different shapes\cite{Budinski-Petkovic2005PRE,Budinski-Petkovic2008PRE,Budinski-Petkovic2012,Budinski-Petkovic2017PRE}, including flexible\cite{Kondrat2002JCP,Adamczyk2008JCP,Adamczyk2009JCP,Zerko2012SM,Sikorski2011JMM} and branched\cite{Dorenbos2015JCP} chains, are also of interest. In this review have omitted  considerations of many common problems related with packing and ordering in mixtures, vibrated and agitated systems, effects of confinement, condensation-evaporation equilibria, phase behaviors, relaxation to equilibrium state, etc. An inquisitive reader can find additional information in the original and review articles\cite{Araujo2014EPJST,Saberi2015PR,Frenkel2014,Gonzalez-Pinto2017,Windows-Yule2017} and books\cite{Stauffer1992,Sahimi1994}.

%==================

In addition, we are paying our main attention to a square lattice when dealing with a discrete space. Jamming and percolation produced by means of the RSA of $k$-mers has also been studied on other kinds of substrates, \emph{e.g.}, on triangular lattices\cite{Budinski-Petkovic2012,Perino2017JSMTE}. A non-monotonic size dependence of the percolation threshold and decrease of the jamming coverage have also been observed on such triangular lattices\cite{Perino2017JSMTE}.

\section*{Acknowledgements}
We acknowledge funding from the National Academy of Sciences of Ukraine, Projects No.~0117U004046 and~43/19-H (N.I.L.), the Ministry of Science and Higher Education of the Russian Federation, Project No.~3.959.2017/4.6 and Russian Foundation for Basic Research, Project No.~18-07-00343 (Yu.Yu.T.).

\bibliographystyle{ws-rv-van}
\bibliography{percolation,Jam}

%\blankpage
%\printindex[aindx]   % to print author index
%\printindex    % to print subject index

\end{document}